\documentclass[twocolumn, aps,
english,
superscriptaddress,
prl,
amsmath,
amssymb,
floatfix,
longbibliography,
10pt
]{revtex4-1}
\usepackage[T1]{fontenc}
\usepackage[utf8]{inputenc}
\usepackage{color}
\usepackage[english]{babel}
\usepackage{amsmath}
\usepackage{mathrsfs}
\usepackage{graphicx}
\usepackage{cancel}
\usepackage{esint}
\usepackage{float}
\usepackage{tikz}
\usepackage{empheq}
\usepackage[export]{adjustbox}
\usepackage[unicode=true,pdfusetitle,bookmarks=true,bookmarksnumbered=false,bookmarksopen=false,breaklinks=false,pdfborder={0 0 0},pdfborderstyle={},backref=false,colorlinks=true]{hyperref}

\makeatletter

\usepackage{dcolumn}
\usepackage{bm}
\usepackage{color}
\usepackage{soul}
\usepackage{amsfonts}
\usepackage{slashed}
\usepackage{enumerate}
\usepackage{mathtools}
\usepackage{physics}
\usepackage{amsthm}
\usepackage{inputenc}
\usepackage[normalem]{ulem}
\usepackage{svg}

\let\old@makecaption=\@makecaption
\usepackage{subcaption}
\let\@makecaption=\old@makecaption

\usepackage[figurename=FIG.]{caption}

\usepackage{epsfig}
\usepackage{dsfont}
\usepackage{xcolor}

\makeatother
\begin{document}

\title{Correlations and Krylov spread for a non-Hermitian Hamiltonian: \\
Ising chain with a complex-valued transverse magnetic field}
\author{E. Medina-Guerra}
\affiliation{Department of Condensed Matter Physics, Weizmann Institute of Science, Rehovot 7610001, Israel}
\author{I. V. Gornyi}
\affiliation{\mbox{Institute for Quantum Materials and Technologies, Karlsruhe Institute of Technology, 76131 Karlsruhe, Germany}}
\affiliation{\mbox{Institut für Theorie der Kondensierten Materie, Karlsruhe Institute of Technology, 76131 Karlsruhe, Germany}}

\author{Yuval Gefen}
\affiliation{Department of Condensed Matter Physics, Weizmann Institute of Science, Rehovot 7610001, Israel}
\date{\today}

\begin{abstract}
Krylov complexity measures the spread of an evolved state in a natural basis, induced by the generator of the dynamics and the initial state, and serves as an effective probe for detecting various properties of many-body systems. Here, we study the spread in Hilbert space of the state of an Ising chain subject to a complex-valued transverse magnetic field, initialized in a trivial product state with all spins pointing down. We demonstrate that Krylov spread reveals structural features of many-body systems, which remain hidden in correlation functions that are traditionally employed to determine phase diagrams. When the imaginary part of the spectrum of the non-Hermitian Hamiltonian is gapped, the system’s state asymptotically approaches the non-Hermitian Bogoliubov vacuum for this Hamiltonian.  We find that the spread of this evolution unravels three different dynamical phases based on how the spread reaches its infinite-time value. We also show that the second derivatives of the spread with respect to the real and imaginary components of the complex magnetic field exhibit algebraic divergence at the transition between gapped and gapless imaginary spectra. Furthermore, we establish a connection between the Krylov spread and the static correlation function for the $z$-components of spins in the underlying non-Hermitian Bogoliubov vacuum, providing a full analytical characterization of correlations across the phase diagram. Specifically, for a gapped imaginary spectrum in a finite magnetic field, we find that the correlation function exhibits an oscillatory behavior that decays exponentially in space. Conversely, for a gapless imaginary spectrum, the correlation function displays an oscillatory behavior with an amplitude that decays algebraically in space; the underlying power law depends on the manifestation of two exceptional points within this phase.
Our findings, showing the potential of the Krylov spread in exposing dynamical transitions, motivate further studies of Krylov's complexity in open many-body systems, including monitored ones.
\end{abstract}

\maketitle

\section{Introduction}
\label{sec:intro}

The description of quantum systems via one-parameter unitary groups---i.e., when the dynamics is generated by a unitary operator governed by a Hermitian Hamiltonian and parametrized by time---is insufficient once the interaction with (possibly unknown) environments
is considered. 
This inevitably more realistic situation requires implementing more general mathematical tools, such as the theory of semigroups of completely positive maps \cite{KRAUS1971311,barchielli1986measurement}, for which the former scenario is just a special case. Within this more general framework, deterministic or stochastic master equations replace the Schr{\"o}dinger equation as the equation governing the dynamics of the system \cite{breuer2002theory} (see Ref.~\cite{Fazio2024} for a recent review). The dynamics of certain classes of systems interacting with Markovian environments are adequately described by the Lindblad equation
\begin{equation}\label{eq:x01}
    \partial_t\rho(t)\!=\!-i[H,\rho(t)] + \sum_{i=1}^{d}\gamma_i\!\left(\! L_i\rho(t)L_i^\dagger - \frac{1}{2}\{L_i^\dagger L_i,\rho(t) \}\! \right),
\end{equation}
where $[.\,,.]$ and $\{.\,,.\}$ denote the commutator and anticommutator, respectively, and the role of the environment is encoded in the jump operators $L_i$ and the dissipation rates $\gamma_i$. The solution of the Lindblad equation can be understood as describing the expectation of a stochastic density matrix $\omega(t)$ of a monitored quantum system, i.e., $\rho(t) = \mathbb{E}[\omega(t)]$ \cite{barchielli1986measurement,breuer2002theory,Fazio2024,attal2010stochastic,pellegrini2009diffusion,yiep}. In particular, if the open system undergoes a counting measurement, $\omega(t)$ corresponds to a quantum trajectory of detections and instants of detections of the form $(j_1,t_1;\ldots;j_n,t_n)$, where $j_i\in \{1,\dots, d\}$ specifies the type of count detected at time $t_i$. Given this trajectory up to time $t>t_n$, the equation determining the a-posteriori state is  given by the Poissonian stochastic master equation 
\begin{multline}\label{eq:xsto}
    \dd \omega (t)\! =\! -i[H,\omega(t)]\dd t -\! \dfrac{1}{2}\!\sum_{i=1}^d\{L_i L_i^\dagger - \expval*{L_i^\dagger L_i}_t,\omega(t) \} \dd t \\
    + \sum_{i=1}^d\Bigg(\dfrac{L_i\omega(t) L_i^\dagger }{\expval*{L_i^\dagger L_i}_t}-\omega(t) \Bigg)\dd N_i(t),
\end{multline}
where $\{\dd N_i(t) \}_i$ are the Poissonian increments describing the detections and $\langle L_i^\dagger L_i \rangle_t\coloneqq \mathrm{Tr}[L_i^\dagger L_i\omega(t)]$. Here, the evolution between any two measurements occurring at $t_{i_1}$ and $t_{i_2}>t_{i_1}$ ($i_2>i_1$) is dictated by the  non-Hermitian Hamiltonian arising from Eq.~\eqref{eq:xsto} when $\dd N_i(t) \equiv 0$ for all $i$ and $t\in [t_{i_1},t_{i_2})$, \begin{equation}H_{\text{eff}} = H - \frac{i}{2}\sum_{i}\gamma_i L_i^\dagger L_i.
\end{equation}
Then, in this time interval, the evolution of the trajectory $\omega(t)$ is given by
\begin{equation}\label{eq:xeff}
   i \partial_t\omega(t) = H_{\text{eff}}\,\omega(t) - \omega(t)H_{\text{eff}}, \qquad t\in [t_{i_1},t_{t_2}].
\end{equation}
Moreover, if $\omega(t_{i_1}) = \ket{\psi(t_{i_1})}\bra{\psi(t_{i_1})}$, the solution to Eq.~\eqref{eq:xeff} is equivalent to
\begin{equation}\label{eq:xeff1}
       \ket{\psi(t)} = \dfrac{\exp[-i H_{\text{eff}}\, (t-t_{i_1})]\ket{\psi(t_{i_1})}}{\norm{\exp[-i H_{\text{eff}}\, (t-t_{i_1})]\ket{\psi(t_{i_1})}}},
\end{equation}
which describes the evolution of a pure system under a non-Hermitian Hamiltonian. The conditions leading to Eq.~\eqref{eq:xeff1} justify the implementation of non-Hermitian Hamiltonians in the problem of monitored systems.

Notwithstanding the fact that describing the evolution of an open system via Eq.~\eqref{eq:xeff1} implies considerable simplifications and assumptions, effective non-Hermitian Hamiltonians provide an enormously rich arena of novel phenomena that are not present in their Hermitian counterparts. 
The immediate difference and cornerstone of non-Hermitian Hamiltonians is the existence of complex eigenvalues and exceptional points in the parameter space where the eigenvalues and eigenstates coalesce \cite{kato2013perturbation}.  For example, such points can be responsible for unconventional phase transitions~\cite{bender_-symmetric_1999, heralded}, the non-Hermitian skin effect \cite{zhang_universal_2022, ent3, lin_topological_2023,kunst1}, or chiral topological conversion in the slow non-Hermitian dynamics (see Ref.~\cite{kumar2025} for a recent overview). 
Moreover, near (and at) exceptional points, correlation functions can display anomalous power-law behavior \cite{Correlations_at_higher} and faster spatial decay than their Hermitian counterparts \cite{correlations_at_pt}.
Gaps in the complex spectrum, where both the real and imaginary parts of the spectrum may be gapless or gapful, also play an essential role in the physics of non-Hermitian systems, such as in, e.g., in topological band theory \cite{many_body_topology, hermitian_bulk_non-hermitian, exceptional_topology}. 
Various critical
phenomena, such as entanglement phase transitions or dynamical phase transitions, among many others \cite{ent1, zerba_measurement_2023, Turkeshi_1, Turkeshi_2, ent2, kattel2024spin, kattel2024dissipation, biorthogonal, dora2023work, mondal_persistent_2024, zhang2024comprehensive,leela1,leela2,kunst3}, can be associated with the peculiarities of the complex spectra in non-Hermitian systems. 
In general, phases of non-Hermitian systems can be classified via symmetry and topology \cite{topological_phases_of,
many_body_topology,
exceptional_topology,
hermitian_bulk_non-hermitian,
Chen2025,kunst2,kunst4,kunst5}. 
We point out that all of the phenomena mentioned above can be found not only in systems governed by non-Hermitian Hamiltonians but also in generic open quantum systems (in particular, monitored ones), see, e.g., Refs.~ \cite{Snizhko2020, pavlov2024topological,
free-fermions-one-dim,
localization,
abo2024liouvillian,
syk-lindbladians,
skin_lindb, kawabata,roccati,lenke}.

Much of the characterization of non-Hermitian systems is in line with their Hermitian counterparts, where, e.g., correlation functions, entanglement entropy, Loschmidt echo \cite{pasta1,pasta2}, among other measures, are used. In addition to these traditional quantities, Krylov's complexity has been recently introduced as an alternative tool (mainly in Hermitian systems). It determines the spread of an evolved state (or operator) in a natural basis, called the Krylov basis, constructed from the initial state and the generator of the dynamics. 
This notion of complexity was initially put forward to study the operator spread in chaotic systems \cite{parker2019universal} and has been implemented as a probe of chaos and integrability \cite{assesing_the_saturation, quantifying_operator, erdmenger_universal_2023, hashimoto_krylov_2023, integrability_to_chaos,rabinovici2022k,Jeong1,Jeong4,Jeong5,Jeong6}, topological phases \cite{caputa_spread_2023,caputa2022quantum}, dynamical phase transitions \cite{krylov_dynamical_phase_bento,xia2024complexity,Afrasiar_2023}, among other phenomena and applications, see, e.g., \cite{Jeong2,Jeong3,ref2bata}. For example, the Krylov complexity of an integrable system is much smaller than that of an ergodic system since states of the latter explore the underlying Krylov space formed by the Krylov basis as much as possible. In general, an important advantage of utilizing such a measure is the low computational cost of finding the subspace in which an initial system evolves under the action of a generator of dynamics. This complexity measure is straightforward to calculate if the generator of the dynamics belongs to a semisimple Lie algebra \cite{caputa2022geometry} or a semidirect sum of semisimple Lie algebras \cite{natural_basis}.
Furthermore, the Krylov basis offers yet another appealing feature. Being the basis that minimizes the spread over all possible bases of an evolved state \cite{quantum_chaos_and_complexity}, it naturally serves as a natural tool to study quantum speed limits \cite{hornedal2022ultimate,Carabba2022quantumspeedlimits,gill2024speed,speed_limits}.

Krylov's complexity has also been studied in open quantum systems \cite{liu2022krylov, op1, op3, bhattacharya2022operator, bhattacharya2023krylov}, where it has been found to be significantly suppressed compared to its behavior in their Hermitian counterparts. 
We note that, in general, the Krylov complexity of systems governed by Eq.~\eqref{eq:xeff1} remains largely unexplored (see, however, Refs.~\cite{bhattacharya_spread_2024, nandy2024krylovspaceapproachsingular, ganguli2024spreadcomplexitynonhermitianmanybody}). 
While the Krylov spread has been employed as a probe for characterizing known dynamical phase transitions, we are unaware of this measure being used as a tool to detect hitherto ``hidden'' dynamical phases of non-Hermitian systems. This is the focus of this work.

Specifically, we demonstrate that the Krylov spread reveals new dynamical phases of the one-dimensional (1D) Ising model with a transverse, single-component, complex-valued magnetic field of the form $h+i\gamma/4$. This model is known to have two main phases that depend on the gap or lack thereof of the imaginary part of the complex spectrum of the non-Hermitian Hamiltonian. The two distinct phases manifest through several features. For instance, the entanglement entropy of the non-Hermitian vacuum obeys a logarithmic law in the gapless phase and an area law in the gapped phase  \cite{Turkeshi_2,
zerba_measurement_2023}. Similar transitions have also been studied for variations of the above models in, e.g., Refs.~\cite{many_body_phase_,entanglement_trns_perido}, where the gapped nature of the spectrum dictates the scaling behavior of the entanglement entropy. Another distinctive feature of the two phases is the asymptotic behavior of correlation functions. Specifically, the static correlation function of the $z$-components of spins, calculated in the non-Hermitian vacuum, exhibits an oscillatory behavior with an algebraic decay in the gapless phase, while in the other phase, it displays an exponential decay. 

Going beyond the two main phases, here, through the employment of \emph{Krylov spread fidelity}, we identify and characterize the previously unnoticed dynamical phases of a non-Hermitian Ising chain. The Krylov spread fidelity is defined as the absolute difference between the infinite-time limit and the time-dependent value of the Krylov spread density in the thermodynamic limit when the imaginary part of the spectrum is gapped. In the regime where the fidelity tends to zero, we identify three specific time scales that arise from the interplay between the real and imaginary parts of the transverse magnetic field. The former, alongside the hopping terms of the Hamiltonian, governs the unitary evolution of the system, while the latter generates the dissipative dynamics. Importantly, the identified characteristic time scales refine the region of the $h$-$\gamma$ plane where the imaginary part of the spectrum is gapped by splitting it into distinct dynamical phases characterized by different time dependences of the Krylov fidelity.

The paper is organized as follows. In Sec.~\ref{sec:spectrum}, we define the model and diagonalize the Hamiltonian, choosing a sign convention such that the imaginary part of the spectrum is always negative. Analogously to the Hermitian case where $\gamma = 0$ \cite{kam_coherent_2023}, we show that the ground state of the non-Hermitian Hamiltonian is a generalized coherent state of the Jordan-Wigner vacuum. We define a stationary-state regime based on how the Jordan-Wigner vacuum reaches the non-Hermitian Bogoliubov vacuum when the imaginary part of the spectrum is gapped. For a gapless spectrum, we demonstrate that the stationary state of the same initial state is a linear combination of time-dependent coefficients of the non-Hermitian vacuum and two non-Hermitian excitations over the same vacuum.   
 
In Sec.~\ref{sec:correlation_functions}, we analyze, both in the gapless and gapped phases, the asymptotic behavior (at large spatial distance $x$) of the static spin-spin correlation function $\texttt{C}^{zz}(x)$ for $z$-components of spins. In the gapless phase, we find that the correlation function decays algebraically, displaying an oscillatory behavior with a power-law exponent that changes depending on the emergence of two exceptional points. The same correlation function was calculated in Ref.~\cite{Turkeshi_1} for the XY model only in one subregion of the gapless phase where the exceptional points are absent.  In the gapped phase, the asymptotic behavior of $\texttt{C}^{zz}(x)$ is characterized by an exponential decay, with oscillations on top of it. To the best of our knowledge, this correlation function was calculated in this phase only for a vanishing real magnetic field in the XY-chain~\cite{heralded}. The correlation functions for other spin components have been numerically calculated in Refs.~\cite{Liu_2021, yang_hidden_2022}.

In Sec.~\ref{sec:krylov_spread}, we establish a connection between the Krylov spread and the correlation function  $\texttt{C}^{zz}(x)$  when the Jordan-Wigner vacuum evolves unitarily to the non-Hermitian Bogoliubov vacuum. 
Furthermore, we demonstrate that the spread undergoes a third-order phase transition when the imaginary part of the spectrum goes from being gapless to gapped at the critical value $h_c$ of the real part of a magnetic field. We also confirm and expand the results presented in Ref.~\cite{caputa_spread_2023}, where the spread in a Kitaev chain via unitary dynamics was calculated. We explicitly calculate the spread in the oscillatory gapless phase where $(h,\gamma)< (J,0)$ and show that its first derivative with respect to $h$ diverges logarithmically as $h$ approaches the value of the hopping term $J$ at $\gamma = 0$.

In Sec.~\ref{Sec:V-DPT}, we analyze the time dependence of the Krylov spread across the entire phase diagram. We define the Krylov spread fidelity density, which reveals three distinct dynamical phases existing in the gapped phase when the Jordan-Wigner vacuum evolves under the non-Hermitian Hamiltonian to the non-Hermitian Bogoliubov vacuum in the infinite-time limit. We demonstrate how this spread is related to the one calculated in the previous section. In Fig.~\ref{fig:4}, we present the refined phase diagram of the non-Hermitian 1D Ising model, including the dynamical ``Krylov phases'', which is one of the key results of our work.

Finally, the insights into the dynamics of non-Hermitian quantum systems, which this paper gained, and the outlook are presented in Sec.~\ref{sec:conclusions}. Technical details of calculations are relegated to Appendices.

\section{Model: Spectrum and non-Hermitian vacuum}\label{sec:spectrum}

The model we study in this work is an Ising chain with $N$ sites and nearest-neighbor XX-interactions, subject to a complex magnetic field in the $z$-direction acting at each site. The non-Hermitian Hamiltonian of this 1D transverse-field Ising model is
\begin{equation}\label{eq:x1}
    H = -\sum_{i= 1}^N\left[J\sigma_i^x \sigma_{i+1}^x + \left( h + i\dfrac{\gamma}{4} \right)\sigma_i^z\right],
\end{equation}
where $J>0$ is the hopping strength and $h+i\gamma/4$ is the complex magnetic field with $h, \gamma \geq 0$. Even with this complex magnetic field, the system is still integrable. It can be easily diagonalized via the Jordan-Wigner transformation, followed by a non-Hermitian Bogoliubov rotation, as we show in the following subsection, cf. Refs.~\cite{zerba_measurement_2023,biella_many-body_2021,Liu_2021,FERNANDEZ2023169429,NguyenBaAn_1988}.  Afterward, in a situation akin to the Hermitian case, we demonstrate that the ground state of $H$, or its vacuum, is a coherent state defined in coset space $\text{SU}(2)^{\otimes N}/\text{U}(1)^{\otimes N}$ \cite{kam_coherent_2023, perelomov_generalized_1986}. At the end of this section, we will find an explicit expression for the time-evolved state
\begin{equation}\label{eq:x1.1}
    \ket{\psi(t)} = \dfrac{\exp(-iHt)\ket{\psi(0)}}{\norm{\exp(-iHt)\ket{\psi(0)}}},
\end{equation}
where $\ket{\psi(0)} = \ket{\downarrow\ldots \downarrow}$ is the state of all $N$ spins pointing down with $\sigma^z\ket{\downarrow} = -\ket{\downarrow}$.

\subsection{\texorpdfstring{Spectrum of $H$}{Spectrum of H}}

Let us start by introducing the conventional Jordan-Wigner (JW) transformation that expresses spin operators via fermionic operators:
\begin{align}\label{eq:x2}
    \sigma_j^x = \exp(i\pi\sum_{l=1}^{j-1}\hat{n}_l)\left(c_j+c_j^\dagger\right), 
    \quad\
     \sigma_j^z = 1-2c_j^\dagger c_j, 
\end{align}
where $\hat{n}_i = c_i^\dagger c_i$, 
$\{c_i,c_j^\dagger \}=\delta_{ij}$ and $\{c_i,c_j \}=\{c_i^\dagger,c_j^\dagger \}=0$. Upon performing the JW transformation of $H$, one gets 
\begin{equation}\label{eq:x3}
    H\!=\!-J\! \sum_{j=1}^{N}(c_j^\dagger c_{j+1}+c_j^\dagger c_{j+1}^\dagger + \text{H.c.}\!) -\left(h +i \frac{\gamma}{4} \right)\!\sum_{j=1}^N(1-2\hat{n}_j),
\end{equation}
where $c_{N+1} = e^{i\phi}c_1$ is the generalized boundary condition. Note that $e^{i\phi} = 1$ or $e^{i\phi} = -1$ for periodic and anti-periodic boundary conditions (PBC and ABC, respectively). For concreteness, we pick the latter. 

To diagonalize $H$, we implement the following Fourier representation of the JW operators
\begin{equation}\label{eq:x5}
   c_j = \frac{e^{-i\pi/4}}{\sqrt{N}}\sum_{k}e^{ik j}c_k,
\end{equation}
where we have set the lattice spacing to unity, and the wave numbers belong to the following set:
\begin{equation}\label{eq:x6}
    \mathcal{K}\!=\!\left\lbrace k_n=\dfrac{2\pi}{N}\left(\!- \Bigg\lceil \frac{N-1}{2}\Bigg\rceil + n +\frac{1}{2}\right)\!: n \in [0,N)\cap\mathbb{N} \right\rbrace\!.
\end{equation}
Here, the ceiling function $\lceil x \rceil$ denotes the least integer that is greater than or equal to $x$.

\begin{widetext}
Using the Fourier representation of the JW operators, after standard manipulations, we get 
\begin{equation}\label{eq:x7}
   H = \sum_{k \in \mathcal{K}^+} \Bigg\lbrace\left[2\left(h-J\cos k\right)+i\dfrac{\gamma}{2} 
 \right]\left(c_k^\dagger c_k - c_{-k}c_{-k}^\dagger \right)+ 2J\sin k\, \left( c_k^\dagger c_{-k}^\dagger +c_{-k}c_k \right) \Bigg\rbrace,
\end{equation}
where $\mathcal{K}^+ \coloneqq  \mathcal  K \cap \mathbb{R}_{\geq 0}$.
Since we are working with ABC, the subspaces associated with each $k\in \mathcal K^+$  are two-dimensional. The Hamiltonian can be written as $ H = \sum_{k \in \mathcal{K}^+}\hat{H}(k)$, where
\begin{equation}\label{eq:x8}
     \hat{H}(k) = \Psi_k^\dagger M(k)\Psi_k =  \begin{pmatrix}
       c^\dagger_k & c_{-k}
   \end{pmatrix} \begin{pmatrix}
       2(h-J\cos k)+i\dfrac{\gamma}{2} & 2J\sin k \\
       2J\sin k &  -2(h-J\cos k)-i\dfrac{\gamma}{2}
   \end{pmatrix}\begin{pmatrix}
       c_k \\ c_{-k}^\dagger
   \end{pmatrix}.
\end{equation}
\end{widetext}
The symmetric matrix $M(k)$ can be diagonalized with a similarity transformation using orthogonal matrices: $V_{k}^{-1}M(k)V_k = \text{diag}[{\Lambda(k)},-{\Lambda(k)}]$. For our particular model, the presence of a complex magnetic field implies that the eigenvalues of $M(k)$ are complex:
\begin{subequations}\label{eq:x09}
\begin{align}
 {\Lambda(k)} &= E(k)+i\Gamma(k)\label{eq:x9.1}\\
 &= \sqrt{4\left( h-J\cos k + i\dfrac{\gamma}{4}\right)^2 + 4J^2\sin^2 k}. \label{eq:x9}
\end{align}
\end{subequations}

\begin{figure*}
\includegraphics[width=0.9\textwidth]
{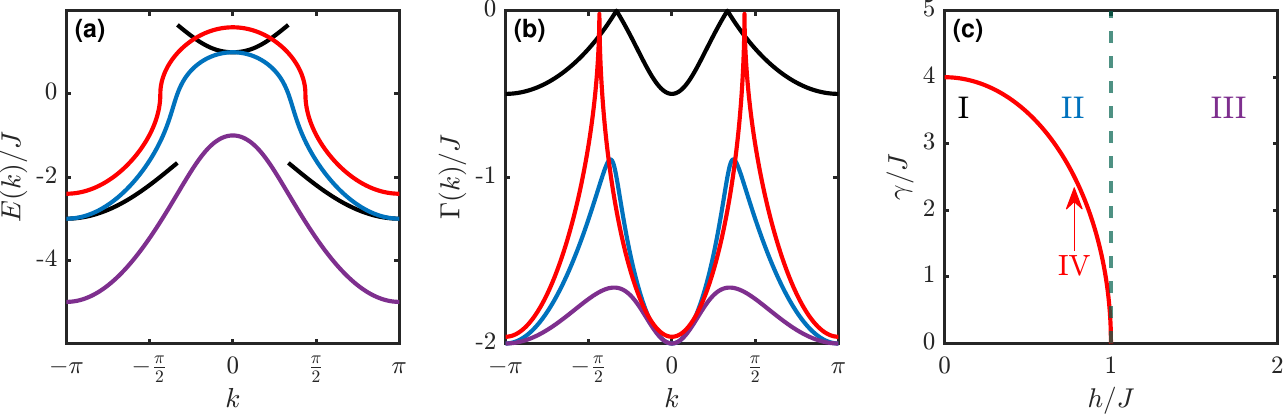}
        \caption{Examples of the real, $E(k)$ (a), and the imaginary, $\Gamma(k)$ (b), parts of the spectrum \eqref{eq:x09} satisfying the convention  $\mathfrak{Im}\,\Gamma(k) \leq 0$ for all $k$ in units of $J$, exemplifying cases (I)-(IV) discussed in the main text. (I): $h= 0.5$, $\gamma = 1$ (black curves). (II): $h = 1/2$, $\gamma = 4$ (blue). (III): $h = 1.5$, $\gamma = 4$ (purple). (IV): $h = 0.2$, $\gamma = \gamma_c(h) = 4\sqrt{1-h^2} \approx 3.91$ (red). Panel (c) shows the $h$-$\gamma$ plane with the four regions corresponding to cases (I)-(IV) indicated. The green dashed line separates regions (II) and (III) of the phase where $\Gamma(k)$ is gapped. The critical curve $\gamma=\gamma_c(h)$ corresponding to case (IV) is shown in red. } 
        \label{fig:1}
\end{figure*}

As in the Hermitian case, the sign of the square root in $\Lambda(k)$ for each $k$ depends on the convention (or ``picture'') chosen to describe the excitations on the vacuum \cite{lieb_two_1961}. In what follows, we choose the sign such that $\Gamma(k) = \mathfrak{Im}\,{\Lambda}(k) \leq 0$ $\forall k\in \mathcal K$. This guarantees that the eigenvalue associated with the ground state of $H$ has the largest imaginary part. This way, this state is the one associated with the slowest decay in time. Our chosen sign convention forces the sign of $E(k)$ to change along the $k$-axis, as shown in Fig.~\ref{fig:1}a. This sign convention can be conveniently cast by choosing the positive root in Eq.~\eqref{eq:x9} for all $k$ and multiplying it with $\text{sgn}(J\cos k - h)$, where $\text{sgn}(0)=1.$ The real and imaginary parts of the spectrum can either individually or jointly vanish at $\pm q = \pm \arccos(h/J)$, where 
\begin{equation}\label{eq:x12}
     \Lambda(\pm q) = \sqrt{4J^2\sin^2q-\dfrac{\gamma^2}{4}} = J \sqrt{1 - \frac{h^2}{J^2}-\frac{\gamma^2}{16J^2}}.
\end{equation}
From this expression, we can identify the critical curve in the $h$-$\gamma$ plane 
\begin{equation}\label{eq:x13}
    \gamma_c(h) = 4J\sqrt{1-\frac{h^2}{J^2}}, \qquad h<J,
\end{equation}
where $\Lambda(\pm q) =0$ ($\gamma_c=0$ for $h>J$). 

The properties of $\Lambda(k)$ define the following phase diagram in the $h$-$\gamma$ plane (with $J$ fixed) characterized by the regions where the real and imaginary parts of the spectrum, $E(k)$ and $\Gamma(k)$, respectively, are gapless or gapful. Specifically, we identify the following four regions, defined by the intervals along the $h$ and $\gamma$ axes $\{h\}\times \{\gamma\}$:
\begin{enumerate}[(I)]
    \item $ [0,J)\times (0,\gamma_c(h))$. Below the critical curve (\ref{eq:x13}),
    $\Gamma(k)$ is gapless at $k = \pm q$, i.e., $\Gamma(\pm q) = 0$, and $E(k)$ is gapped, i.e., $E(k) \neq 0$ $\forall k\in \mathcal{K}$.
    \item $[0,J]\times (\gamma_c(h),\infty)$.  Above the critical curve, $\Gamma(k)$ is gapped, and $E(k)$ is gapless at $k =\pm q$. 
    \item  $(J,\infty)\times(0,\infty)$. Both $\Gamma(k)$ and $E(k)$ are gapped.
    \item $[0,J)\times \gamma_c(h)$. Exactly on the critical curve, $\Gamma(k)$ and $E(k)$ are gapless, and thus the spectrum has two exceptional points at $\pm q$. 
    \end{enumerate}
If $\gamma = 0$, then $\Gamma(k) \equiv 0$, and $E(k)$ is gapped for $h \in [0,1)\cup (1,\infty)$ and gapless for $h = J$.

Henceforth, we refer to the cases $\gamma > \gamma_c(h)$ and $0<\gamma \leq \gamma_c(h)$ as the gapped and gapless phases, respectively. In each of these phases, with a slight abuse of language, we say that the spectrum of $H$ is gapped and gapless, respectively. The characteristic examples of the behavior of $E(k)$ and $\Gamma(k)$ in the four cases (I)-(IV), as well as the ``phase diagram'' comprising these regions, are presented in Fig.~\ref{fig:1}.

Continuing with the diagonalization of $H$,
let 
\begin{equation}\label{eq:x14}
    \Vec{R}(k) = \begin{pmatrix}
        R_x(k) \\
        R_y(k) \\
        R_z(k)
    \end{pmatrix} =\begin{pmatrix}
        2J\sin k \\
        0\\
        2(h-J\cos k) + i\dfrac{\gamma}{2}
    \end{pmatrix}
\end{equation}
be the complex Bloch vector associated with $M(k)$ from Eq.~\eqref{eq:x8}. Furthermore, let \begin{equation}\label{eq:x15}
    V_k = \begin{pmatrix}
        u_k & -v_k \\
        v_k & u_k
    \end{pmatrix}
\end{equation}
be an orthogonal matrix such that $V_k^{-1}\hat{H}(k)V_k = \text{diag}[\Lambda(k),-\Lambda(k)]$ with 
\begin{equation}\label{eq:x16}
    u_k = \frac{\Lambda(k) + R_z(k)}{N(k)}, \quad\text{and}\quad v_k = \frac{R_x(k)}{N(k)},
\end{equation}
where $N(k)$ is a complex normalization factor such that $u_k^2+v_k^2=1$. With the aid of $V_k$, we get
\begin{equation}\label{eq:x17}
    \hat{H}(k) = \begin{pmatrix}
        c^\dagger_k & c_{-k}
    \end{pmatrix}\, V_k\, \text{diag}[\Lambda(k),-\Lambda(k)] \, V_k^{-1}\begin{pmatrix}
        c_k\\ c_{-k}^\dagger
    \end{pmatrix}.
\end{equation}
The introduction of the orthogonal matrix $V_k$ allows us to define the non-Hermitian fermionic operators
\begin{equation}\label{eq:x18}
  \begin{pmatrix}
      \eta_k \\
      \eta_{-k}^*
  \end{pmatrix} = \begin{pmatrix}
        u_k & v_k \\
        -v_k & u_k
    \end{pmatrix}\begin{pmatrix}
        c_k \\
        c_{-k}^\dagger
    \end{pmatrix} = \begin{pmatrix}
        u_k c_k + v_kc^\dagger_{-k}\\
        -v_kc_k+u_kc_{-k}^\dagger \end{pmatrix}.
\end{equation}
These operators satisfy
\begin{equation}\label{eq:x19}
    \{\eta_k ^*,  \eta_q \} = \delta_{k q},\quad \{\ \eta_k, \eta_q \} = \{ {\eta }^*_k ,  \eta^*_q \} = 0
\end{equation}
and $\eta_k^* \neq \eta^\dagger_k$.
Thus, $H$ can be written in a diagonal form in terms of these operators as
\begin{equation}\label{eq:x20}
    H = \sum_{k\in \mathcal{K}^+}\Lambda(k)\left( \eta_k^*\eta_k-\eta_{-k}\eta^*_{-k}\right).
\end{equation}

Our next goal is to find the state satisfying $\eta_k \ket{\Omega} = 0$, which is the \emph{right} non-Hermitian Bogoliubov vacuum or (right) ground state of $H$. Similarly, one can find the \emph{left} non-Hermitian Bogoliubov vacuum, which satisfies $(\eta_k^*)^\dagger \ket{\Xi} = 0.$ It is worth recalling that in non-Hermitian systems, the left and right eigenvalues are different from each other and need not be orthogonal. This implies that expectation values can be calculated over the different vacua or their excitations; namely, when calculating the expectation value of an operator $\hat{O}$ with respect to the vacua of $H$, one can consider $\bra{\Xi}\hat{O}\ket{\Omega}$, $\bra{\Xi}\hat{O}\ket{\Xi}$ or $\bra{\Omega}\hat{O}\ket{
\Omega}$ \cite{meden_pt-symmetric_2023,yamamoto1,yamamoto2,Brody_2014,castro-alvaredo_spin_2009}. The latter two are called biorthogonal correlation functions. For definiteness, we will calculate expectation values only in the right vacuum. 

\subsection{Non-Hermitian Bogoliubov vacuum}\label{ssec:non_hermitian_vacuum}

We proceed to demonstrate that the right vacuum $\ket{\Omega}$ is a generalized coherent state of the JW vacuum $\ket{0}$ (i.e., $c_i\ket{0}=0)$ defined in the coset space $\text{SU}(2)^{\otimes N}/\text{U}(1)^{\otimes N}$.
To this end, 
let us return to Eq.~\eqref{eq:x7} and define the following operators:
\begin{equation}\label{eq:x21}
    J_{z}(k)\coloneqq \frac{1}{2}\left( c_k^\dagger c_k-c_{-k}c_{-k}^\dagger \right),
\end{equation}
and 
\begin{equation}\label{eq:x22}
   J_{+}(k)\coloneqq c_k^\dagger c_{-k}^\dagger,\quad  J_{-}(k)\coloneqq c_{-k}c_k.
\end{equation}
By implementing the anticommutation relations obeyed by the fermionic operators $\{c_k \}_{k\in \mathcal{K}}$, it is straightforward to demonstrate that the above operators satisfy the multiplication table of the $\mathfrak{su}(2)$ Lie algebra for each $k$
\begin{align}\label{eq:x23}
    [J_z(k),J_{\pm}(q)] &= \pm \delta_{kq}J_{\pm}(k), \\
    [J_+(k),J_-(q)]&=2\delta_{kq}J_z(k).
\end{align}
Hence, Eq.~\eqref{eq:x7} can be written as an element of $\mathfrak{su}(2)^{\otimes N}$,
\begin{align}\label{eq:x24}
     H &=  \sum_{k\in \mathcal K ^+}\hat{H}(k)\notag \\ &= \sum_{k\in \mathcal K ^+}\big\lbrace 2R_z(k)J_z(k)+ R_x(k)\left[ J_+(k)+J_-(k)\right] \big\rbrace.
\end{align}
Given this Lie-algebraic representation of the Hamiltonian, we proceed to find the right non-Hermitian vacuum.

Let $\mathsf g \in \mathsf{G} =  \mathsf{SU}(2)^{\otimes N}$ be an arbitrary group element in an appropriate unitary representation, and 
\begin{equation}\label{eq:x25}
    \ket{0} = \bigotimes_{i=1}^N\ket{\downarrow_i} = \bigotimes_{k \in \mathcal K}\ket{1/2,-1/2}_k
\end{equation}
be the JW vacuum defined by $c_k\ket{0}=0$ ($\forall k \in \mathcal{K}$) with $J_-(k)\ket{1/2,-1/2}_k=0$. Then, if the action of $\mathsf g$ on $\ket{0}$ satisfies $\ket{\Phi} = \mathsf g \ket{0} = \Omega \, \mathsf h \ket{0} = e^{i\varphi(\mathsf h)}\Omega \ket {0}$, where $\mathsf h \in \mathsf{H}$ is an element of the isotropy group of $\ket{0}$ and $\Omega \in \mathsf{G}/\mathsf{H}$ is a unitary operator, then $\Omega$ defines a \emph{generalized coherent state} of $\ket{0}$ via
\begin{equation}\label{eq:x26}
    \ket{\Omega} \coloneqq \Omega \ket{0}.
\end{equation}
In our case, the isotropy group is $\mathsf{H} = U(1)^{\otimes N}$, as the action of an arbitrary element of this group, such as, e.g., 
\begin{equation}\label{eq:x27}
\prod_{k\in \mathcal K^+}\exp[\alpha_z(k)J_z(k)] \in U(1)^{\otimes N},  
\end{equation}
on $\ket{0}$ produces only a global phase. Therefore, the generic form of our coherent state is given by
\begin{subequations}
\label{eq:x28}
\begin{align}
    \ket{\Omega} &= \Omega\ket{0}  = \prod_{k \in \mathcal K^+}\Omega_k\ket{0}\label{eq:x28.2} \\
    &=  \prod_{k \in \mathcal K^+}\exp\big[\alpha_+(k) J_+(k) + \alpha_-(k)J_-(k)\big]\ket{0},\label{eq:x28.3}
\end{align}
\end{subequations}
where $\alpha_{\pm}(k)$ are complex functions of $k$. This coherent state coincides with the right non-Hermitian vacuum of \eqref{eq:x20} if (see Appendix~\ref{sec:equivalence_formulas})
\begin{equation}\label{eq:x29}
    \alpha_+(k) = -\alpha_-^*(k) = -\exp(i \text{Arg} \, \tau_k)\,\arccos\dfrac{1}{\sqrt{1+\abs{\tau_k}^2}},
\end{equation}
where 
\begin{equation}
\tau_k \coloneqq \dfrac{v_k}{u_k}.
\label{eq:tauk}
\end{equation} 
The reason for this choice is that the vacuum found traditionally is 
\begin{subequations}
\label{eq:x30}
\begin{align}
    \ket{\Omega} &= \prod_{k\in \mathcal K^+} \frac{\eta_{-k}\eta_k}{\bra{0}\eta^\dagger _k \eta^\dagger_{-k}\eta_{-k}\eta_k\ket{0}^{1/2}}\ket{0}\label{eq:x30.1}\\
    &= \prod_{k\in \mathcal{K}^+}\dfrac{1-\tau_k c_k^\dagger c_{-k}^\dagger}{\sqrt{1+\abs{\tau_k}^2}}\ket{0}. \label{eq:x30.2}
\end{align}
\end{subequations}
Here, the order of the products of the form $\eta_{-k}\eta_k$ does not matter since they commute for different $k$. Owing to this, it is straightforward to verify that the non-Hermitian right vacuum satisfies the following conditions:
\begin{equation}\label{eq:x31}
    \eta_k\ket{\Omega} = 0,  \quad \bra{\Omega}\eta^*_k \neq 0, \quad \bra{\Omega}\eta_{-k} \neq 0.
\end{equation}
Also, the left vacuum $\ket{\Xi}$ can be constructed following similar steps as above. It is worth noting that the first condition can also be obtained by noting that (Appendix \ref{sec:equivalence_formulas})
\begin{equation}\label{eq:x32}
    \hat{c}_k \coloneqq \Omega c_k \Omega^{-1} = \Omega_k c_k \Omega^{-1}_k = \tilde u _k c_k + \tilde v_k c^\dagger_{-k},
\end{equation}
where 
\begin{align}\label{eq:x33}
    \tilde u_k = \frac{1}{\sqrt{1+\abs{\tau_k}^2}}, \quad \tilde v_k = \dfrac{\tau_k}{\sqrt{1+\abs{\tau_k}^2}}.
\end{align}
Then,  
 $\hat{c}_k\ket{\Omega} = \Omega_k c_k\ket{0}=0$ implies that $\eta_{k}$ also annihilates $\ket{\Omega}$, since
\begin{equation}\label{eq:x34}
    \eta_k = \sqrt{1+\abs{\tau_k}^2}\,u_k\,  \hat{c}_k.
\end{equation}
With this relation, the state $\ket{\Omega}$ can be thought of as a state populated either by Hermitian or non-Hermitian quasiparticles created by products of the form $\hat{c}_{-k}\hat{c}_k$ or $\eta_{-k}\eta_k$, respectively, acting on  $\ket{0}$. Moreover, the excited non-Hermitian particle states created on the right vacuum $\ket{\Omega}$ are of the form \begin{equation}
\ket{\Psi({\{n_k\}})} = \prod_{k \in \mathbb{K}}M_k\hat \eta^*_k \ket{\Omega}
\end{equation}
with $\mathbb{K} = \{k \in \mathcal K: n_k = 1 \}$, and $\sum_{k\in \mathbb K} n_k = N_F$ with $N_F \in 2\mathbb{Z}_{>0}$. These are thus the right eigenstates of $H$ with eigenvalues 
\begin{equation}
E(\{n_k \}) = \sum_{ k \in \mathbb{K}}\Lambda(k) +E_\Omega,
\end{equation}
where  $E_\Omega = - \sum_{k\in \mathcal{K}^+}\Lambda(k)$ is the energy of the non-Hermitian right vacuum. Note that the sign convention we adopted for $\Gamma(k)$ implies $\mathfrak{Im}\, E_\Omega \geq 0$.

\subsection{\texorpdfstring{Time evolution of JW vacuum $\ket{0}$}{Time-evolution of \#0}}
\label{eq:ssec:time_evolution_of}

In this section, we explicitly find the time evolution of $\ket{0}$ under $H$ and demonstrate that if $\Gamma(k)$ is gapped, the stationary state (to be defined below) is $$\ket{\psi_{\text{st}}(t)} \simeq \ket{\Omega}.$$ 
If, however, $\Gamma(k)$ is gapless, then $\ket{\psi_{\text{st}}(t)}$ does not reduce to the Bogoliubov vacuum and involves the excited state 
$\ket{q,-q}\propto \eta_{-q}^*\eta_q^*\ket{\Omega}$ populated with the non-decaying quasiparticles corresponding to the modes 
\begin{equation}
    \pm q = \pm \arccos \dfrac{h}{J}.
    \label{eq:pmq}
\end{equation}   
Above, the sign $\simeq$ indicates leading-order contributions (see below).

Let $\hat{H}(k)$ be as in Eq.~\eqref{eq:x24}, and let us consider $\exp[-i\hat{H}(k)t]$ acting on $\ket{0}$. Applying the normal-order decomposition formula of $\mathfrak{su}(2)$ \cite{Ban:93} yields
\begin{widetext}
\begin{equation}
    \exp[-i\hat{H}(k)t]\!\ket{0}\!= \exp\!\Big(\!-i \left\{2R_z(k) J_z(k) + R_x(k)[J_+(k)+J_-(k)] \right\}t\Big)\!\ket{0}\! = \exp[-\frac{1}{2}\ln A_z(k;t)\!]\exp[A_+(k;t)J_+(k)]\!\ket{0}, \label{eq:x35}
\end{equation}
\end{widetext}
where
\begin{equation}
    A_z(k;t) = \left[ \cos \Lambda(k)t  +i\dfrac{R_z(k)}{\Lambda(k)}\sin \Lambda(k)t \right]^{-2} 
    \label{eq:Az}
\end{equation}   
and
\begin{equation}
    A_+(k;t) = -\dfrac{i[R_x(k)/\Lambda(k)]\sin \Lambda(k)t }{\cos \Lambda(k) t +i[R_z(k)/\Lambda(k)]\sin \Lambda(k)t }. \label{eq:x36}
\end{equation}
Hence, the normalized evolved state $\ket{0}$ is, up to a phase, 
\begin{subequations}
\label{eq:x37}
\begin{align}
    \ket{\psi(t)} &= \frac{\exp(-i H t)\ket{0}}{\norm{\exp(-i H t)\ket{0}}} \label{eq:x37.1}\\ &= \prod_{k\in \mathcal K^+}\frac{\exp[A_+(k;t)J_+(k)]}{\sqrt{1+\abs{A_+(k;t)}^2}}\ket{0}. \label{eq:x37.2}
\end{align}
\end{subequations}
From this expression, we define the stationary state for times satisfying  (see Appendix~\ref{sec:stationary_state})
\begin{equation}\label{eq:x38}
  t \gg   t_\star \coloneqq \max_{k \in \mathcal{K}^+\setminus\{q \}} \dfrac{1}{2\abs{\Gamma(k)}}\ln\left( m_k^2 + \sqrt{m_k^2 - \beta_k} \right),
\end{equation}
where
\begin{equation}\label{eq:x39}
     \beta_k = \dfrac{4\left[1-w_k(\Omega)\right]\left(1 - \abs{f_k}\cos\vartheta_k\right)}{\abs{f_k}^2} + 1,
\end{equation}
with 
\begin{equation}
    f_k =\abs{f_k}e^{i\vartheta_k} 
    =1+R_z(k)/\Lambda(k),
    \label{eq:fk}
\end{equation}
\begin{equation}\label{eq:x40}
    m_k \coloneqq \left\lbrace 1 + \dfrac{4[1-w_k(\Omega)]}{\abs{f_k}} \left( \dfrac{[1-w_k(\Omega)]}{\abs{f_k}} - \cos\vartheta_k\right)\right\rbrace^{1/2},
\end{equation}
and
\begin{equation}
w_k(\Omega) = \frac{\abs{\tau_k}^2}{1+\abs{\tau_k}^2}.
\end{equation}
Clearly, if there are non-decaying modes, i.e., $\Gamma(\pm q) = 0$, then $t_\star \rightarrow \infty$ as $k \rightarrow \pm q$.  Returning to Eq.~\eqref{eq:x37}, if the spectrum is gapped, then (see Appendix~\ref{sec:stationary_state})
\begin{align}
 & \ket{\psi_{\text{st}}(t)}\coloneqq   \ket{\psi(t\gg t_\star)}\notag  \\
    &\quad = \prod_{k\in \mathcal{K}^+}\dfrac{\exp[-\tau_k J_+(k)]}{\sqrt{1+\abs{\tau_k}^2}}\ket{0} + \mathcal{O}\left(\prod_{k\in \mathcal{K}^+}l_k(t)\right),\label{eq:x41}
\end{align}
where 
\begin{equation}
    l_k(t) = \frac{2}{f_k\left(1-\exp[2i\Lambda(k) t]\right)}.
    \label{eq:gk}
\end{equation}
It is guaranteed that in the stationary state, $\abs{l_k(t)}<1$ and $l_k(t) \rightarrow 0$ as $t \rightarrow \infty$. Comparing Eq.~\eqref{eq:x41} with Eq.~\eqref{eq:x30.2} indicates that, in the gapped phase, the state $\ket{0}$ reaches the non-Hermitian Bogoliubov right vacuum in the stationary-state regime. We denote this as $\ket{\psi_{\text{st}}(t)} \simeq \ket{\Omega}$, where the sing $\simeq$ indicates that we keep the leading contributions to Eq.~\eqref{eq:x41} only. 

If the spectrum is gapless, in the stationary state regime, the evolved state is a superposition of two states (Appendix~\ref{sec:stationary_state}): the vacuum $\ket{\Omega}$ and the states of two excited non-Hermitian quasiparticles over the vacuum. More precisely,
 \begin{equation}\label{eq:x42}
     \ket{\psi_{\text{st}}(t)} \simeq \mathcal{A}(t)\ket{\Omega} + \mathcal{A}^*(t)\ket{q,-q},
 \end{equation}
 where 
 \begin{equation}
      \mathcal{A}(t) =
      \dfrac{e^{i\lambda_q} + ia(t)}{\sqrt{2}\sqrt{1+a(t)^2}\cos\lambda_q}
      \label{eq:x43}
 \end{equation}
with
 \begin{equation}\label{eq:x45}
     a(t) = -\dfrac{\sin E(q)t }{\cos[E(q) t + \lambda_q]}
 \end{equation}
 and $\lambda_q$ satisfies 
 $$\tan\lambda_q = \gamma\,[2E(q)]^{-1}.$$ 
 In contrast to what was reported in Ref.~\cite{zerba_measurement_2023}, the coefficients in Eq.~\eqref{eq:x42} are time-dependent. Let us remark that for any initial state of the form $$\ket{\psi(0)}\propto \prod_{k \in \mathcal{K}}\Big(d^{-}(k)\ket{1/2,-1/2}_k + d^{+}(k) \ket{1/2,1/2}_k \Big)$$ 
 with a non-zero overlap with $\ket{\Omega}$ and $\ket{q,-q}$, has a stationary state wave function similar to Eq.~\eqref{eq:x42}, with the appropriate (and possibly) time-dependent coefficients. Nonetheless, we focus only on the initial state $\ket{0}$ as it is canonically related to $\ket{\Omega}$.

\section{Correlation functions}\label{sec:correlation_functions}
After having precisely defined the link between the vacua $\ket{0}$ and $\ket{\Omega}$, we proceed to broaden the characterization of the latter state by calculating the asymptotic behavior of the correlation function between $z$-components of spins,
\begin{equation}\label{eq:x46}
    \texttt{C}^{zz}(n-m)\coloneqq \expval{\sigma_n^z\sigma_m^z} -\expval{\sigma_m^z}\expval{\sigma_n^z},
\end{equation}
where the expectation value is in the state $\ket{\Omega}$ \footnote{In principle, there is no restriction on the choice of state for computing the correlation function \eqref{eq:x46}, provided it is a valid state in the Hilbert space of the quantum spin chain we study. However, our physical motivation is to use $\ket{\Omega}$, which can be reached from the state $\ket{0}$ via the unitary evolution \eqref{eq:x28}, independent of the spectrum of $H$.}. The asymptotic behavior of this correlation function has been calculated for the XY model in Ref.~\cite{heralded} for $h = 0$ in the gapped phase and in Ref.~\cite{Turkeshi_1} for any $h$ in the gapless phase but for $\gamma <\gamma_c(h)$. 

\begin{table}
\centering
\caption{Asymptotic behavior of the correlation function \eqref{eq:x46} for $m=0$ and $n=x\rightarrow \infty$. The functions $\mu(x)$ and $\chi(x)$ are given by Eqs.~\eqref{eq:mu} and \eqref{eq:chi1}, respectively. The correlation length $\xi$ is given by Eq.~\eqref{xi-z1} and $\xi_+=\lim_{\gamma \to 0}\xi$.}
\begin{tabular}[t]{|l|c|c}
\hline
\phantom{\Big(...\Big)} Region: $\{h\}\times\{\gamma\}$ & $\texttt{C}^{zz}(x)$
\\
\hline
\rule{0pt}{1\normalbaselineskip}Oscillatory gapless phase: $ [0,J)\!\times\![0,\gamma_c(h))$ &  $x^{-2}\cos^2 qx$ \\[1ex]
Oscillatory critical line: $[0,J)\!\times\! \{\gamma_c(h) \}$  & $x^{-4}\mu(x)$\\[1ex]
Ising critical point: $\{J\}\!\times\! \{0\}$  &  $x^{-2}$ \\[1ex]
Disordered Hermitian: $(J,\infty)\!\times\!\{0\}$ & $x^{-2}e^{-x/\xi_+}$\\[1ex]
Disordered non-Hermitian: $[0,\infty)\!\times\!(\gamma_c,\infty)$ & \ $x^{-3}\chi(x)e^{-x/\xi}$\ \ 
\\[1ex]
\hline
\end{tabular}
\label{tab1}
\end{table}%

In this section, we complete the characterization of the asymptotic behavior of Eq.~\eqref{eq:x46} for our non-Hermitian Ising chain in the gapped phase and in the full region of the gapless phase $\gamma \leq \gamma_c(h)$, as well as in the limiting case $\lim_{\gamma \to 0}H$. As we show below, in this limit, for $n=x$ and $m=0$, the asymptotics of $\texttt{C}^{zz}(x)$ in the region $h\leq J$ and $\gamma = 0$ differs from the one found in the standard ferromagnetic phase. This is because the ground state $\lim_{\gamma\to 0}\ket{\Omega}$ is an excited state of the ground state of the Hermitian $H$ (the one with $\gamma \equiv 0$ ab-initio). A schematic overview of the asymptotics of $\texttt{C}^{zz}(x)$ as $x\rightarrow \infty$ is displayed in Table~\ref{tab1} and Fig.~\ref{fig:1.2}. The detailed calculations are shown in Appendix~\ref{App:Czz}.

Following the standard procedure as in, e.g., Ref.~\cite{heralded}, we express the correlation function \eqref{eq:x46} in terms of the operators $\texttt{A}_n = c_n^\dagger + c_n$ and $\texttt{B}_n = c^\dagger_n-c_n$:
\begin{equation}\label{eq:x47}
    \texttt{C}^{zz}(n-m) = -\expval{\texttt{A}_m \texttt{B}_n}\expval{\texttt{A}_n\texttt{B}_m} - \expval{\texttt{A}_m \texttt{A}_n}\expval{\texttt{B}_m\texttt{B}_n},
\end{equation}
where 
\begin{align}
    &\expval{\texttt{A}_m \texttt{A}_n}\! =\! \delta_{mn}\!+\!\frac{1}{\pi}\!\int_{0}^\pi\!\!\dd k\, \sin [k(n-m)] \, \frac{u_kv_k^*-u_k^*v_k}{\abs{u_k}^2+\abs{v_k}^2}, \label{eq:x48}\\
    &\expval{\texttt{B}_m \texttt{B}_n}\! 
    =\! -\delta_{mn}\!+\!\frac{1}{\pi}\!\int_{0}^\pi\!\!\dd k\,  \sin [k(n-m)]  \,\frac{u_kv_k^*-u_k^*v_k}{\abs{u_k}^2+\abs{v_k}^2}
        \notag 
    \\
    &\qquad\quad\, = \! -2\delta_{mn}\!+\! \expval{\texttt{A}_m \texttt{A}_n},\label{eq:x49}
\end{align}
and
\begin{align}
    \expval{\texttt{B}_m \texttt{A}_n} &= -\expval{\texttt{A}_n\texttt{B}_m}\notag \\  &= -\frac{1}{\pi}\int_{0}^\pi\dd k \, \cos[k (n-m)] \,\frac{\abs{u_k}^2-\abs{v_k}^2}{\abs{u_k}^2+\abs{v_k}^2} \notag \\
    &\,\,\,\,\,\,+\frac{1}{\pi}\int_{0}^\pi \dd k\, \sin [k (n-m)] \,\frac{u_k v_k^*+u_k^*v_k}{\abs{u_k}^2+\abs{v_k}^2}.\label{eq:x50}
\end{align}
The functions $u_k$ and $v_k$ are given by Eq.~\eqref{eq:x16}. Let $m = 0$ and $n = x \in 2\mathbb{Z}^+$. Then,   $$\expval{\texttt{A}_0 \texttt{B}_x} = -\expval{\texttt{B}_0 \texttt{A}_{-x}}, \quad \expval{\texttt{A}_0\texttt{A}_x} = \expval{\texttt{B}_0\texttt{B}_x},$$ and so 
\begin{equation}\label{eq:x51}
    \texttt{C}^{zz}(x) =- \expval{\texttt{B}_0 \texttt{A}_{-x}}\expval{\texttt{B}_0 \texttt{A}_{x}}-\expval{\texttt{A}_0\texttt{A}_x}^2.
\end{equation}
We note that $\expval{\texttt{A}_0\texttt{A}_x} = 0$ for $h = 0$, which was obtained in Ref.~\cite{heralded} only for $\gamma >4J$. In the following subsections, we first discuss the asymptotic behavior of $\texttt{C}^{zz}(x)$ in the gapped phase and then in the gapless phase. We will finish this section by discussing the limiting Hermitian case $\gamma \rightarrow 0$ while keeping the sign convention of $\Gamma(k)$.

\subsection{Correlation function in the gapped phase}
\label{subsec:3A}

Under our convention regarding the sign of $\mathfrak{Im}\, \Lambda(k)$,  the spectrum \eqref{eq:x09} can be written as 
\begin{align}\label{eq:a0x1}
     \Lambda(k) &= \text{sgn}(J\cos k-h) \,\sqrt{R_z(k)^2+R_x(k)^2},
\end{align}
where the positive branch of the square root is taken and $\text{sgn}(0) =1$. For $\gamma >0$, we have the following useful relations involving the real and imaginary parts of the spectrum:
\begin{equation}\label{eq:ax2}
    E(k) \Gamma(k)  = \gamma\,(h-J\cos k) ,
\end{equation}
and
\begin{equation}
    \Gamma(k)  = \dfrac{\gamma}{2}\,\dfrac{\abs{u_k^2}-\abs{v_k}^2}{\abs{u_k}^2+\abs{v_k}^2}.
    \label{eq:Gamma-gamma}
\end{equation}
\begin{widetext}
\noindent
These allow us to greatly simplify the contractions \eqref{eq:x48}-\eqref{eq:x50} (cf. Eq.~(D8) of Ref.~\cite{heralded}):
 \begin{equation}\label{eq:ax6.1}
        \expval{\texttt{A}_0 \texttt{A}_x} = \dfrac{i}{\gamma \pi}\int_0^\pi \dd k\, \sin kx\, \dfrac{\gamma^2-4\Gamma^2(k)}{4J \sin k}\, 
    \end{equation}
and
    \begin{equation}\label{eq:ax6}
        \expval{\texttt{B}_0 \texttt{A}_{\pm x}} = \mathfrak{Re}\left\lbrace \dfrac{1}{\gamma \pi}\int_{-\pi}^\pi \dd k \,\left[i\cos kx \pm \sin kx \, \dfrac{\gamma + i 4(h-J\cos k)}{4J\sin k}\right]\Lambda(k) 
     \right\rbrace.
    \end{equation}

To find the asymptotic behavior of these contractions as $x\rightarrow \infty$, the integrals are analytically extended to the complex plane. Then,  the integration contours are deformed, such that they end up surrounding the branch cuts lying outside the unit circle [see Fig.~\ref{fig:contours}]. It turns out that the dominant contributing points in the deformed integration contour lie on the closest branch points to the unit circle from the outside. This is known as the Darboux method \cite{jones_asymptotic_2019,Temme_AsymptoticMethods}. Next, a series expansion around the external branch points is performed, and the resulting divergent series is integrated term-wise, giving
   \begin{equation}\label{eq:b0asx}
    \expval{\texttt{B}_0\texttt{A}_{s x}} \simeq \begin{cases}
       \dfrac{4}{\sqrt{\pi}}\,\dfrac{J}{\gamma }\, (-1)^{x/2}\, x^{-3/2}\,\abs{z_1}^{-x}\kappa_+(x) + \mathcal{O}\left(\abs{z_1}^{-x}x^{-5/2}\right), &\quad s = +, \\[0.5cm]
        \dfrac{4}{\sqrt{\pi}}\,\dfrac{J}{\gamma } \, (-1)^{x/2}\,x^{-3/2}\,\abs{z_1}^{-x} \kappa_-(x) + \mathcal{O}\left(\abs{z_1}^{-x}x^{-5/2}\right), &\quad s = -,
    \end{cases}
\end{equation}
where $\kappa_{\pm}(x)$ are oscillatory functions of $x$ given by Eqs.~\eqref{eq:kappa+} and \eqref{eq:kappa-}, respectively, and $z_1 = (-\gamma+i4 h)/(4J)$ is the above-mentioned external branch point at which the series expansion is performed. Returning to Eq.~\eqref{eq:ax6.1}, the Darboux method gives
\begin{equation}\label{eq:a0axg}
    \expval{\texttt{A}_0\texttt{A}_x} \simeq \dfrac{4i}{\sqrt{\pi}}\,\dfrac{J}{\gamma}\,(-1)^{x/2}\, x^{-3/2}\, \abs{z_1}^{-x}\nu(x) + \mathcal{O}\left(\abs{z_1}^{-x}x^{-5/2}\right),
\end{equation}
\end{widetext}
where $\nu(x)$ is an oscillatory function of $x$ given by Eq.~\eqref{eq:nu}. 

The asymptotic behavior of the correlation function in the gapped phase, which we also call the disordered non-Hermitian phase (see Table~\ref{tab1}), is 
\begin{equation}\label{eq:czz1}
      \texttt{C}^{zz}(x)\simeq  - \dfrac{16}{\pi}\left(\dfrac{J}{\gamma}\right)^2x^{-3}e^{-x/\xi}\left[ \chi(x) +\mathcal{O}(x^{-1}) \right],
\end{equation}
where 
\begin{equation}\label{eq:chi1}
\chi(x)\coloneqq \kappa_+(x)\kappa_-(x) - \nu^2(x),    
\end{equation}
and 
\begin{equation}
    \xi = (2\ln \abs{z_1})^{-1}
    \label{xi-z1}
\end{equation} 
is the correlation length. Let us note that for a non-vanishing real part of the magnetic field, an oscillatory function of $x$ emerges not only from $\expval{\texttt{B}_0\texttt{A}_{\pm x}}$ but also from $\expval{\texttt{A}_0\texttt{A}_x}$. For $h=0$, the latter contraction vanishes, and we find (cf. Eq.~(D12) of Ref.~\cite{heralded}):
\begin{equation}\label{eq:beta}
    \chi(x) = \kappa_+(x)\kappa_-(x) = -\dfrac{1}{8}\left(\dfrac{\gamma}{J}\right)^2 \dfrac{\gamma^2-16J^2}{\gamma^2+16J^2}.
\end{equation}
Hence, Eq.~\eqref{eq:czz1} does not display any oscillatory behavior in $x$  when $h=0$ and $\gamma>4J$. From Eqs.~\eqref{eq:beta} and \eqref{eq:czz1}, it is clear that the $\mathcal{O}(x^{-3})$ term tends to zero as $\gamma \rightarrow 4J$. In this limit, the next contributing term of Eq.~\eqref{eq:czz1} (not shown here) coincides with the leading-order term in the gapless phase, for which $\gamma = \gamma_c(0) = 4J$ [cf. Eq.~\eqref{eq:tcor1}]. In contrast, when $h>0$ and $\gamma \rightarrow \gamma_c(h)$ from above, only the algebraic terms of Eqs.~\eqref{eq:czz1} and \eqref{eq:tcor1} coincide but not the prefactors. Regardless of this issue, which we will address in brief, for $h\in [0,J)$ and $\gamma \rightarrow \gamma_c(h)^+$, the correlation length diverges as 
\begin{equation}\label{eq:x58}
    \xi \simeq \dfrac{2J^2}{\sqrt{J^2-h^2}\left[\gamma-\gamma_c(h)\right]} + \mathcal{O}(1).
\end{equation}

The mismatch of the prefactors occurs because the Darboux method we implemented to calculate the correlation function \eqref{eq:czz1} ``does not know'' about the sign discontinuities that were on the unit circle in the complex plane before the contour was deformed off the unit circle. In addition to this, the internal and external branch cuts to the unit circle emerging from the analytical continuation of the contraction integrals tend to merge as $\gamma\rightarrow \gamma_c(h)$, rendering the Darboux method inapplicable. A way to circumvent this issue occurring in the critical gapless phase $\gamma = \gamma_c(h)$ is presented in Sec.~\ref{subsec:3B}. 

It is important to provide some additional comments regarding the gapped phase. The asymptotic behavior of $\texttt{C}^{zz}(x)$ in the standard Hermitian case, i.e., $\gamma \equiv 0$, in the ferromagnetic and paramagnetic phases is (see Appendix~\ref{ssec:correlation_function_standard}) $$\texttt{C}^{zz}_0(x)\sim x^{-2}e^{-x/\xi_{\pm}},$$ 
where the correlation lengths of the ferromagnetic and paramagnetic phases are 
$$\xi_- =  [2\ln (J/h)]^{-1}\quad \text{and} \quad \xi_+ = [2\ln (h/J)]^{-1},$$ respectively. The exponential decay occurs at all orders in $x^{-n}$ ($n>0)$ in the two phases, and the argument of the logarithm in the correlation function changes from $J/h$ to $h/J$ as the Ising critical point $(h,\gamma) = (J,0)$ is crossed. In contrast, in the gapped phase, as long as $\gamma >\gamma_c(h)$, this inversion in the correlation function does not happen: it is always $\xi = (2\ln \abs{z_1})^{-1}$ and not $\xi = (2\ln \abs{1/z_1})^{-1}$. However, as we discuss in the following section, if the critical line $\gamma = \gamma_c(h)$ is crossed and we stay in this gapless region even for $\gamma \rightarrow 0$, the inversion in $\xi$ does occur, but the term containing the exponential decay in the asymptotic expansion becomes subdominant. Despite this fact, the correlation function is continuous for $h>J$ and all $\gamma$. 

For example, if $h>J$, $\lim_{\gamma\to 0}\xi = \xi_+$, so that it behaves as
\begin{equation}
    \xi \simeq \dfrac{1}{2}\dfrac{J}{h-J} + \mathcal{O}(1)
\end{equation}
for $\gamma \rightarrow 0$ and then $h\rightarrow J^+$. Conversely, if $h=J$, 
\begin{equation}
    \xi \simeq 16\left(\dfrac{J}{\gamma}\right)^2 + \mathcal{O}(1)
\end{equation}
as $\gamma \rightarrow 0$. Nevertheless, for $h>J$, $\lim_{\gamma\to 0}\texttt{C}^{zz}(x) \neq \texttt{C}^{zz}_0(x)$, as the former goes as $x^{-3}$ and the latter as $x^{-2}$ [compare Eq.~\eqref{eq:corrher} with Eq.~\eqref{eq:czz1}]. This discrepancy of the powers can be explained as follows: when the integrals in the contractions $\expval{\texttt{B}_0\texttt{A}_{\pm x}}$ and $\expval{\texttt{A}_0 \texttt{A}_x}$ are analytically extended into the complex plane, in the ``disordered non-Hermitian'' and ``disordered Hermitian'' phases, as well as in the paramagnetic phase where $\gamma \equiv 0$ (see Appendixes~\ref{ssec:hermitian_case} and \ref{ssec:correlation_function_standard}, and Table~\ref{tab1}), there is one branch point outside the unit circle, which is dominant when performing the asymptotic expansion. In the disordered non-Hermitian phase, the branch point $z_1$ appears in the numerator of a function containing $\sqrt{z-z_1}$, whereas in the paramagnetic and disordered Hermitian phases, it appears in the denominator. This leads to the contractions to behave as $\abs{z_1}^{-x}x^{-3/2}$ for the former case and as $\abs{z_1}^{-x}x^{-1/2}$ for the latter as $x\rightarrow \infty$, see, e.g., Eqs.~(73) and (74) of Ref.~\cite{jones_asymptotic_2019}.

\begin{figure}
\tikzset{every picture/.style={line width=0.75pt}} 

\begin{tikzpicture}[x=0.75pt,y=0.75pt,yscale=-1.5,xscale=1]

\draw    (100,200.67) -- (100,50.57) ;
\draw    (100,200.67) -- (201.07,199.93) ;
\draw  [draw opacity=0] (100.07,94.32) .. controls (121.76,94.31) and (142.41,97.36) .. (161.16,102.86) -- (100,200.67) -- cycle ; \draw   (100.07,94.32) .. controls (121.76,94.31) and (142.41,97.36) .. (161.16,102.86) ;  
\draw [color={rgb, 255:red, 1; green, 1; blue, 1 }  ,draw opacity=1 ]   (274.86,170.44) -- (256.27,198.56) ;
\draw [shift={(255.17,200.23)}, rotate = 303.46] [color={rgb, 255:red, 0; green, 0; blue, 0 }  ,draw opacity=1 ][line width=0.75]    (6.56,-1.97) .. controls (4.17,-0.84) and (1.99,-0.18) .. (0,0) .. controls (1.99,0.18) and (4.17,0.84) .. (6.56,1.97)   ;
\draw    (201.07,199.93) -- (267.2,199.93) ;
\draw  [draw opacity=0] (219.56,132.8) .. controls (241.7,151.11) and (255.05,174.6) .. (255.17,200.23) -- (100,200.67) -- cycle ; \draw   (219.56,132.8) .. controls (241.7,151.11) and (255.05,174.6) .. (255.17,200.23) ;  
\draw    (345.4,199.93) -- (367.67,199.93) ;

\draw (84.33,58.8) node [anchor=north west][inner sep=0.75pt]    [font=\Large]{$\gamma $};
\draw (355.67,201.8) node [anchor=north west][inner sep=0.75pt] [font=\Large]   {$h$};
\draw (262.67,155.13) node [anchor=north west][inner sep=0.75pt]  [font=\normalsize] {$\mathcal{O}(x^{-2})$};
\draw (120,170.33) node [anchor=north west][inner sep=0.75pt]  [font=\normalsize]  {$\mathcal{O}\textbf{(} x^{-2}\cos^{2} qx \textbf{)}$};
\draw (268.33,186.67) node [anchor=north west][inner sep=0.75pt] [font=\normalsize] {$\mathcal{O}\left(\dfrac{e^{-x/\xi_+}}{x^2}\right)$};
\draw (103.33, 130.73) node [anchor=north west][inner sep=0.75pt]   [align=left] {{Gapless phase}};
\draw (185.33,65.07) node [anchor=north west][inner sep=0.75pt]   [align=left] {{ Gapped phase}};
\draw (168,98.69) node [anchor=north west][inner sep=0.75pt]  [rotate=-38.41]  [font=\normalsize] {$\mathcal{O}\textbf{(} x^{-4}\mu (x)\textbf{)}$};
\draw (259.88,89.11) node [anchor=north west][inner sep=0.75pt]  [font=\normalsize]  {$\mathcal{O}\textbf{(}  x^{-3}\chi(x) e^{-x/\xi}\textbf{)}$};
\draw (251.67,202.8) node [anchor=north west][inner sep=0.75pt]  [font=\normalsize]  {$J$};
\draw (83,90.27) node [anchor=north west][inner sep=0.75pt]    {$4J$};

\end{tikzpicture}
    \caption{Schematic phase diagram displaying the asymptotic behavior of $\texttt{C}^{zz}(x)$ calculated in the non-Hermitian vacuum $\ket{\Omega}$, in the gapped and gapless regions [cf. Fig.~\ref{fig:1}] separated by the critical curve $\gamma = \gamma_c(h)$. The functions $\mu(x)$ and $\chi(x)$ are given by Eqs.~\eqref{eq:mu} and \eqref{eq:chi1}, respectively. The correlation length $\xi$ is given by \eqref{xi-z1} and $\xi_+=\lim_{\gamma \to 0}\xi$. The correlation function is continuous throughout the subregion $0\leq \gamma <\gamma_c(h)$ and the approach to the segment  $0\leq h <J$ at $\gamma = 0$ must be understood as a limiting procedure at the level of the non-Hermitian Hamiltonian \eqref{eq:x1}; that is, the sign-convention for the eigenvalues of $H$ is chosen, and then, one takes the limit $\gamma \rightarrow 0$. As a consequence, $\texttt{C}^{zz}(x)$ does not display an exponential decay that would correspond to the standard (Hermitian, $\gamma\equiv 0$) ferromagnetic phase for $0\leq h<J$. }
    \label{fig:1.2}
\end{figure}

\subsection{Correlation function in the gapless phase}
\label{subsec:3B}

Calculating the asymptotics of $\texttt{C}^{zz}(x)$ in the gapless phase via contour integration is somewhat more involved than in the previous case, as more sophisticated methods must be invoked (for gapless Hermitian models, see, e.g., Refs.~\cite{jones_asymptotic_2019,t.t.wu}). To avoid this issue, we opt for a more conventional approach, such as performing a series expansion of the integrands of $\expval{\texttt{A}_0 \texttt{A}_x}$ and $\expval{\texttt{B}_0 \texttt{A}_{\pm x}}$, not involving $x$, at relevant $k$ points such as $x \rightarrow \infty$, and adding appropriate regulators to the integrals. In this way, the integration domains can be extended to $\infty$ or $-\infty$, depending on the point at which the Taylor expansion is taken, to facilitate the calculation. 

As shown in Appendix~\ref{ssec:gapless_phase}, a jump discontinuity occurs at $k = q$. Therefore, this point contributes the most to the asymptotic expansion of the contractions. First, for $0<\gamma<\gamma_c(h)$, we find $\expval{\texttt{A}_0\texttt{A}_x} \sim \mathcal{O}\textbf{(}x^{-2}\sin qx \textbf{)}$, and 
\begin{equation}
    \expval{\texttt{B}_0\texttt{A}_{\pm x}}\!=\!\mp \dfrac{1}{2\pi} \sqrt{\dfrac{16(J^2-h^2)-\gamma^2}{J^2-h^2}}\dfrac{\cos q x}{x}\left[1\!+\!\mathcal{O}(x^{-1})\right].
\end{equation}
Thus, for $0<\gamma<\gamma_c(h) = 4\sqrt{J^2-h^2}$, we obtain
    \begin{equation}\label{eq:tcor}
        \texttt{C}^{zz}(x)\simeq \dfrac{1}{4\pi^2} \dfrac{16(J^2-h^2)-\gamma^2}{J^2-h^2}\, \dfrac{\cos^2 q x}{x^2}\, \left[1 \!+\! \mathcal{O}(x^{-1})\right].
    \end{equation}
Let us note that in this phase, which we also call ``oscillatory gapless phase'' (see Table~\ref{tab1}), there are subdominant exponentially decaying terms similar to Eq.~\eqref{eq:czz1} with correlation length $\xi = \left(2\ln\abs{1/z_1}\right)$.
If $\gamma = \gamma_c(h)$ for $h\neq 0$, the leading term in Eq.~\eqref{eq:tcor} vanishes and the correlation function yields
\begin{equation}\label{eq:tcor1}
    \texttt{C}^{zz}(x) \simeq \dfrac{1}{\pi}x^{-4}\mu(x) + \mathcal{O}\textbf{(}\theta(x) x^{-5}\textbf{)}, 
\end{equation}
 where $\mu(x)$ is given by Eq.~\eqref{eq:mu}, and $\theta(x)$ is not relevant to our analysis.
Finally, if $h=0$ and $\gamma = \gamma_c(0) = 4J$, $\mu(x)$ and $\theta(x)$ become constant, and
\begin{equation}\label{eq:tcor2}
  \texttt{C}^{zz}(x) \simeq \dfrac{3}{2\pi}x^{-4} + \mathcal{O}(x^{-5}).
\end{equation}
The change in the power-law decay from $\sim x^{-2}$ to $x^{-4}$ is due to the emergence of two exceptional points at $\pm q$ (see Fig.~\ref{fig:1} in red). These exceptional points do not represent critical points between a $\mathcal{PT}$-symmetric and symmetry-broken phases of Eq.~\eqref{eq:x1} as this Hamiltonian is not $\mathcal{PT}$-symmetric. Nevertheless, the existence of these exceptional points seems to translate into a higher level of suppression in the correlation function.
We emphasize that to our knowledge, $\texttt{C}^{zz}(x)$ had not previously been calculated for $\gamma = \gamma_c(h)$.

\subsection{Hermitian limiting case}
\label{subsec:3C}

The next natural step is to calculate the same correlation function in the limiting Hermitian case; that is, $\lim_{\gamma \to 0}\Lambda(k) = E(k)$. 
Since, following our convention dictated by the sign of $\Gamma(k)$, $E(k)$ changes sign with $k$, the state $\lim_{\gamma \to 0}\ket{\Omega}$ is an \textit{excited state} of the \emph{true} ground state $\ket{\text{GS}}$ of the corresponding Hermitian Hamiltonian (i.e., $H$ with $\gamma \equiv 0$ ab-initio). Specifically,  
$$\lim_{\gamma\to 0}\ket{\Omega} = \prod_{k>\pi/2}\hat{d}_{k}^\dagger \hat{d}_{-k}^\dagger \ket{\text{GS}},$$ 
where, for a given $k$, $\hat{d}_k$ is the Hermitian Bogoliubov fermionic operator and $\ket{\text{GS}}\propto \prod_{k>0}\hat{d}_{k} \hat{d}_{-k}\ket{0}$ is the ground state of $H\vert_{\gamma \equiv 0}$. 

Thus, starting from the non-Hermitian model, the limit $\gamma\to 0$ is not expected to always reproduce the results for the Hermitian case ($\gamma\equiv 0$), as the vacua of the two models do not coincide. Based on the consideration of Sec.~\ref{eq:ssec:time_evolution_of}, this reflects the fact that, for an arbitrarily small but finite $\gamma$, evolving the initial state with the non-Hermitian Hamiltonian for times much longer than the time scale set by $\gamma^{-1}$, the specific characteristics of the non-Hermitian Bogoliubov vacuum and of the corresponding steady state will always reveal themselves. We remind the reader that in the gapless region (I), there are excitations at $\pm q$ in the steady state for arbitrary, whatever small $\gamma$.
A similar situation is encountered in the context of measurement-induced transitions. There, the properties of the monitored systems in the steady state at an arbitrary small but finite measurement rate are drastically different from those in non-monitored systems; see, e.g., Refs.~\cite{Poboiko2023a, Poboiko2025}.

Upon implementing the same techniques as for the non-Hermitian case (see Appendix~\ref{ssec:hermitian_case}), we get for the correlation function in the limit $\gamma\to 0$:
\begin{equation}\label{eq:corrher}
    \texttt{C}^{zz}(x)\simeq \begin{cases}
        \dfrac{1}{2\pi}\dfrac{e^{-x/\xi_+}}{x^2}\left[1 + \mathcal{O}(x^{-1})\right], \quad &h>J, \, \gamma \rightarrow  0,\\[2ex]
        \dfrac{1}{\pi^2}\dfrac{1}{x^2} + \mathcal{O}(x^{-4}), \quad &h = J,\, \gamma \rightarrow 0, \\[2ex]
      \dfrac{4}{\pi^2}\dfrac{\cos^2 qx}{x^2}\left[1 + \mathcal{O}(x^{-2}) \right], \quad&h<J,\, \gamma \rightarrow 0, 
    \end{cases}
\end{equation}
where $\xi_+ = [2\ln(h/J)]^{-1}$ is the correlation length in the ``disordered Hermitian phase.'' The correlation function in this phase, as well as in the ``critical Ising'' phase ($h=J, \gamma\rightarrow 0$), coincides with the one calculated in $\ket{\text{GS}}$ [cf.~Eq.~\eqref{eq:ph}]. In contrast, in the ``oscillatory phase'' ($h>J$, $\gamma \rightarrow 0$), the leading-order term displays oscillatory and algebraic decay $x$ instead of exponential behavior characteristic of the Hermitian ferromagnetic phase. Similarly to the gapless phase, the leading term in Eq.~\eqref{eq:ph} for $h<J$ appears as subleading.

\section{Krylov spread}\label{sec:krylov_spread}

In this section, we first briefly review the generalities of Krylov spread complexity (henceforth, \textit{Krylov spread}) of states to establish our notation. We proceed to calculate the Krylov spread of the JW vacuum $\ket{0}$ when it evolves unitarily to $\ket{\Omega}$ by means of the Hermitian Hamiltonian (as opposed to the non-unitary evolution addressed in Sec.~\ref{eq:ssec:time_evolution_of}), corresponding to the natural relation between the two vacua. We find the critical behavior of the derivatives of the spread density with respect to $h$ or $\gamma$ across the boundary corresponding to the critical value $\gamma = \gamma_c(h)$ and across the Ising critical point. These results will set the stage for the analysis of dynamical phase transitions in the case of non-unitary evolution, revealed via the Krylov spread in Sec.~\ref{Sec:V-DPT}. 

\subsection{Generalities}
\label{subsec:4A}

The span of the vectors obtained from the repeated action of an operator on a seed vector creates what is called a Krylov subspace. For example, for an initial state $\ket{\psi(0)}$ and a Hermitian, time-independent Hamiltonian $H$, the Krylov subspace is given by
\begin{equation}\label{eq:kk1}
    \mathfrak{K}\big(\psi(0), H\big)\!= \text{span}\left\lbrace \ket{\psi(0)}\!, H\ket{\psi(0)}\!, \ldots , H^n\ket{\psi(0)}\!,\ldots\right\rbrace.
\end{equation}
For a time-dependent generator of the dynamics, see, e.g.,~Ref.~\cite{Kazu1}.
The orthonormal basis of such a space obtained via the Gram-Schmidt process is called the Krylov basis and is denoted as $\{ \ket{K_n}\}_{n=0}^{D},$ where $D = \text{dim}\, \mathfrak{K}\big(\psi(0), H\big)\leq \text{dim}\,\mathcal{H}$. Here, $\mathcal
H$ is the Hilbert space. Given this basis, we can expand the evolved state $\ket{\psi(t)}$  as
\begin{equation}\label{eq:kk2}
    \ket{\psi(t)} = e^{-iHt}\ket{\psi(0)} = \sum_{n=0}^D\varphi_n(t)\ket{K_n},
\end{equation}
where $\{\varphi_n(t) = \bra{K_n}\ket{\psi(t)}\}_{n=0}^D$ are the Krylov coefficients. The Krylov spread is then defined as 
\begin{equation}\label{eq:kk3}
    C\left(\psi(0), H; t \right) = \sum_{n=0}^D n\, \abs{\varphi_n(t)}^2,
\end{equation}
which is the mean value of the distance of the evolved state $\ket{\psi(t)}$ on the Krylov basis from the origin. In the absence of ambiguity, we write $C(t) = C(\psi(0),H;t)$. We remark that the spread can be interpreted as the expectation value of the operator
\begin{equation}\label{eq:kk3.1}
    \hat{C} = \sum_{n=0}^Dn\ket{K_n}\bra{K_n}
\end{equation}
in the state $\rho(t) = \ket{\psi(t)}\bra{\psi(t)}$.

A key advantage of using the Krylov basis to describe quantum dynamics~\cite{parker2019universal} is its natural emergence from the unitary evolution, which, in a way, is the repeated action of the Hamiltonian $H$ on $\ket{\psi(0)}$. Furthermore, the above definitions can be extended to operators instead of states and to systems governed by non-Hermitian Hamiltonians, where some nuances must be considered, see Sec.~\ref{Sec:V-DPT} below. 
For a recent review of Krylov complexity, see Refs.~\cite{nandy2024quantumdynamicskrylovspace}. 

Krylov spread becomes especially tractable when the Hamiltonian is an element of a Lie algebra, and the initial state is the lowest (or highest) weight state of the algebra. For example, if 
$$H = (r\, J_++ r^* J_-) + s\, J_z + p\, \mathbb{I}\, \in\, \mathfrak{su}(2),$$ 
and $\ket{\psi(0)} = \ket{1/2,-1/2}$ with $J_-\ket{1/2,-1/2} =0$, the spread  reads \cite{caputa2022geometry,quantum_chaos_and_complexity}:
\begin{subequations}\label{eq:kr1}
    \begin{align}
         C(t) &= \abs{\bra{1/2,1/2}\ket{\psi(t)}}^2\\
         &= \abs{\bra{1/2,1/2}e^{-iHt}\ket{1/2,-1/2}}^2 \\
         &= \dfrac{\abs{r}^2}{\abs{r}^2+{s^2}/{4}}\sin^2\left( t \,\sqrt{\abs{r}^2+{s^2}/{4}}\right).
    \end{align}
\end{subequations}

Having outlined the basics of Krylov spread, we return to our problem. Let us assume that our initial state is the JW vacuum $\ket{0}$. Then, we calculate its Krylov spread per unit site (spread density) with respect to two different Hamiltonians by taking advantage of the fact that it is the lowest weight state of $\mathfrak{su}(2)^{\otimes N}$. The two spreads are calculated as follows: first, by implementing a single unitary evolution operator that takes the state to the non-Hermitian vacuum $\ket{\Omega}$; second, by evolving $\ket{0}$ in time via the non-Hermitian Hamiltonian \eqref{eq:x1} until the stationary state is reached. By doing so, we find that the spreads ``coincide'' in the gapped phase, whereas in the gapless phase, the spread obtained via a unitary operator coincides with the time-averaged (in the steady state) spread. Additionally, we demonstrate that, at least for the non-Hermitian 1D Ising model, the Krylov spread is encoded in the same-site pair contraction $\expval{\texttt{B}_n \texttt{A}_n}$, which provides an alternative and more direct approach to calculating the spread (cf. Ref.~\cite{caputa_spread_2023}).

\medskip

\subsection{Krylov spread via unitary dynamics}\label{eq:ssec:circuit_spread}
Since $\ket{0}$ and $\ket{\Omega}$ in Eq.~\eqref{eq:x28} are related via a unitary operator, there exists a Hermitian Hamiltonian 
\begin{equation}\label{eq:n1}
    \tilde H = \sum_{k\in \mathcal{K}^+}\tilde H_k=\sum_{k\in \mathcal{K}^+}\Big[\,\abs{\alpha_+(k)}e^{i \text{Arg}\, \tau_k - i\frac{\pi}{2}}J_+(k) + \text{H.c.} \Big],
\end{equation}
such that  $\ket{\Omega}$ is obtained from $\ket{0}$ with the unitary evolution governed by $\tilde H$ from time $t=0$ to $t=1$:
    \begin{equation}\label{eq:n2}
        \ket{\Omega} = 
        \left.\exp(-i \tilde H \,t)\ket{0}\right|_{t=1}.
    \end{equation}
Let us write the JW vacuum as in Eq.~\eqref{eq:x25}. Since each mode in Eq.~\eqref{eq:n2} evolves unitarily, i.e., $e^{-i t\tilde H_k}\ket{1/2,-1/2}_k\vert_{t=1}$, we can use $J_+(k)\ket{1/2,-1/2}_k = \ket{1/2,1/2}_k$ together with Eq.~\eqref{eq:kr1} to find the Krylov spread per mode:
\begin{equation}\label{eq:n3}
    C(t=1;k) = \frac{\abs{\tau_k}^2}{1+\abs{\tau_k}^2}.
\end{equation}

Following Ref.~\cite{caputa_spread_2023}, since the Hermitian Hamiltonian \eqref{eq:n1} is an element of the semisimple Lie algebra $\mathfrak{su}(2)^{\otimes N}$ and can be written in an upper Hessenberg form due to the Gram-Schmidt process applied to $\{\tilde{H}^n\ket{0}\}_{n\geq 0}$ to find $\{\ket{K_n} \}_{n\geq 0}$, one can demonstrate that, in the thermodynamic limit, the $n$-th Krylov vector can be written as \begin{equation}\label{eq:Kn}
\ket{K_n} = \mathcal{N}_n\left(\sum_{k\in\mathcal{K}^+} \alpha_+(k)J_+(k) \right)^n\ket{0},
\end{equation}
where $\mathcal{N}_n$ is a normalization factor. Because of this, one also finds that $\sum_k\hat{C}(k)\ket{K_n} = n\ket{K_n}$ holds in the thermodynamic limit, where $\hat{C}(k) = J_z(k) + 1/2$ is the spread operator per $k$ mode [cf.~Eq.~\eqref{eq:kk3.1}].
Hence, the spread per site (or density) can be written as a sum of the individual spreads per mode plus corrections:
\begin{subequations}\label{eq:x63}
\begin{align}
\mathcal{C}_\Omega &\coloneqq \dfrac{1}{N}\sum_{n}n\abs{\bra{\Omega(t=1)}\ket{K_n}}^2\\
&= \frac{1}{N}\sum_{k\in \mathcal{K}}C(t = 1;k) + \mathcal{O}\left(N^{-2}\right),
\end{align}
\end{subequations}
which, in the thermodynamic limit, yields
\begin{equation}\label{eq:x64}
  \mathcal{C}_\Omega = \frac{1}{\pi}\int_0^\pi \dd k\, \dfrac{\abs{v_k}^2}{\abs{u_k}^2+\abs{v_k}^2}.
\end{equation}

\begin{figure}
\includegraphics[width=0.95\columnwidth,center]{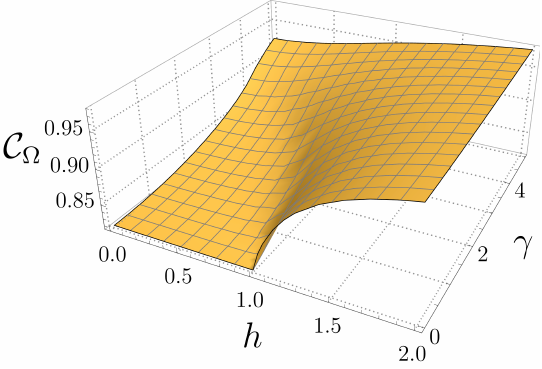}
    \caption{ Krylov spread complexity density \eqref{eq:x68} for $\ket{0}$ evolving to $\ket{\Omega}$ via unitary dynamics ($h$ and $\gamma$ are in units of $J$).}
    \label{fig:3}
\end{figure}

Quite interestingly, this spread density is related to the same-site contraction
\begin{equation}\label{eq:x65}
\mathcal{C}_\Omega= \frac{1+\expval{\texttt{B}_n \texttt{A}_n}}{2}
\end{equation}
or, in terms of the same-site correlation function,
\begin{equation}\label{eq:x66}
   \mathcal{C}_\Omega= \dfrac{1+\sqrt{1-\texttt{C}^{zz}(0)}}{2}.
\end{equation} 
Equation \eqref{eq:x65} offers us the great advantage of being able to analytically perform the integration in Eq.~\eqref{eq:x64}. Thus, with the aid of Eq.~\eqref{eq:ax6}, we can rewrite the spread density \eqref{eq:x64} as 
\begin{equation}\label{eq:x67}
  \mathcal{C}_\Omega = \dfrac{1}{2} + \mathfrak{Re}\left\lbrace \dfrac{i}{\gamma \pi}\int_{0}^\pi \dd k\, \Lambda(k) \right\rbrace,
\end{equation}
yielding
\begin{widetext}
    \begin{equation}\label{eq:x68}
       \mathcal{C}_\Omega = \begin{cases}
            \dfrac{1}{2} - \dfrac{1}{\pi}\mathfrak{Re}\left\lbrace\left[ 1 - i\dfrac{4(J-h)}{\gamma}  \right]\left[ E(\zeta) - 2 E\left(\frac{1}{2}\arccos \frac{h}{J}\Big\lvert\zeta
 \right) \right] 
 \right\rbrace, \quad &\gamma \leq \gamma_c(h), \\[9pt]
  \dfrac{1}{2} + \dfrac{1}{\pi}\mathfrak{Re}\left\lbrace 
\left[1 + i\dfrac{4(J-h)}{\gamma}\right]E(\zeta)\right\rbrace,\quad &\gamma \geq \gamma_c(h),
        \end{cases}
    \end{equation}
where $\zeta = -16J(4h+i\gamma)\left[ 4(h-J)+i\gamma\right]^{-2}$, and $E(\phi\vert m )$ and $E(\phi)$
are the incomplete and complete elliptic integrals of the second kind, respectively. Note that the spread density is continuous across the critical curve $\gamma_c(h)$ (see Fig.~\ref{fig:3}). Furthermore, in the absence of dissipation and keeping the sign convention of the spectrum  as $\lim_{\gamma\to 0}\Lambda(k)$, the spread density is constant for $h \leq J$ and monotonically increases for $h>J$, tending to unity as $h\rightarrow \infty$ \footnote{A similar result was obtained in Ref.~\cite{caputa_spread_2023}.}:
\begin{equation}\label{eq:x69}
   \mathcal{C}_\Omega = \begin{cases}
    \dfrac{1}{2} + \dfrac{1}{\pi}, \quad & h\leq J,\, \gamma = 0,\\[2ex]
    \dfrac{1}{2}  + \dfrac{1}{2\pi}\left\lbrace \dfrac{h-J}{h} E\left(-\dfrac{4hJ}{(h-J)^2}  \right) + \dfrac{h+J}{h}K\left(-\dfrac{4hJ}{(h-J)^2}  \right)\right\rbrace, \quad & h\geq J,\, \gamma = 0,
    \end{cases}
\end{equation}
\end{widetext}
where $K(\phi)$ is the complete elliptic integral of the first kind.  In the limit $\gamma \rightarrow 0$, Eq.~\eqref{eq:x68} coincides with Eq.~\eqref{eq:x69}; hence, $\mathcal{C}_\Omega$ is continuous for $(h,\gamma)\in [0,\infty]\times[0,\infty)$.

Given the analytical expressions for $\mathcal{C}_\Omega$, we find that, in the Hermitian limit, its first derivative with respect to $h$ diverges logarithmically as $h\rightarrow J^+$, i.e.,
\begin{equation}\label{eq:x70}
    \pdv{\mathcal{C}_\Omega}{h} \simeq  \frac{1}{2\pi J}\ln \left(\dfrac{2J}{{h-J}}\right) + \mathcal{O}(1),
\end{equation}
and it vanishes from the left. Therefore, in the same spirit as in Ref.~\cite{caputa_spread_2023}, we can assert that the spread density witnesses the quantum phase transition occurring at $(h,\gamma)=(J,0)$ by displaying a second-order phase transition. 

In the presence of dissipation, for a fixed $0<\gamma < 4J$, the critical real magnetic field is $h_c = \sqrt{J^2-\gamma^2/16}$ and, as $h\rightarrow h_c^-$, the second derivative of $\mathcal{C}_\Omega$ with respect to $h$ diverges as
\begin{equation}\label{eq:x71}
    \dfrac{\partial^2\mathcal{C}_\Omega}{\partial h^2} \simeq \dfrac{1}{\sqrt{2}\pi J^2}\frac{h_c}{\sqrt{J^2-h_c^2}}\sqrt{\dfrac{h_c}{h_c-h}} + \mathcal{O}\left(\sqrt{h_c-h}\right)
\end{equation}
and tends to a constant as $h\rightarrow h_c^+$. Hence, there is a third-order phase transition in the spread density once $\ket{0}$ evolves to $\ket{\Omega}$ via unitary dynamics [Eq.~\eqref{eq:n2}]. In a similar manner, if $h=0$, the second derivative of the spread  with respect to $\gamma$ behaves as
\begin{equation}\label{eq:c2g}
     \dfrac{\partial^2\mathcal{C}_\Omega}{\partial \gamma^2} \simeq \dfrac{4}{\pi\gamma^2 }\dfrac{J}{\sqrt{16J^2-\gamma^2}} + \mathcal{O}(1)
\end{equation}
for $\gamma \rightarrow 4J^-$.  

It is worth recalling that the entanglement entropy of a partition of the system in the state $\ket{\Omega}$ undergoes a phase transition across $\gamma =\gamma_c(h)$ [or, equivalently, $h= h_c(\gamma)$] from a logarithmic law in the gapless phase to an area law in the gapped phase \cite{zerba_measurement_2023,Turkeshi_2}. Thus, since $\mathcal{C}_\Omega$ also exhibits a phase transition at this boundary, we argue that the spread density may serve as a probe for various exotic phase transitions. We will elaborate on this expectation in the following section, where we consider the time dependence of the Krylov spread when the state is evolved by the non-Hermitian Hamiltonian \eqref{eq:x1}, as in Sec.~\ref{eq:ssec:time_evolution_of}.

\section{Dynamical phase transition and Krylov spread}
\label{Sec:V-DPT}

In this section, we calculate the Krylov spread of the state $\ket{0}$ evolving under the non-Hermitian Hamiltonian $H$, Eq.~\eqref{eq:x1}. Similarly to the Hermitian case [cf. Eq.~\eqref{eq:kk1}], the Krylov space for a non-Hermitian $H$ is $\mathfrak{K}\big(\ket{0}, H\big)\!=\text{span}\{H^n\ket{0}\}_{n\geq 0}$. Once the Gram-Schmidt procedure is applied to this set of vectors, one obtains the Krylov basis $\{\ket{K_n}\}_{n\geq 0}^D$,  where $D={\text{dim}\, \mathfrak{K}\big(\ket{0}, H\big)}$. The non-Hermitian Hamiltonian can also be written in a Hessenberg form with respect to the Krylov basis. 
Furthermore, the Krylov coefficients are $\{ \phi_n(t) = \bra{K_n}\ket{\psi(t)}\}_{n=0}^D$, where $\ket{\psi(t)}$ is the state \eqref{eq:x1.1}, which is evolved by the non-Hermitian Hamiltonian and contains the extra normalization factor not appearing in Eq.~\eqref{eq:kk2}. Then, the Krylov spread of this evolution is 
\begin{equation}
    C(\ket{0},H;t)  \equiv C(t) = 
 \sum_{n=0}n\abs{\phi_n(t)}^2.
\end{equation} 
Owing to the $\mathfrak{su}(2)$ structure of $H$ [see Eq.~\eqref{eq:x24}], the Krylov vectors can be written in the thermodynamic limit as in Eq.~\eqref{eq:Kn}  with $R_x(k)$ instead of $\alpha_+(k)$.

We will show that the infinite-time limit of $C(t)$ in the gapped phase coincides with the spread found in the previous section in the same phase. We will also demonstrate that the infinite-time average and the subsequent thermodynamic limit yield the spread found via unitary dynamics. 

Further, we introduce a dynamical quantity called the \emph{Krylov spread fidelity}, which we use to define the stationary state of the time-dependent Krylov spread density in the thermodynamic limit. With the aid of this measure, we present one of our main findings, namely, the identification of three dynamical phases found in $\ket{\Omega}$ when the spectrum is gapped, each one with an associated characteristic time.

\begin{figure}[b!]
\includegraphics[width=.35\textwidth,center]{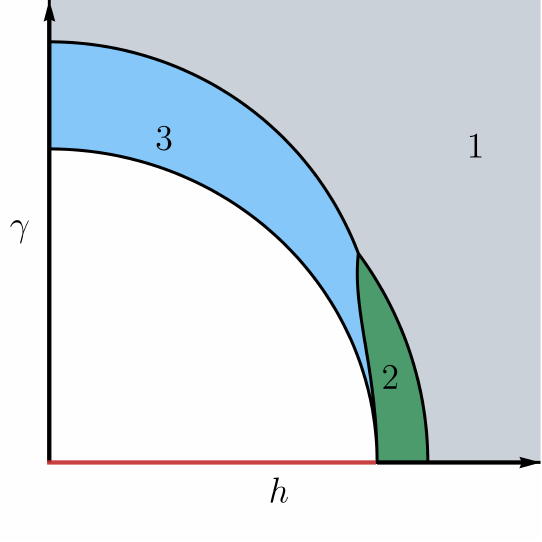}
    \caption{Schematic phase diagram defined by the fidelity \eqref{eq:fi2}. The white and red regions correspond to the gapless phase, where $\Gamma(\pm q) = 0$, and do not have any associated characteristic time of the form given by Eq.~\eqref{eq:time}. The dynamical phases and their characteristic times $t^*$ are: 1 (gray)  characterized by $t^*\simeq \gamma^{-1}$; 2 (green) by $t^*\simeq (4\abs{\Gamma(\bar k)})^{-1}$; and 3 (blue) by $t^* \simeq \abs{\gamma_Y}^{-1}$, where $\pm \bar k$ are the modes given by Eq.~\eqref{eq:kbar}.}
    \label{fig:4}
\end{figure}

\subsection{Time-dependent Krylov spread in the gapped phase and Krylov fidelity}
\label{subsec:5A}
Since the non-Hermitian Hamiltonian $H$ is also an element of $\mathfrak{su}(2)^{\otimes N}$ containing the raising and lowering operators per mode $J_{\pm}(k)$, one can apply the same reasoning as that used in Eqs.~\eqref{eq:x63} and \eqref{eq:x64}. Hence, in the thermodynamic limit, the time-dependent spread when $\ket{0}$ evolves by a non-Hermitian Hamiltonian with a gapped spectrum is 
\begin{equation}\label{eq:x72}
    \mathcal{C}(t) \coloneqq \dfrac{1}{\pi}\int_0^\pi
    \dd k\, \dfrac{\abs{A_+(k;t)}^2}{1+\abs{A_+(k;t)}^2} ,
\end{equation}
where $A_+(k;t)$ is given by Eq.~\eqref{eq:x36}.
To characterize the approach to the steady state in the thermodynamic limit, we define the \emph{spread fidelity}
\begin{equation}\label{eq:fi1}
    \mathcal{F}(t)\coloneqq \abs{\mathcal{C}(t)-\mathcal{C}_\Omega}.
\end{equation}
We denote the time for which the inequality 
\begin{equation}\label{eq:fi2}
    \mathcal{F}(t) < \epsilon
\end{equation}
holds for a given $\epsilon \in (0,1)$, as $t(\epsilon)$. Then, we denote $\mathcal{C}_{\text{st}}(t) = \mathcal{C}(t>t(\epsilon)) \simeq \mathcal{C}_\Omega$ as the spread density in the stationary-state regime, in the gapped phase. For an $\epsilon$-independent description, we define the time
\begin{equation}\label{eq:time}
    t^* \coloneqq \lim_{\epsilon \to 0}\dfrac{t(\epsilon)}{\abs{\ln \epsilon}}.
\end{equation}
Then, the stationary state holds for $t \gg t^*$. Since this time is defined for a vanishing $\epsilon$, it serves as a probe of dynamical phase transitions, as we show next.

For sufficiently small $\epsilon$ and large $t$, there are four main contributions to $\mathcal{F}(t)$,
\begin{equation}\label{eq:fi2.1}
    \mathcal{F}(t)\approx \abs{I_0(t)+I_\pi(t) + I_{b,1}(t)+I_{b,2}(t)},
\end{equation}
where the first two arise from expanding the integrand of $\mathcal{F}(t)$ around $k=0$ and $k=\pi$, and the other two from the bulk of the integration interval (see Appendix \ref{ssec:stationary_state_continuous}). The fidelity \eqref{eq:fi2.1} allows us to identify three different dynamical phases that characterize how $\mathcal{C}(t)$ approaches $\mathcal{C}_\Omega$.
These phases are shown in Fig.~\ref{fig:4} in gray, green, and blue, where the derivatives of $t^*$ with respect to $h$ or $\gamma$ are discontinuous across the boundaries separating the regions.  Note that the white and red regions in Fig.~\ref{fig:4} do not have an associated time $t^*$, as they lie in the gapless phase. For a given $\epsilon$, the characteristic times, which we denote with a slight abuse of notation as $t_i(\epsilon)$ and $t^*_i$ with $i=1,2,3$, are:
\begin{widetext}
\begin{alignat}{3}
    t_\text{1}(\epsilon) &\simeq \dfrac{1}{\gamma}\ln\left[ \dfrac{4J\epsilon}{\sqrt{16(h+J)^2+\gamma^2}+\sqrt{16(h-J)^2+\gamma^2}} \right], \qquad  &&t_{1}^*= \gamma^{-1}, \label{eq:t1}\\
    t_{\text{2}}(\epsilon) &\simeq  \dfrac{1}{8\abs{\Gamma(\bar k)}}W_0\left( \dfrac{\pi X^2(\bar k )\abs{\Gamma(\bar k)}}{2\epsilon^2\abs{\Gamma''(\bar k )}}\right) \ 
    \underbrace{\rightarrow}_{\epsilon\to 0} \ \dfrac{1}{4\abs{\Gamma(\bar k)}} \ln\dfrac{1}{\epsilon}, \qquad  &&t_{\text{2}}^*=  \left(4\abs{\Gamma(\bar k)}\right)^{-1},\\
     t_{\text{3}}(\epsilon) &\simeq  \dfrac{1}{2\abs{\gamma_Y}}W_0\left( \dfrac{8\pi \abs{Y(\bar k )}^2\abs{\gamma_Y}}{\epsilon^2 \abs{\Lambda''(\bar k )}}\right) \ \underbrace{\rightarrow}_{\epsilon\to 0} \ \dfrac{1}{\abs{\gamma_Y}} \ln\dfrac{1}{\epsilon},\qquad  &&t_{\text{3}}^*= \left|\gamma_Y\right|^{-1}. \label{eq:t3}
\end{alignat}
\end{widetext}
Here, 
\begin{equation}\label{eq:kbar}
    \bar k = \arccos\left(\dfrac{16hJ}{16h^2+\gamma^2} \right)
\end{equation}
is the mode with the slowest decay, i.e., $\text{min}_{k\in \mathcal{K}}\abs{\Gamma(k)} = \abs{\Gamma(\pm \bar k)}$.
The function $W_0(z)$ is the Lambert $W$ function, which has the asymptotics $$W_0(z\to \infty)\simeq \ln z -\ln(\ln z),$$ and $X(k)$ and $Y(k)$ are functions extracted from appropriate approximations of Eq.~\eqref{eq:fi1} in the bulk of the integration interval [see Eqs.~\eqref{eq:X} and \eqref{eq:Y}]. 
Finally, 
\begin{equation}
    \gamma_Y = 2\Gamma(\bar k ) + \frac{E'(\bar k )^2\Gamma''(\bar k )}{\abs{\Lambda''(\bar k)}^2}
\end{equation}
is an effective decay obtained from the condition \eqref{eq:an0}.

Let us recall that the real part of the spectrum goes from gapless ($h\leq J$) to gapped  ($h>J$) at $\pm q = \pm \arccos(h/J)$ [see Fig.~\ref{fig:1}], and thus one could naturally expect that the long-time dynamics would be different across the line $h=J$ for any $\gamma>0$. Nevertheless, this does not occur, and the modes $\pm q$ appear to be not that relevant for the dynamical phases shown in Fig.~\ref{fig:4}. Instead, the intricate and unexpected regions of the diagram indicate that the modes at $k=0,\pi$ and $k \approx \bar k$ are those dictating the long-time dynamics.

\subsection{Time-dependent Krylov density in the gapless phase}
In the gapless phase  $\gamma\leq\gamma_c(h)$, before taking the thermodynamic limit, let us assume that the modes $\pm q\in \mathcal{K}$ such that $\Gamma(\pm q ) =0$ exist. Then, following Eq.~\eqref{eq:x63}, the leading-order contribution to the complexity is 
\begin{multline}\label{eq:x73}
    {C(t)} =\dfrac{2\sin^2[E(q)t]}{\sin^2[E(q) t] + \cos^2[E(q) t + \lambda_q]}\\ + \sum_{k\in \mathcal{K}\setminus\{\pm q \}}C(t;k) + \mathcal{O}\left(N^{-1}\right).
\end{multline}
Taking the time average and then the thermodynamic limit yields
\begin{equation}\label{eq:x74}
    \lim_{N\to \infty}\dfrac{1}{N}\lim_{T\to \infty}\dfrac{1}{T-t_\star}\int_{t_\star}^T \dd t\,  C(t) \simeq \mathcal C_\Omega.
\end{equation} 
Note that the lower integration limit is given by Eq.~\eqref{eq:x38}; this time is found for the gapped modes in $\mathcal{K}$. We observe that, in the thermodynamic limit, the time average of the spread density of the state evolving from $\ket{0}$ to $\mathcal{A}(t)\ket{\Omega} + \mathcal{A}^*(t)\ket{q,-q}$ with the non-Hermitian Hamiltonian $H$ coincides with the spread density \eqref{eq:x69} found via the unitary dynamics induced by $\tilde H$. 
 
We emphasize that we calculated the spread of the evolution of $\ket{0}$ as it is the state that, at (appropriately defined) long times, evolves to $\ket{\Omega}$ when the spectrum of $H$ is gapped and to the state \eqref{eq:x42}, when it is gapless. On top of this, as we discussed in Sec.~\ref{eq:ssec:time_evolution_of}, any initial state that has a non-zero overlap with $\ket{\Omega}$ and the state possessing two non-decaying excitations will always have the structure comprising the gapless or gapped stationary states of $\ket{0}$. Consequently, the spreads we found behave universally for such states in the long-time limit.

\section{Summary and discussion}\label{sec:conclusions}

Krylov subspace methods offer an efficient approach to simulating quantum systems, which requires only an initial state, a generator of dynamics, and an inner product. One such method, based on Krylov spread complexity, quantifies the spread of an evolved state within the Krylov basis constructed from these three ingredients. This seemingly straightforward measure has proven valuable in exploring diverse quantum phenomena, including entanglement and dynamical phase transitions, in both Hermitian and non-Hermitian systems. Furthermore, connections have been established between Krylov spread and other complexity measures (e.g., circuit complexity, Nielsen complexity) and dynamical quantities (e.g., Loschmidt echo, quantum Fisher information). However, the potential of Krylov spread complexity to reveal quantum dynamical phase transitions beyond the reach of conventional methods remains largely unexplored. This work addressed this gap.

In this paper, we have investigated the spin-spin correlations and their relation with Krylov spread for a non-Hermitian Hamiltonian describing an Ising chain with a complex-valued transverse magnetic field. We found that (i) the infinite-time Krylov spread is related to the same-site spin correlation function, whereas (ii) its time dependence upon approaching the asymptotic value reveals dynamical phase transitions when the imaginary part of the spectrum is gapped.

First, by employing asymptotic methods, we fully characterized the static correlation function of the spin $z$-component in the vacuum state of the non-Hermitian Hamiltonian, covering the entire phase diagram defined by the real and imaginary parts of the magnetic field ($h,\gamma$). Then, we established a connection between the Krylov spread of the all-spin-down state evolving unitarily toward the Bogoliubov vacuum of the non-Hermitian Hamiltonian, which allowed us to calculate the spread analytically in terms of elliptic integrals. One surprising result was that the spread density in the thermodynamic limit displayed a third-order phase transition across the critical line separating the gapless phase from the gapful one. Lastly, for the same initial state evolving under the non-Hermitian Hamiltonian itself, we found that the spread density could uncover dynamical phase transitions in the gapped region of the phase diagram, which remained hidden from other conventional measures. In what follows, we provide an expanded summary of our results and future directions. 

More specifically, we addressed a non-Hermitian 1D Ising model by choosing the complex eigenvalues \eqref{eq:x09} of the non-Hermitian Hamiltonian  \eqref{eq:x1} to always have a negative imaginary part for each momentum mode. We derived analytically the asymptotics of the spin-spin correlation function \eqref{eq:x46} in the right-ground state (i.e., the non-Hermitian Bogoliubov vacuum) and the Krylov spread complexity [Eqs.~\eqref{eq:x67} and \eqref{eq:x72}] of the state that evolves from the Jordan-Wigner vacuum to the right-ground state. 
The asymptotic behavior of the correlation function is of the form $x^{-3}\exp(-x/\xi)\times\left(\text{oscillatory term in $x$}\right)$ as $x\rightarrow \infty$ when the imaginary part of the complex spectrum is gapped [Eq.~\eqref{eq:czz1}]. In the gapless phase, the generic behavior of the correlation function is of the form  $x^{-n}\times\left(\text{oscillatory term in $x$}\right)$ [Eqs.~\eqref{eq:tcor}-\eqref{eq:tcor1}]. If $\gamma < \gamma_c(h)$, no exceptional points emerge and only $\Gamma(\pm q) = 0$ and $n=2$, while if $\gamma = \gamma_c(h)$, $n=4$ and two exceptional points develop at $k = \pm q = \pm \arccos(h/J)$, where $\Lambda(\pm q ) =0$. What is more, in the limit $\gamma \rightarrow 0$, the correlation function is oscillatory and decays algebraically as $x^{-2}$ instead of displaying an exponential decay as in the ferromagnetic phase. The correlation functions at $(h,\gamma)=(J,0)$ and $(h,\gamma)>(J,0)$ are the same as in the Ising critical phase and paramagnetic phase, respectively [see Eq.~\eqref{eq:corrher} and Fig.~\ref{fig:1.2}]. 

Moving to Krylov complexity, similar to what was done in Ref.~\cite{caputa_spread_2023} for the non-dissipative limiting case ($\gamma \rightarrow  0$), the thermodynamic limit of the Krylov spread complexity density \eqref{eq:x69} of the state evolving unitarily from $\ket{0}$ to $\ket{\Omega}= \Omega\ket{0}$, 
was found to be constant for $h\leq J$. In contrast, for $h>J$, we found that the spread density displayed a monotonically increasing behavior and its asymptotic behavior tending to unity as $h\rightarrow \infty$. In the limit of no dissipation, the first derivative of the spread density with respect to $h$ diverges logarithmically as $h$ approaches $J$ from the right. 
In the non-Hermitian case ($\gamma>0$), we obtained the relationship between the spin-spin correlation function and the Krylov spread density [Eq.~\eqref{eq:x65}]. By exploiting this relation, we found that the second derivative with respect to $\gamma$ of the Krylov spread density (in the thermodynamic limit) 
exhibits a power-law behavior near the real critical field $h_c = (J^2-\gamma^2/16)^{-1/2}$, diverging as $(h_c-h)^{-1/2}$ when $h_c$ is approached from the left [Eq.~\eqref{eq:c2g}]. It is well-known \cite{zerba_measurement_2023,Turkeshi_1,Turkeshi_2} that entanglement phase transitions occur in the non-Hermitian vacuum $\ket{\Omega}$ when real part of the magnetic field (or the imaginary part) crosses the critical value, where the entanglement entropy obeys an area law for $\gamma >\gamma_c(h)$, and a logarithmic law for $\gamma\leq \gamma_c(h).$ We argue that the Krylov spread also serves as a probe of such a transition, and it might be possible that the Krylov spread density and its derivatives can serve as a probe for quantum-phase transitions in general. 

In the same gapped phase, we found that the spread density of the state that evolves under unitary dynamics from $\ket{0}$ to $\ket{\Omega}$  coincides with the spread density of the state that evolves under the non-Hermitian Hamiltonian from $\ket{0}$ to $\ket{\Omega}$ in the infinite-time limit. In addition, to have a well-defined stationary state regime in the thermodynamic limit, we defined a fidelity based on the absolute value of the difference between the infinite-time and time-dependent Krylov spread densities in the gapped phase. This \emph{Krylov spread fidelity} yielded one of our key results: refining the gapped region $\gamma > \gamma_c(h)$ of the $h$-$\gamma$ plane into three subregions (see Fig.~\ref{fig:4}). That is to say, we found that for long times and vanishing fidelity, there are three characteristic timescales after which the stationary state regime is reached [Eqs.~\eqref{eq:t1}-\eqref{eq:t3}]: one given by the imaginary part of the magnetic field $\gamma$, one proportional to the imaginary part of the complex spectrum of Hamiltonian $H$ evaluated at the slowest decay mode, and another, highly nontrivial one obtained from a saddle point of the spectrum belonging to the complex $k$-space, involving thus the spectrum and its derivatives. This indicates that the non-trivial dynamical features are not only mainly dictated by the modes for which the real part of the spectrum vanishes in the gapped phase and the slowest decay mode, but also by a broader set of modes, including a continuum of them (see Appendix~\ref{ssec:stationary_state_continuous}). In fact, our results showed that the modes, in which the real part of the spectrum vanishes, did not have a special role in the long-time behavior of the Krylov spread density. Therefore, our results suggest that the Krylov spread density broadens the characterization of dynamical phases via the Krylov spread fidelity. That said, we are unaware of whether the standard time-dependent correlation function, the time-dependent entanglement entropy, or any other dynamical measure can easily discern these phases when the imaginary part of the spectrum is gapped. In any case, the spread in Krylov subspace serves, by construction, as a natural, powerful tool for unraveling such hidden dynamical transitions.  

For a gapless spectrum, we demonstrated that when the zero modes $\pm q$ exist, $\ket{0}$ evolves to $\ket{\psi_{\text{st}}(t)} \simeq \mathcal{A}(t)\ket{\Omega} + \mathcal{A}^*(t)\eta^*_{-q}\eta^*_q\ket{\Omega}$ for sufficiently long times. This is a more precise result than the one shown in Ref.~\cite{zerba_measurement_2023}, where the coefficient of $\ket{\Omega}$ was guessed to be a constant. In the same phase and for the same initial state, we demonstrated that the time average of the time-dependent spread obtained via dissipative dynamics coincides with the one obtained via unitary dynamics. This is a result akin to the one found in Refs.~\cite{zerba_measurement_2023, Turkeshi_2}, where it was found that the properties of the entanglement entropy are encoded in the non-Hermitian vacuum $\ket{\Omega}$. In the context of Krylov spread in the stationary state regime, even if the state $\ket{0}$ evolves to the one containing the two zero modes, it is sufficient to study the spread via the natural relation between $\ket{0}$ and $\ket{\Omega}$, implying that $\ket{\Omega}$ encodes all the pertinent physical properties. However, as we discussed above, it does matter how the state $\ket{\Omega}$ is reached. 

A promising albeit more challenging future direction of our work is to compute additional canonical correlation functions in the state $\ket{\Omega}$, such as $\texttt{C}^{xx}(x)$ and $\texttt{C}^{yy}(x)$, along with the time-dependence of such correlation functions, including $\texttt{C}^{zz}$. Based on our results on $\texttt{C}^{zz}(x)$ and on what was reported in Ref.~\cite{Liu_2021} mainly via numerical methods, we expect that these exhibit distinct asymptotic behaviors compared to their Hermitian counterparts \cite{BI,BII,BIII,BIV}.
In addition, all the above correlations and spreads can also be calculated in the left-ground states and left and right, i.e., ${}_{L}\expval{\ldots}_L$ and ${}_{L}\expval{\ldots}_R$, respectively. Further, the critical behavior of the spread and its derivatives with respect to physical parameters can be examined in other non-Hermitian models, potentially revealing novel phase transitions.  Another interesting direction would be to continue studying and exploiting the relationship we found (at least for this model) between the contractions and the spread, as calculating one quantity might facilitate obtaining the other. It is also worth exploring other Krylov-based characteristics of the models to understand whether they can reveal other unexpected transitions. In particular, it is interesting to investigate whether there is an analog of Eq.~\eqref{eq:x63}, where the Krylov entropy (see, e.g., Ref.~\cite{Barb_n_2019}) in the thermodynamic limit can be written as the integral over the momentum modes of the Krylov entropy associated to a single $\mathfrak{su}(2)$ copy. 

Our study of non-Hermitian chains can be naturally applied to, e.g., the extended Kitaev chain \cite{rahul_topological_2022,rahul_unconventional_2023} or the Su–Schrieffer–Heeger (SSH) model, among other non-Hermitian systems.
Further, our approach can be generalized to systems governed by the Lindblad equation, Eq.~\eqref{eq:x01}, or by stochastic master equations (\ref{eq:xsto}), as well as to hybrid dynamics, which is the interpolation between non-Hermitian Hamiltonian and Lindbladian dynamics (cf. Refs.~\cite{liu2022krylov, op1, op3, bhattacharya2022operator, bhattacharya2023krylov} where Krylov spread was studied in open systems).
Finally, our study paves the way for exploring the hidden dynamical phase transitions
of monitored systems in terms of Krylov fidelity. Importantly, Krylov methods appear to be extremely efficient in characterizing the degree of chaos in many-body systems, while the measurements naturally introduce quantum stochasticity of evolution.

\section*{Acknowledgements}
One of the authors (E.M.G.) wishes to thank Xhek Turkeshi and Horacio M. Pastawski for helpful discussions. Y.G. was supported by the Deutsche Forschungsgemeinschaft (DFG,
German Research Foundation) grant SH 81/8-1 and by
a National Science Foundation (NSF)–Binational Science Foundation (BSF) grant 2023666. 

\textit{Note added:} While finalizing this manuscript, we became aware of a related work \cite{chakrabarti2025}, where dynamics of  a non-Hermitian SSH model was analyzed in Krylov space.

\begin{appendix}
\begin{widetext}

\section{The Bogoliubov vacuum as a generalized coherent state}\label{sec:equivalence_formulas}

In this Appendix, we show that the coherent state  $\ket{\Omega}$ \eqref{eq:x30} introduced in Sec.~\ref{ssec:non_hermitian_vacuum}  coincides with the non-Hermitian  Bogoliubov vacuum state \eqref{eq:x28} when the coefficients $\alpha_\pm(k)$ are given by Eq.~\eqref{eq:x29}. 
Further, we also demonstrate the relation \eqref{eq:x32} between the fermionic operators.

Start with the normal-ordering decomposition formula of $\mathfrak{su}(2)$ 
\cite{Ban:93}
\begin{equation}\label{eq:app1}
    \exp(\alpha_+ J_+ + \alpha_z J_z + \alpha_- J_-) = \exp(A_+J_+)\exp\big[\ln (A_z)  J_z\big] \exp(A_- J_-),
\end{equation}
with $\alpha_z=0$ and
\begin{align}
    A_{\pm} = \dfrac{ (\alpha_{\pm}/\phi)\sinh \phi}{\cosh \phi - (\alpha_z/2\phi)\sinh\phi },\qquad 
    A_z = \left[\cosh\phi - (\alpha_z/2\phi)\sinh \phi \right]^{-2}, \qquad
    \phi = \sqrt{(\alpha_z/2)^2+\alpha_+\alpha_-}\, .
    \label{eq:app2}
\end{align}
By using Eq.~\eqref{eq:x29}, we get 
\begin{equation}
 A_+ = -A_-^*=-\tau_k, \qquad A_z =\sqrt{{1+\abs{\tau_k}^2}},\qquad \phi = i\arccos\, (1 + \abs{\tau_k}^2 )^{-1/2}.
\end{equation} 
Recalling that $J_-(k)\ket{1/2,-1/2}_k = 0$ together with $J_z(k)\ket{1/2,-1/2}=-(1/2)\ket{1/2,-1/2}_k$, we have
\begin{subequations}
    \label{eq:app4}
    \begin{align}
     \Omega_k\ket{1/2,-1/2}_k & = \exp[\alpha_+(k) J_+(k) + \alpha_-(k)J_-(k)]\ket{1/2,-1/2}_k\label{eq:app4.1}\\
     &= \exp[-\tau_k J_+(k)]\exp[\ln\left(1+\abs{\tau_k}^2\right)^{-1/2}J_z(k)]\exp[\tau_k^*J_-(k)]\ket{1/2,-1/2}_k   \label{eq:app4.2}\\
     &= \dfrac{\exp[-\tau_k J_+(k)]}{\sqrt{1+\abs{\tau_k}^2}}\ket{1/2,-1/2}_k. 
    \end{align}
\end{subequations}
Thus, Eq.~\eqref{eq:x28.3} coincides with Eq.~\eqref{eq:x30.2}. Note that when normalizing the equation that yields the state \eqref{eq:x30.2}, a global phase $\exp[i\varphi(\Omega)] = \prod_{k\in \mathcal{K}^+}u_k/\abs{u_k}$ is obtained. Since these states belong to the same ray, the global phase is unimportant for the coherent state $\ket{\Omega}$.

To demonstrate relation \eqref{eq:x32}, we utilize the following identities: 
\begin{equation}\label{eq:app6}
    \exp[\beta_- J_-(k)]\,c_k \exp[-\beta_- J_-(k)] = c_k, \qquad \exp[\beta_+ J_+(k)]\,c_k \exp[-\beta_+ J_+(k)] = c_k - \beta_+ c_{-k}^\dagger,
\end{equation}
and 
\begin{equation}\label{eq:app7}
    \exp[\beta_z J_z(k)]\, c_k \exp[-\beta_z J_z(k)] = \exp(-\frac{1}{2}\beta_z)c_k,
\end{equation}
where $\beta_{\pm},\beta_z \in \mathbb{C}$. Using these formulas together with
\begin{subequations}
\label{eq:appp9}
\begin{align}
    \Omega &=  \prod_{k \in \mathcal K^+}\Omega_k = \prod_{k \in \mathcal K^+}\exp[A_+(k)J_+(k)]\exp[\log (A_z(k))  J_z(k)]\, \exp[A_-(k) J_-(k)], \label{eq:app9.2}
\end{align}
\end{subequations}
we have
\begin{subequations}
\begin{align}
    \Omega c_k \Omega^{-1} &= \Omega_k c_k \Omega_k^{-1} = A_z(k)\left(c_k - A_+(k)c_{-k}^\dagger  \right)  = \dfrac{c_k+\tau_k c_{-k}^\dagger}{\sqrt{1+\abs{\tau_k}^2}},
\end{align}
\end{subequations}
which coincides with Eq.~\eqref{eq:x32}.

\section{Stationary state}\label{sec:stationary_state}

In this Appendix, we determine the time $t_\star$ \eqref{eq:x38} that defines the onset of the stationary-state regime, based on how $\ket{0}$ reaches $\ket{\Omega}$ via $H$ when the imaginary part of the spectrum is gapped. Once we find $t_\star$, we demonstrate that $\ket{0}$ evolves to Eq.~\eqref{eq:x42} when the imaginary part of the spectrum is gapless.

We start with Eq.~(\ref{eq:x36}), which we identically rewrite as 
\begin{align}
    A_+(k;t)  = -\tau_k \dfrac{1}{1-l_k(t)},
    \label{eq:y1}
\end{align}
where $\tau_k$ is defined in Eq.~\eqref{eq:tauk} and $l_k(t)$ is given by Eq.~\eqref{eq:gk}.
Clearly, if $\Gamma(k)$ is gapped, then $l_k(t)\rightarrow 0$ as $t\rightarrow \infty$. Let us pick any operator from the product in Eq.~\eqref{eq:x37.2} corresponding to an arbitrary mode leading to a gapped imaginary spectrum, and let us demand from
\begin{equation}\label{eq:y3}
   \left(1+\abs{A_+(k;t)}^2\right)^{-1/2}= \left(1+\abs{\tau_k}^2\right)^{-1/2}\left[1- w_k(\Omega) \dfrac{\abs{1-l_k(t)}^2-1}{\abs{1-l_k(t)}^2}\right]^{-1/2},
\end{equation}
where $w_k(\Omega) = \abs{\tau_k}^2(1+\abs{\tau_k}^2)^{-1}$, that
\begin{equation}\label{eq:y4}
  \abs{L_k(t)} \coloneqq w_k(\Omega)  \abs{\dfrac{\abs{1-l_k(t)}^2-1}{\abs{1-l_k(t)}^2}} < 1
\end{equation}
holds. Given this condition, we have
\begin{subequations}
\label{eq:y4.1}
    \begin{align}
       F_k(t) &\coloneqq \dfrac{\exp[A_+(k;t)J_+(k)]}{\sqrt{1+\abs{A_+(k;t)}^2}}
       = \dfrac{\exp[-\tau_k J_+(k)]}{\sqrt{1+\abs{\tau_k}^2}\sqrt{1+L_k(t)}} - \sqrt{\dfrac{w_k(\Omega)}{1+L_k(t)}}l_k(t)e^{i\text{Arg}\,\tau_k}J_+(k) \\
       &= \dfrac{\exp[-\tau_k J_+(k)]}{\sqrt{1+\abs{\tau_k}^2}} - \dfrac{l_k(t)}{1-l_k(t)}\sqrt{w_k(\Omega)}e^{i\text{Arg}\,\tau_k}J_+(k) + \mathcal{O}\textbf{(}L_k(t)\textbf{)}.
       \end{align}
\end{subequations}
Note that if $\abs{L_k(t)} < 1$, then $\sqrt{w_k(\Omega)}\abs{l_k(t)(1-l_k(t))^{-1}} < 1$ and $\abs{l_k(t)}<1$; therefore, the magnitude of the second and remaining $\mathcal{O}\textbf{(}l_k(t)\textbf{)}$ terms in the above equation are smaller than unity for the times that fulfill the inequalities and tend to zero as $t \rightarrow \infty$. Given all this, we proceed to calculate bounds of $t$ such that $\abs{L_k(t)} < 1$ holds. To this end, 
\begin{align}\label{eq:y5}
    \abs{L_k(t)} <1\ \Leftrightarrow \ w_k(\Omega)\abs{2\, \mathfrak{Re}\,l_k(t) - \abs{l_k(t)}^2} < 1-2\,\mathfrak{Re}\, l_k(t) + \abs{l_k(t)}^2\ \Rightarrow \ 1 + \left[1 - w_k(\Omega) \right]\left( \abs{l_k(t)}^2  -2 \, \mathfrak{Re}\, l_k(t)\right) >0.
\end{align}
Note that we have chosen the right-hand side of the inequality, i.e., $w_k(\Omega)\left[2\, \mathfrak{Re}\, l_k(t)-\abs{l_k(t)}^2\right] < \abs{1-l_k(t)}^2,$ instead of the other one since the latter gives negative times. We thus have 
\begin{equation}\label{eq:y6}
    e^{4\abs{\Gamma(k)}t} + 2e^{2\abs{\Gamma(k)}t}\left[ \dfrac{2\left( 1- w_k(\Omega)  \right)}{\abs{f_k}}\cos[2E(k) t + \vartheta_k] - \cos(2E(k)t) \right] + \dfrac{4\left(1-w_k(\Omega)\right)\left(1 - \abs{f_k}\cos\vartheta_k\right)}{\abs{f_k}^2} + 1 > 0.
\end{equation}

Finding the tightest bound is difficult, if not impossible, so it is enough to find the times for which the lower envelope of the left-hand side lies above zero. To find it, we need the amplitude of the oscillatory function that multiplies the second exponential, which has the following generic form:
\begin{equation}\label{eq:y7}
    b \cos(a t + \phi) - \cos(at) = \text{sgn}(b\cos\phi - 1)\sqrt{1-2b\cos\phi + b^2}\cos\left[ at + \arctan\left( \dfrac{2b\sin\phi}{1-b\cos\phi} \right) \right].
\end{equation}
Let $m_k = \sqrt{1-2b\cos\phi + b^2}$ thus be the amplitude with $b = 2\left[1- w_k(\Omega)\right]\abs{f_k}^{-1}$ and $\phi = \vartheta_k$, and then consider the easier inequality
\begin{equation}\label{eq:y8}
    e^{4\abs{\Gamma(k)}t} - 2\, m_k\, e^{2\abs{\Gamma(k)}t} + \beta_k >0,
\end{equation}
where $\beta_k$ is given by Eq.~(\ref{eq:x39}).
The times satisfying Eq.~\eqref{eq:y8} are 
\begin{equation}\label{eq:y10}
    t_k > \dfrac{1}{2\abs{\Gamma(k)}}\ln\left( m_k + \sqrt{m_k^2-\beta_k} \right).
\end{equation}
Therefore, we define the stationary state regime for $t \gg t_\star =  \max_{k\in \mathcal{K}^+\setminus \{ q\}}t_k$.

The immediate conclusion is that if the imaginary part of the spectrum $\Lambda(k)$ is gapped, then $\ket{0}$ reaches the Bogoliubov right vacuum in the stationary state regime, i.e.,  $\ket{\psi(t \gg t_\star)} = \ket{\psi_{\text{st}}(t)} \simeq \ket{\Omega}$, where $\simeq$ means that we keep the leading-order terms only. Let us see what occurs when $\Gamma(\pm q) =0$, where $q = \pm \arccos(h/J)$. 

Starting from Eq.~\eqref{eq:x37}, we know that for all the modes different than $\pm q$ [see Eq.~\eqref{eq:y4.1}], 
\begin{equation}\label{eq:y11}
    F_k(t) = \dfrac{\exp[-\tau_k J_+(k)]}{\sqrt{1+\abs{\tau_k}^2}} + \mathcal{O}(l_k(t)).
\end{equation}
In short, $F_k(t)\simeq \Omega_k$ and
\begin{equation}\label{eq:y12}
    \Omega_{\slashed{q}} \coloneqq \prod_{k\neq q}\Omega_k.
\end{equation}
Therefore, in the stationary state regime, we can write 
\begin{equation}\label{eq:y13}
    \ket{\psi_{\text{st}}(t)} = F_q(t)\Omega_{\slashed{q}}\ket{0} + \mathcal{O}\left( \prod_{k\in \mathcal K \setminus \{\pm q\}}l_k(t)
 \right)
\end{equation}
where
\begin{equation}\label{eq:y14}
    F_q(t) = \frac{\exp[A_+(q;t)J_+(q)]}{\sqrt{1+\abs{A_+(q;t)}^2}},
\end{equation}
and
    \begin{align}
        A_+(q;t) = -\dfrac{iR_x(q)}{E_q\cot [E(q)t] + iR_z(q)} = -i\dfrac{2J\sin q}{E(q)\cot[E(q) t]-\gamma/2} = -i\dfrac{\sin E(q) t}{\cos[E(q) t + \lambda_q]} = -ia(t),
        \label{eq:y14.3}
    \end{align}
where $\tan\lambda_q = \gamma[2E(q)]^{-1}$.  Continuing with Eq.~\eqref{eq:y13}, we can add to it the zero $\mathcal{A}(t)(\Omega_q - \Omega_q) =0$ and keep only the leading terms:
\begin{align}\label{eq:y15}
    \ket{\psi_{\text{st}}(t)} &\simeq \mathcal{A}(t)\ket{\Omega} + \Omega_{\slashed{q}}\left[F_q(t)-\mathcal{A}(t)\Omega_q\right]\ket{0},
\end{align}
where $\mathcal{A}(t)$ is an arbitrary complex function. Let us note that we have also used $\Omega_{\slashed{q}}\Omega_q\ket{0}= \Omega_q\ket{0}=\ket{\Omega}$. By inserting $\Omega^\dagger\Omega = I$ in the last term of Eq.~~\eqref{eq:y15} and using $[F_q,\Omega_{\slashed{q}}] = 0$, we get
\begin{equation}\label{eq:y16}
    \ket{\psi_{\text{st}}(t)} \simeq \mathcal{A}(t)\ket{\Omega} + \left[F_q(t)\Omega_q^\dagger - \mathcal{A}(t)I \right]\ket{\Omega}.
\end{equation}

The two-quasiparticle state created from $\ket{\Omega}$ reads 
\begin{equation}\label{eq:y17}
    \ket{q,-q} = N_q\eta^*_{-q}\eta_{q}^*\ket{\Omega}= \dfrac{1}{\sqrt{2}}\left( 1
+e^{i\lambda_q}J_+(q)  \right)\Omega_{\slashed{q}}\ket{0}.
\end{equation}
We want the term in brackets of Eq.~\eqref{eq:y16}, i.e.,
\begin{equation}\label{eq:y18}
    \left[  F_q(t)\Omega_q^\dagger -\mathcal{A}(t)I\right]\ket{\Omega} = \left[ \dfrac{1}{\sqrt{1+a(t)^2}} - \dfrac{\mathcal{A}(t)}{\sqrt{2}}+\left( \dfrac{\mathcal{A}(t)e^{-i\lambda_q}}{\sqrt{2}} - i\dfrac{a(t)}{\sqrt{1+a(t)^2}}\right)J_+(q) \right]\Omega_{\slashed{q}}\ket{0},
\end{equation}
to be equal to $\mathcal{B}(t)\ket{q,-q}$, where $\mathcal{B}(t)$ is some complex function. Hence, we must have
\begin{equation}\label{eq:y19}
    \mathcal{A}(t) = \mathcal{B}^*(t) = \dfrac{e^{i\lambda_q} + ia(t)}{\sqrt{2}\sqrt{1+a(t)^2}\cos\lambda_q}
\end{equation}
With the above identification, if there are non-decaying modes such that $\Gamma(\pm q)  = 0$, the state $\ket{0}$ evolves to $\ket{\psi_{\text{st}}(t)} \simeq \mathcal{A}(t)\ket{\Omega} + \mathcal{A}^*(t)\ket{q,-q}$ in the stationary state regime, which is Eq.~\eqref{eq:x42} of the main text.

\section{Calculation of the spin-spin correlation function}
\label{App:Czz}
In this appendix, we present a detailed calculation of the asymptotic expansion of the correlation function \eqref{eq:x51} as  $x\rightarrow \infty$. In Sec.~\ref{ssec:app_gapped_phase}, we calculate the asymptotics of $\texttt{C}^{zz}(x)$ in the gapped phase, Eq.~\eqref{eq:czz1}, by applying tools from complex analysis. In Sec.~\ref{ssec:gapless_phase}, we treat the same correlation function in the gapless phase. However, we implement more conventional methods using exponential regulators and find Eqs.~\eqref{eq:tcor}-\eqref{eq:tcor2}. In Sec.~\ref{ssec:hermitian_case}, we demonstrate Eq.~\eqref{eq:corrher}, which is the asymptotics of $\texttt{C}^{zz}(x)$ when the limit $\gamma\rightarrow 0$ is taken at the level of the non-Hermitian Hamiltonian \eqref{eq:x1}. Finally, in Sec.~\ref{ssec:correlation_function_standard}, we calculate Eq.~\eqref{eq:ph}, which is the asymptotic expansion $\texttt{C}^{zz}(x)$ in ground state of $H\vert_{\gamma \equiv 0}$. 

\subsection{Gapped phase}\label{ssec:app_gapped_phase}
In this section, we implement the Darboux method to calculate the asymptotic behavior of $\texttt{C}^{zz}(x)$ when the imaginary part of the spectrum \eqref{eq:x09} is gapped. In summary, we define contour integrals in the complex plane containing the contractions $\expval{\texttt{A}_0 \texttt{A}_x}$ and $\expval{\texttt{B}_0 \texttt{A}_{\pm x}}$ (see Eqs.~\eqref{eq:cintegral} and \eqref{eq:cont1}, respectively), and perform suitable series expansion around the branch points that contribute the most to the integrals. In what follows, we identify some useful relations that simplify the complicated integrands of Eqs.~\eqref{eq:x48} and \eqref{eq:x50}.

By using Eqs.~\eqref{eq:ax2} and \eqref{eq:Gamma-gamma},
we can rewrite the functions of $u_k$ and $v_k$ appearing in Eqs.~\eqref{eq:x48} and \eqref{eq:x50} as
\begin{equation}\label{eq:appp2}
       -\dfrac{\abs{u_k^2}-\abs{v_k}^2}{\abs{u_k}^2+\abs{v_k}^2} = \mathfrak{Re}\left\lbrace   \dfrac{2i }{\gamma}
 \Lambda(k)\right\rbrace,
\end{equation}
and
\begin{equation}\label{eq:app3}
     \mathfrak{Re} \left\lbrace \dfrac{u_kv_k^*}{\abs{u_k}^2+\abs{v_k}^2} \right\rbrace = \mathfrak{Re}\left\lbrace \dfrac{u_k}{v_k} \right\rbrace\dfrac{\gamma-2\Gamma(k)}{2\gamma}.
\end{equation}
Now, by noting that 
\begin{equation}\label{eq:appp4}
    \dfrac{u_k}{v_k} = \dfrac{E(k) + 2(h-J \cos k)}{2 J \sin k} + i\dfrac{\gamma + 2\Gamma(k)}{4 J \sin k},
\end{equation}
we have
\begin{equation}\label{eq:appp5.1}
      \mathfrak{Re}\left\lbrace \dfrac{u_k}{v_k}\right\rbrace\, \dfrac{\gamma - 2\Gamma(k)}{\gamma} = \mathfrak{Re}\left\lbrace 
 \dfrac{\gamma + i 4(h-J\cos k)}{\gamma 4 J \sin k} \, \Lambda(k)\right\rbrace.
\end{equation}
Hence, the contractions $\expval{\texttt{A}_0 \texttt{A}_x}$ and $\expval{\texttt{B}_0 \texttt{A}_{\pm x}}$ can be written as in Eqs.~\eqref{eq:ax6.1} and \eqref{eq:ax6}, respectively.
Recall that the gapped phase corresponds to  $\gamma > \gamma_c(h).$ In what follows, we first address $\expval{\texttt{A}_0 \texttt{A}_x}$ and then $\expval{\texttt{B}_0 \texttt{A}_{\pm x}}$.

\subsubsection{\texorpdfstring{Contraction \textnormal{\(\expval{\texttt{A}_0\texttt{A}_x}\)}}{Contraction}}

The explicit form of $\Gamma^2(k)$ is
\begin{equation}\label{eq:app9}
    \Gamma^2(k) = -2(J^2+h^2)+ \dfrac{\gamma^2}{8} + 4hJ\cos k+ \dfrac{1}{2}\left( 4\gamma^2\left( h-J \cos k \right)^2 + \left[ 4(J^2+h^2) - \dfrac{\gamma^2}{4} - 8hJ \cos k \right]^2\right)^{1/2},
\end{equation}
and after replacing it into Eq.~\eqref{eq:ax6.1} and using the fact that  $\int_{0}^\pi \dd k \, \sin kx\, \csc k \, (a_0 + a_1 \cos k) = a_1 \pi$ for $x\in 2\mathbb{Z}^+$ and $a_{0,1}\in \mathbb{C}$, we get 
\begin{equation}\label{eq:app10}
    \expval{\texttt{A}_0 \texttt{A}_x} = -i\dfrac{4h}{\gamma} - i\int_{0}^\pi \, \dd k\sin kx \, \mathcal{G}(k),
\end{equation}
where 
\begin{equation}\label{eq:app11}
    \mathcal{G}(k) = \dfrac{1}{2J\gamma \pi }\csc k \left[4\gamma^2(h-J\cos k)^2 + \left( 4(J^2+h^2) -\dfrac{\gamma^2}{4} - 8hJ\cos k
 \right)^2 \right]^{1/2}.
\end{equation}
For convenience, let us write
\begin{equation}
    \mathcal{I}(x) = -i\int_0^\pi \, \dd k \sin kx\, \mathcal{G}(k) = i\, \mathfrak{Im}\left\lbrace \text{p.v.}\int_0^\pi \, \dd k\, e^{-ikx}\tilde{\mathcal{G}}(k) \right\rbrace,
\end{equation}
where p.v. denotes the Cauchy principal value and the singular points are $k = 0$ and $\pm \pi$. To attack $\mathcal{I}(x)$, we first consider the contour integral in the complex plane 
\begin{equation}\label{eq:cintegral}
    \int_C \dd z  f(z) = \int_C\dd z \, z^{-x}\tilde{\mathcal{G}} (z) ,
\end{equation}
where 
\begin{equation}
    \tilde{\mathcal{G}}(z) = -\dfrac{4i^{x-1}}{ \pi}\dfrac{J}{\gamma}\dfrac{\abs{z_1}}{z^2+1}\sqrt{\dfrac{(z_1-z)(z+z_1^*)(z+1/z_1)(z-1/z_1^*)}{z^2}}
\end{equation}
and the contour $C$ is depicted in Fig.~\ref{fig:contours}(a) in blue. Here, $z_1 = (-\gamma + i 4h)/(4J)$ is the initial point of the branch cut extending radially outward from the origin. The other branch cut extending to infinity starts at $-z_1^*$, and we connect the horizontal branch cut between the points $-1/z_1$ and $1/z_1^*$. The poles of $f(z)$ are located at $z=\pm i$ and $z=0$. The contour integral \eqref{eq:cintegral} is zero, for the contour does not enclose any singularity. Let us note that $C  = C_1(R)\cup C_2\cup C_3(\epsilon) \cup C_4 \cup C_5(\epsilon)\cup C_6$, where $0<R<\abs{z_1}$. Later, we set $R\rightarrow \infty$ and $\epsilon \rightarrow 0$. $C_1(R)$ is the semicircle of radius $R$, and $C_4(\epsilon)$ is the arc of unity radius that intersects with the upper and lower arcs of radius $\epsilon$ centered at $z=i$ and $z=-i$, respectively. The upper one is $C_5(\epsilon)$ and the lower one is $C_2(\epsilon)$. The upper and lower segments lying on the $\mathfrak{Im}\, z$-axis are $C_2$ and $C_6$, respectively.

By setting $z = ie^{ik}$, it is evident that the imaginary part of the integral over $C_4$ is $-\mathcal{I}(x)$. Hence, 
\begin{equation}
    \mathcal{I}(x) = i\, \mathfrak{Im}\left\lbrace \int_{C\setminus C_4}\dd z  f(z) \right\rbrace = i\, \mathfrak{Im}\left\lbrace \int_{C_2\cup C_6}\dd z f(z) + \int_{C_1}\dd z f(z) - i\dfrac{\pi}{2}\left[\text{Res}(f(z),i) + \text{Res}(f(z),-i) \right] \right\rbrace.
\end{equation}
The sum of the residues gives $-8h/(\gamma \pi)$ and the integral over $C_2\cup C_6$ is purely real, which leaves us with
\begin{equation}
    \mathcal{I}(x) = i\dfrac{4h}{\gamma} +i\,\mathfrak{Im}\left\lbrace \int_{C_1(R)}\dd z \, f(z) \right\rbrace.
\end{equation}
The next step is to utilize the Darboux method \cite{Temme_AsymptoticMethods} to find the leading-order contribution to $\mathcal{I}(x)$ as $x\rightarrow  \infty$. More precisely,  we extend the radius $R$ to infinity so the deformed contour $\tilde C$ surrounds the external branch cut located on the left, as shown in Fig.~\ref{fig:contours}(a) in red. Then, the dominant contribution to this contour integral comes from the branch point $z = z_1$ as $x \rightarrow \infty$, and the integral vanishes at infinity due to the dominant term $z^{-x}$. Therefore, we perform a series expansion of $\tilde{\mathcal{G}}(z_1y)$ around $y=1$ and integrate the resulting divergent series term-by-term and gather only the leading order terms. This way, we obtain the asymptotic expansion of $\mathcal{I}(x)$, where, without loss of generality, we assume that $x$ is even. We have thus
\begin{subequations}
\begin{align}
    \int_{\tilde C}\dd z \,f(z) &\simeq -2\dfrac{-4i^{x}J\abs{z_1}}{\gamma \pi z_1^{x-1}}\dfrac{1}{z_1^2+1}\sqrt{\dfrac{2\,\mathfrak{Re} \{z_1\}(z_1+1/z_1)(z_1-1/z_1^*)}{z_1^2}}\int_1^\infty \dd y \,(y-1)^{1/2}y^{-x-1} + \mathcal{O}(\abs{z_1}x^{-5/2}) \\
    &= \dfrac{4 J i^x}{\gamma \sqrt{\pi} }z_1^{-x+1}x^{-3/2}\sqrt{\dfrac{2\, \mathfrak{Re}\{z_1 \}(\abs{z_1}^2-1)}{z_1(z_1^2+1)}} + \mathcal{O}(\abs{z_1}^{-x}x^{-5/2}). 
\end{align}
\end{subequations}
Let 
\begin{equation}\label{eq:nu}
    \nu(x)\coloneqq \mathfrak{Im}\left\lbrace z_1\exp(-ix\, \text{Arg}\, z_1)\sqrt{\dfrac{2\,\mathfrak{Re}\{z_1 \}(\abs{z_1}^2-1)}{z_1(z_1^2+1)}}\right\rbrace
\end{equation}
with $\text{Arg} z_1 \in (-\pi,\pi]$.
Then, the asymptotic expansion of $\expval{\texttt{A}_0\texttt{A}_x}$ is given by Eq.~\eqref{eq:a0axg}. 

\begin{figure}
\tikzset{every picture/.style={line width=0.75pt}} 
\begin{tikzpicture}[x=0.75pt,y=0.75pt,yscale=-1,xscale=1]
\draw    (180.89,297.11) -- (180.89,46.44) ;
\draw [shift={(180.89,44.44)}, rotate = 90] [color={rgb, 255:red, 0; green, 0; blue, 0 }  ][line width=0.75]    (10.93,-3.29) .. controls (6.95,-1.4) and (3.31,-0.3) .. (0,0) .. controls (3.31,0.3) and (6.95,1.4) .. (10.93,3.29)   ;
\draw    (54.33,170.78) -- (305.44,170.78) ;
\draw [shift={(307.44,170.78)}, rotate = 180] [color={rgb, 255:red, 0; green, 0; blue, 0 }  ][line width=0.75]    (10.93,-3.29) .. controls (6.95,-1.4) and (3.31,-0.3) .. (0,0) .. controls (3.31,0.3) and (6.95,1.4) .. (10.93,3.29)   ;
\draw  [color={rgb, 255:red, 255; green, 0; blue, 0 }  ,draw opacity=1 ] (140.49,129.9) .. controls (141.14,130.99) and (141.77,132.03) .. (142.49,132.03) .. controls (143.21,132.03) and (143.84,130.99) .. (144.49,129.9) .. controls (145.14,128.82) and (145.77,127.78) .. (146.49,127.78) .. controls (147.21,127.78) and (147.84,128.82) .. (148.49,129.9) .. controls (149.14,130.99) and (149.77,132.03) .. (150.49,132.03) .. controls (151.21,132.03) and (151.84,130.99) .. (152.49,129.9) .. controls (153.14,128.82) and (153.77,127.78) .. (154.49,127.78) .. controls (155.21,127.78) and (155.84,128.82) .. (156.49,129.9) .. controls (157.14,130.99) and (157.77,132.03) .. (158.49,132.03) .. controls (159.21,132.03) and (159.84,130.99) .. (160.49,129.9) .. controls (161.14,128.82) and (161.77,127.78) .. (162.49,127.78) .. controls (163.21,127.78) and (163.84,128.82) .. (164.49,129.9) .. controls (165.14,130.99) and (165.77,132.03) .. (166.49,132.03) .. controls (167.21,132.03) and (167.84,130.99) .. (168.49,129.9) .. controls (169.14,128.82) and (169.77,127.78) .. (170.49,127.78) .. controls (171.21,127.78) and (171.84,128.82) .. (172.49,129.9) .. controls (173.14,130.99) and (173.77,132.03) .. (174.49,132.03) .. controls (175.21,132.03) and (175.84,130.99) .. (176.49,129.9) .. controls (177.14,128.82) and (177.77,127.78) .. (178.49,127.78) .. controls (179.21,127.78) and (179.84,128.82) .. (180.49,129.9) .. controls (181.14,130.99) and (181.77,132.03) .. (182.49,132.03) .. controls (183.21,132.03) and (183.84,130.99) .. (184.49,129.9) .. controls (185.14,128.82) and (185.77,127.78) .. (186.49,127.78) .. controls (187.21,127.78) and (187.84,128.82) .. (188.49,129.9) .. controls (189.14,130.99) and (189.77,132.03) .. (190.49,132.03) .. controls (191.21,132.03) and (191.84,130.99) .. (192.49,129.9) .. controls (193.14,128.82) and (193.77,127.78) .. (194.49,127.78) .. controls (195.21,127.78) and (195.84,128.82) .. (196.49,129.9) .. controls (197.14,130.99) and (197.77,132.03) .. (198.49,132.03) .. controls (199.21,132.03) and (199.84,130.99) .. (200.49,129.9) .. controls (201.14,128.82) and (201.77,127.78) .. (202.49,127.78) .. controls (203.21,127.78) and (203.84,128.82) .. (204.49,129.9) .. controls (205.14,130.99) and (205.77,132.03) .. (206.49,132.03) .. controls (207.21,132.03) and (207.84,130.99) .. (208.49,129.9) .. controls (209.14,128.82) and (209.77,127.78) .. (210.49,127.78) .. controls (211.21,127.78) and (211.84,128.82) .. (212.49,129.9) .. controls (213.14,130.99) and (213.77,132.03) .. (214.49,132.03) .. controls (215.21,132.03) and (215.84,130.99) .. (216.49,129.9) .. controls (217.14,128.82) and (217.77,127.78) .. (218.49,127.78) .. controls (219.21,127.78) and (219.84,128.82) .. (220.49,129.9) .. controls (220.61,130.1) and (220.72,130.29) .. (220.83,130.47) ;
\draw  [color={rgb, 255:red, 255; green, 0; blue, 0 }  ,draw opacity=1 ][fill={rgb, 255:red, 255; green, 0; blue, 0 }  ,fill opacity=1 ] (139.47,129.09) .. controls (140,128.67) and (140.77,128.76) .. (141.19,129.29) .. controls (141.61,129.81) and (141.52,130.58) .. (141,131) .. controls (140.47,131.43) and (139.7,131.34) .. (139.28,130.81) .. controls (138.86,130.29) and (138.94,129.52) .. (139.47,129.09) -- cycle ;
\draw  [color={rgb, 255:red, 255; green, 0; blue, 0 }  ,draw opacity=1 ][fill={rgb, 255:red, 255; green, 0; blue, 0 }  ,fill opacity=1 ] (220.02,129.28) .. controls (220.54,128.85) and (221.31,128.94) .. (221.73,129.47) .. controls (222.15,129.99) and (222.07,130.76) .. (221.54,131.19) .. controls (221.01,131.61) and (220.25,131.52) .. (219.82,130.99) .. controls (219.4,130.47) and (219.49,129.7) .. (220.02,129.28) -- cycle ;
\draw [color={rgb, 255:red, 255; green, 0; blue, 0 }  ,draw opacity=1 ]   (177.64,96.88) -- (183.87,103.11) ;
\draw [color={rgb, 255:red, 255; green, 0; blue, 0 }  ,draw opacity=1 ]   (183.87,97.21) -- (177.64,103.44) ;
\draw [color={rgb, 255:red, 255; green, 0; blue, 0 }  ,draw opacity=1 ]   (177.64,236.96) -- (183.87,243.19) ;
\draw [color={rgb, 255:red, 255; green, 0; blue, 0 }  ,draw opacity=1 ]   (183.87,236.96) -- (177.64,243.19) ;
\draw  [draw opacity=0][line width=1.5]  (181.37,250.03) .. controls (181.11,250.05) and (180.84,250.06) .. (180.58,250.06) .. controls (175.17,250.06) and (170.79,245.66) .. (170.79,240.23) .. controls (170.79,239.6) and (170.85,238.98) .. (170.96,238.38) -- (180.58,240.23) -- cycle ; \draw  [color={rgb, 255:red, 13; green, 107; blue, 219 }  ,draw opacity=1 ][line width=1.5]  (181.37,250.03) .. controls (181.11,250.05) and (180.84,250.06) .. (180.58,250.06) .. controls (175.17,250.06) and (170.79,245.66) .. (170.79,240.23) .. controls (170.79,239.6) and (170.85,238.98) .. (170.96,238.38) ;  
\draw  [draw opacity=0][line width=1.5]  (170.97,101.46) .. controls (170.92,101.04) and (170.89,100.61) .. (170.89,100.17) .. controls (170.89,94.75) and (175.25,90.36) .. (180.64,90.34) -- (180.68,100.17) -- cycle ; \draw  [color={rgb, 255:red, 13; green, 107; blue, 219 }  ,draw opacity=1 ][line width=1.5]  (170.97,101.46) .. controls (170.92,101.04) and (170.89,100.61) .. (170.89,100.17) .. controls (170.89,94.75) and (175.25,90.36) .. (180.64,90.34) ;  
\draw [color={rgb, 255:red, 13; green, 107; blue, 219 }  ,draw opacity=1 ][line width=1.5]    (180.64,79.9) -- (180.64,91.32) ;
\draw  [draw opacity=0][line width=1.5]  (171.74,239.47) .. controls (137.66,235) and (111.33,205.69) .. (111.33,170.21) .. controls (111.33,134.68) and (137.72,105.34) .. (171.88,100.93) -- (180.83,170.21) -- cycle ; \draw  [color={rgb, 255:red, 13; green, 107; blue, 219 }  ,draw opacity=1 ][line width=1.5]  (171.74,239.47) .. controls (137.66,235) and (111.33,205.69) .. (111.33,170.21) .. controls (111.33,134.68) and (137.72,105.34) .. (171.88,100.93) ;  
\draw  [draw opacity=0][line width=1.5]  (180.59,259.97) .. controls (180.57,259.97) and (180.56,259.97) .. (180.54,259.97) .. controls (131.09,259.97) and (91,219.88) .. (91,170.43) .. controls (91,120.98) and (131.09,80.89) .. (180.54,80.89) .. controls (180.72,80.89) and (180.91,80.89) .. (181.09,80.89) -- (180.54,170.43) -- cycle ; \draw  [color={rgb, 255:red, 6; green, 108; blue, 223 }  ,draw opacity=1 ][line width=1.5]  (180.59,259.97) .. controls (180.57,259.97) and (180.56,259.97) .. (180.54,259.97) .. controls (131.09,259.97) and (91,219.88) .. (91,170.43) .. controls (91,120.98) and (131.09,80.89) .. (180.54,80.89) .. controls (180.72,80.89) and (180.91,80.89) .. (181.09,80.89) ;  
\draw [color={rgb, 255:red, 13; green, 107; blue, 219 }  ,draw opacity=1 ][line width=1.5]    (180.46,249.07) -- (180.46,260.96) ;
\draw  [color={rgb, 255:red, 255; green, 0; blue, 0 }  ,draw opacity=1 ] (100.18,90.35) .. controls (100.52,89.1) and (100.85,87.9) .. (100.34,87.39) .. controls (99.83,86.87) and (98.63,87.17) .. (97.38,87.49) .. controls (96.12,87.81) and (94.92,88.11) .. (94.41,87.59) .. controls (93.91,87.08) and (94.23,85.88) .. (94.58,84.63) .. controls (94.92,83.38) and (95.25,82.19) .. (94.74,81.67) .. controls (94.24,81.15) and (93.04,81.46) .. (91.78,81.78) .. controls (90.52,82.09) and (89.32,82.4) .. (88.81,81.88) .. controls (88.31,81.36) and (88.63,80.17) .. (88.98,78.92) .. controls (89.32,77.67) and (89.65,76.47) .. (89.14,75.96) .. controls (88.64,75.44) and (87.44,75.74) .. (86.18,76.06) .. controls (84.92,76.38) and (83.72,76.68) .. (83.22,76.16) .. controls (82.71,75.65) and (83.04,74.45) .. (83.38,73.2) .. controls (83.73,71.95) and (84.05,70.76) .. (83.55,70.24) .. controls (83.04,69.73) and (81.84,70.03) .. (80.58,70.35) .. controls (79.32,70.67) and (78.12,70.97) .. (77.62,70.45) .. controls (77.11,69.93) and (77.44,68.74) .. (77.78,67.49) .. controls (78.13,66.24) and (78.45,65.04) .. (77.95,64.53) .. controls (77.44,64.01) and (76.24,64.31) .. (74.98,64.63) .. controls (73.73,64.95) and (72.53,65.25) .. (72.02,64.73) .. controls (71.51,64.22) and (71.84,63.02) .. (72.18,61.77) .. controls (72.53,60.52) and (72.86,59.33) .. (72.35,58.81) .. controls (71.84,58.3) and (70.64,58.6) .. (69.39,58.92) .. controls (68.13,59.24) and (66.93,59.54) .. (66.42,59.02) .. controls (65.91,58.5) and (66.24,57.31) .. (66.59,56.06) .. controls (66.93,54.81) and (67.26,53.61) .. (66.75,53.1) .. controls (66.24,52.58) and (65.04,52.88) .. (63.79,53.2) .. controls (62.53,53.52) and (61.33,53.82) .. (60.82,53.31) .. controls (60.41,52.88) and (60.55,52) .. (60.81,51.01) ;
\draw  [color={rgb, 255:red, 255; green, 0; blue, 0 }  ,draw opacity=1 ][fill={rgb, 255:red, 255; green, 0; blue, 0 }  ,fill opacity=1 ] (99.75,89.09) .. controls (100.27,88.67) and (101.04,88.76) .. (101.47,89.29) .. controls (101.89,89.81) and (101.8,90.58) .. (101.27,91) .. controls (100.75,91.43) and (99.98,91.34) .. (99.56,90.81) .. controls (99.13,90.29) and (99.22,89.52) .. (99.75,89.09) -- cycle ;
\draw  [draw opacity=0][line width=1.5]  (105.9,84.65) .. controls (107.29,86.04) and (108.15,87.97) .. (108.14,90.1) .. controls (108.11,94.31) and (104.68,97.7) .. (100.46,97.68) .. controls (98.42,97.66) and (96.57,96.85) .. (95.21,95.53) -- (100.51,90.05) -- cycle ; \draw  [color={rgb, 255:red, 255; green, 0; blue, 0 }  ,draw opacity=1 ][line width=1.5]  (105.9,84.65) .. controls (107.29,86.04) and (108.15,87.97) .. (108.14,90.1) .. controls (108.11,94.31) and (104.68,97.7) .. (100.46,97.68) .. controls (98.42,97.66) and (96.57,96.85) .. (95.21,95.53) ;  
\draw [color={rgb, 255:red, 255; green, 0; blue, 0 }  ,draw opacity=1 ][line width=1.5]    (66.92,45.67) -- (105.9,84.65) ;
\draw [color={rgb, 255:red, 255; green, 0; blue, 0 }  ,draw opacity=1 ][line width=1.5]    (56.23,56.55) -- (95.21,95.53) ;
\draw    (465.82,297.14) -- (465.82,46.47) ;
\draw [shift={(465.82,44.47)}, rotate = 90] [color={rgb, 255:red, 0; green, 0; blue, 0 }  ][line width=0.75]    (10.93,-3.29) .. controls (6.95,-1.4) and (3.31,-0.3) .. (0,0) .. controls (3.31,0.3) and (6.95,1.4) .. (10.93,3.29)   ;
\draw    (339.26,170.8) -- (590.38,170.8) ;
\draw [shift={(592.38,170.8)}, rotate = 180] [color={rgb, 255:red, 0; green, 0; blue, 0 }  ][line width=0.75]    (10.93,-3.29) .. controls (6.95,-1.4) and (3.31,-0.3) .. (0,0) .. controls (3.31,0.3) and (6.95,1.4) .. (10.93,3.29)   ;
\draw  [color={rgb, 255:red, 255; green, 0; blue, 0 }  ,draw opacity=1 ] (465.91,170.05) .. controls (467.22,170.41) and (468.47,170.75) .. (468.97,170.23) .. controls (469.48,169.71) and (469.09,168.47) .. (468.69,167.17) .. controls (468.28,165.88) and (467.89,164.64) .. (468.4,164.12) .. controls (468.9,163.6) and (470.15,163.93) .. (471.46,164.29) .. controls (472.77,164.65) and (474.02,164.99) .. (474.52,164.47) .. controls (475.03,163.94) and (474.64,162.71) .. (474.24,161.41) .. controls (473.83,160.11) and (473.44,158.88) .. (473.95,158.36) .. controls (474.45,157.83) and (475.7,158.17) .. (477.01,158.53) .. controls (478.32,158.89) and (479.57,159.23) .. (480.07,158.7) .. controls (480.58,158.18) and (480.19,156.95) .. (479.79,155.65) .. controls (479.38,154.35) and (478.99,153.12) .. (479.5,152.59) .. controls (480,152.07) and (481.25,152.41) .. (482.56,152.77) .. controls (483.87,153.13) and (485.12,153.46) .. (485.62,152.94) .. controls (486.13,152.42) and (485.74,151.18) .. (485.34,149.89) .. controls (484.93,148.59) and (484.55,147.35) .. (485.05,146.83) .. controls (485.55,146.31) and (486.8,146.65) .. (488.11,147.01) .. controls (489.42,147.36) and (490.67,147.7) .. (491.17,147.18) .. controls (491.68,146.66) and (491.29,145.42) .. (490.89,144.13) .. controls (490.48,142.83) and (490.1,141.59) .. (490.6,141.07) .. controls (491.1,140.55) and (492.35,140.89) .. (493.66,141.25) .. controls (494.97,141.6) and (496.22,141.94) .. (496.72,141.42) .. controls (497.23,140.9) and (496.84,139.66) .. (496.44,138.36) .. controls (496.03,137.07) and (495.65,135.83) .. (496.15,135.31) .. controls (496.65,134.79) and (497.9,135.13) .. (499.21,135.48) .. controls (500.52,135.84) and (501.77,136.18) .. (502.28,135.66) .. controls (502.78,135.14) and (502.39,133.9) .. (501.99,132.6) .. controls (501.58,131.31) and (501.2,130.07) .. (501.7,129.55) .. controls (502.2,129.03) and (503.45,129.36) .. (504.76,129.72) .. controls (505.43,129.91) and (506.09,130.08) .. (506.64,130.14) ;
\draw  [color={rgb, 255:red, 255; green, 0; blue, 0 }  ,draw opacity=1 ][fill={rgb, 255:red, 255; green, 0; blue, 0 }  ,fill opacity=1 ] (505.28,128.97) .. controls (505.81,128.55) and (506.58,128.63) .. (507,129.16) .. controls (507.42,129.69) and (507.33,130.46) .. (506.81,130.88) .. controls (506.28,131.3) and (505.51,131.21) .. (505.09,130.69) .. controls (504.67,130.16) and (504.75,129.39) .. (505.28,128.97) -- cycle ;
\draw [color={rgb, 255:red, 255; green, 0; blue, 0 }  ,draw opacity=1 ]   (462.57,96.9) -- (468.8,103.13) ;
\draw [color={rgb, 255:red, 255; green, 0; blue, 0 }  ,draw opacity=1 ]   (468.8,97.24) -- (462.57,103.47) ;
\draw [color={rgb, 255:red, 255; green, 0; blue, 0 }  ,draw opacity=1 ]   (462.57,236.99) -- (468.8,243.22) ;
\draw [color={rgb, 255:red, 255; green, 0; blue, 0 }  ,draw opacity=1 ]   (468.8,236.99) -- (462.57,243.22) ;
\draw  [draw opacity=0][line width=1.5]  (475.3,240.19) .. controls (475.3,240.21) and (475.3,240.23) .. (475.3,240.25) .. controls (475.3,245.68) and (470.91,250.09) .. (465.51,250.09) .. controls (460.1,250.09) and (455.72,245.68) .. (455.72,240.25) .. controls (455.72,239.62) and (455.78,239.01) .. (455.89,238.41) -- (465.51,240.25) -- cycle ; \draw  [color={rgb, 255:red, 13; green, 107; blue, 219 }  ,draw opacity=1 ][line width=1.5]  (475.3,240.19) .. controls (475.3,240.21) and (475.3,240.23) .. (475.3,240.25) .. controls (475.3,245.68) and (470.91,250.09) .. (465.51,250.09) .. controls (460.1,250.09) and (455.72,245.68) .. (455.72,240.25) .. controls (455.72,239.62) and (455.78,239.01) .. (455.89,238.41) ;  
\draw  [draw opacity=0][line width=1.5]  (455.91,101.48) .. controls (455.85,101.06) and (455.82,100.63) .. (455.82,100.2) .. controls (455.82,94.77) and (460.21,90.36) .. (465.61,90.36) .. controls (471.02,90.36) and (475.4,94.77) .. (475.4,100.2) .. controls (475.4,100.61) and (475.37,101.01) .. (475.33,101.41) -- (465.61,100.2) -- cycle ; \draw  [color={rgb, 255:red, 13; green, 107; blue, 219 }  ,draw opacity=1 ][line width=1.5]  (455.91,101.48) .. controls (455.85,101.06) and (455.82,100.63) .. (455.82,100.2) .. controls (455.82,94.77) and (460.21,90.36) .. (465.61,90.36) .. controls (471.02,90.36) and (475.4,94.77) .. (475.4,100.2) .. controls (475.4,100.61) and (475.37,101.01) .. (475.33,101.41) ;  
\draw  [draw opacity=0][line width=1.5]  (456.68,239.5) .. controls (422.59,235.02) and (396.26,205.72) .. (396.26,170.23) .. controls (396.26,134.7) and (422.66,105.37) .. (456.81,100.95) -- (465.76,170.23) -- cycle ; \draw  [color={rgb, 255:red, 13; green, 107; blue, 219 }  ,draw opacity=1 ][line width=1.5]  (456.68,239.5) .. controls (422.59,235.02) and (396.26,205.72) .. (396.26,170.23) .. controls (396.26,134.7) and (422.66,105.37) .. (456.81,100.95) ;  
\draw  [draw opacity=0][line width=1.5]  (465.52,260) .. controls (465.5,260) and (465.49,260) .. (465.47,260) .. controls (416.02,260) and (375.93,219.91) .. (375.93,170.46) .. controls (375.93,121) and (416.02,80.91) .. (465.47,80.91) .. controls (514.93,80.91) and (555.02,121) .. (555.02,170.46) .. controls (555.02,219.85) and (515.03,259.9) .. (465.66,260) -- (465.47,170.46) -- cycle ; \draw  [color={rgb, 255:red, 13; green, 107; blue, 219 }  ,draw opacity=1 ][line width=1.5]  (465.52,260) .. controls (465.5,260) and (465.49,260) .. (465.47,260) .. controls (416.02,260) and (375.93,219.91) .. (375.93,170.46) .. controls (375.93,121) and (416.02,80.91) .. (465.47,80.91) .. controls (514.93,80.91) and (555.02,121) .. (555.02,170.46) .. controls (555.02,219.85) and (515.03,259.9) .. (465.66,260) ;  
\draw  [color={rgb, 255:red, 255; green, 0; blue, 0 }  ,draw opacity=1 ] (385.11,90.37) .. controls (385.45,89.12) and (385.78,87.93) .. (385.27,87.41) .. controls (384.77,86.9) and (383.57,87.2) .. (382.31,87.52) .. controls (381.05,87.84) and (379.85,88.14) .. (379.34,87.62) .. controls (378.84,87.1) and (379.16,85.91) .. (379.51,84.66) .. controls (379.85,83.41) and (380.18,82.21) .. (379.67,81.7) .. controls (379.17,81.18) and (377.97,81.48) .. (376.71,81.8) .. controls (375.45,82.12) and (374.25,82.42) .. (373.75,81.9) .. controls (373.24,81.39) and (373.57,80.19) .. (373.91,78.94) .. controls (374.26,77.69) and (374.58,76.5) .. (374.08,75.98) .. controls (373.57,75.47) and (372.37,75.77) .. (371.11,76.09) .. controls (369.85,76.41) and (368.65,76.71) .. (368.15,76.19) .. controls (367.64,75.67) and (367.97,74.48) .. (368.31,73.23) .. controls (368.66,71.98) and (368.98,70.78) .. (368.48,70.27) .. controls (367.97,69.75) and (366.77,70.05) .. (365.51,70.37) .. controls (364.26,70.69) and (363.06,70.99) .. (362.55,70.48) .. controls (362.04,69.96) and (362.37,68.77) .. (362.71,67.51) .. controls (363.06,66.26) and (363.39,65.07) .. (362.88,64.55) .. controls (362.37,64.04) and (361.17,64.34) .. (359.92,64.66) .. controls (358.66,64.98) and (357.46,65.28) .. (356.95,64.76) .. controls (356.44,64.24) and (356.77,63.05) .. (357.12,61.8) .. controls (357.46,60.55) and (357.79,59.36) .. (357.28,58.84) .. controls (356.77,58.32) and (355.57,58.62) .. (354.32,58.94) .. controls (353.06,59.26) and (351.86,59.56) .. (351.35,59.05) .. controls (350.85,58.53) and (351.17,57.34) .. (351.52,56.08) .. controls (351.86,54.83) and (352.19,53.64) .. (351.68,53.12) .. controls (351.18,52.61) and (349.98,52.91) .. (348.72,53.23) .. controls (347.46,53.55) and (346.26,53.85) .. (345.75,53.33) .. controls (345.34,52.91) and (345.48,52.03) .. (345.74,51.03) ;
\draw  [color={rgb, 255:red, 255; green, 0; blue, 0 }  ,draw opacity=1 ][fill={rgb, 255:red, 255; green, 0; blue, 0 }  ,fill opacity=1 ] (384.68,89.12) .. controls (385.21,88.7) and (385.98,88.78) .. (386.4,89.31) .. controls (386.82,89.84) and (386.73,90.61) .. (386.21,91.03) .. controls (385.68,91.45) and (384.91,91.37) .. (384.49,90.84) .. controls (384.07,90.31) and (384.15,89.54) .. (384.68,89.12) -- cycle ;
\draw  [draw opacity=0][line width=1.5]  (390.83,84.67) .. controls (392.22,86.07) and (393.08,88) .. (393.07,90.12) .. controls (393.04,94.33) and (389.61,97.73) .. (385.4,97.7) .. controls (383.35,97.69) and (381.5,96.87) .. (380.14,95.56) -- (385.44,90.07) -- cycle ; \draw  [color={rgb, 255:red, 255; green, 0; blue, 0 }  ,draw opacity=1 ][line width=1.5]  (390.83,84.67) .. controls (392.22,86.07) and (393.08,88) .. (393.07,90.12) .. controls (393.04,94.33) and (389.61,97.73) .. (385.4,97.7) .. controls (383.35,97.69) and (381.5,96.87) .. (380.14,95.56) ;  
\draw [color={rgb, 255:red, 255; green, 0; blue, 0 }  ,draw opacity=1 ][line width=1.5]    (351.85,45.69) -- (390.83,84.67) ;
\draw [color={rgb, 255:red, 255; green, 0; blue, 0 }  ,draw opacity=1 ][line width=1.5]    (341.16,56.58) -- (380.14,95.56) ;
\draw  [color={rgb, 255:red, 255; green, 0; blue, 0 }  ,draw opacity=1 ] (261.2,89.98) .. controls (262.45,90.33) and (263.64,90.66) .. (264.16,90.16) .. controls (264.68,89.65) and (264.38,88.45) .. (264.07,87.19) .. controls (263.75,85.93) and (263.45,84.73) .. (263.97,84.23) .. controls (264.49,83.72) and (265.68,84.05) .. (266.93,84.4) .. controls (268.18,84.75) and (269.37,85.08) .. (269.89,84.58) .. controls (270.41,84.07) and (270.11,82.87) .. (269.8,81.61) .. controls (269.48,80.35) and (269.18,79.15) .. (269.7,78.65) .. controls (270.22,78.14) and (271.41,78.47) .. (272.66,78.82) .. controls (273.91,79.17) and (275.11,79.5) .. (275.62,79) .. controls (276.14,78.49) and (275.85,77.29) .. (275.53,76.03) .. controls (275.22,74.77) and (274.92,73.57) .. (275.44,73.07) .. controls (275.95,72.56) and (277.15,72.89) .. (278.4,73.24) .. controls (279.65,73.59) and (280.84,73.92) .. (281.36,73.42) .. controls (281.88,72.91) and (281.58,71.71) .. (281.26,70.45) .. controls (280.95,69.19) and (280.65,67.99) .. (281.17,67.49) .. controls (281.69,66.98) and (282.88,67.31) .. (284.13,67.66) .. controls (285.38,68.01) and (286.57,68.34) .. (287.09,67.84) .. controls (287.61,67.33) and (287.31,66.13) .. (287,64.87) .. controls (286.68,63.61) and (286.38,62.41) .. (286.9,61.91) .. controls (287.42,61.4) and (288.61,61.73) .. (289.86,62.08) .. controls (291.11,62.43) and (292.3,62.76) .. (292.82,62.26) .. controls (293.34,61.75) and (293.04,60.55) .. (292.73,59.29) .. controls (292.41,58.03) and (292.11,56.83) .. (292.63,56.33) .. controls (293.15,55.82) and (294.34,56.15) .. (295.59,56.5) .. controls (296.84,56.85) and (298.04,57.18) .. (298.55,56.68) .. controls (299.07,56.17) and (298.78,54.97) .. (298.46,53.71) .. controls (298.14,52.45) and (297.85,51.25) .. (298.37,50.75) .. controls (298.79,50.33) and (299.67,50.48) .. (300.66,50.74) ;
\draw  [color={rgb, 255:red, 255; green, 0; blue, 0 }  ,draw opacity=1 ][fill={rgb, 255:red, 255; green, 0; blue, 0 }  ,fill opacity=1 ] (259.41,89.43) .. controls (259.94,89.01) and (260.71,89.09) .. (261.13,89.62) .. controls (261.55,90.15) and (261.47,90.92) .. (260.94,91.34) .. controls (260.41,91.76) and (259.64,91.67) .. (259.22,91.15) .. controls (258.8,90.62) and (258.89,89.85) .. (259.41,89.43) -- cycle ;
\draw [color={rgb, 255:red, 255; green, 0; blue, 0 }  ,draw opacity=1 ]   (177.49,167.57) -- (183.73,173.8) ;
\draw [color={rgb, 255:red, 255; green, 0; blue, 0 }  ,draw opacity=1 ]   (183.73,167.9) -- (177.49,174.14) ;
\draw  [color={rgb, 255:red, 74; green, 74; blue, 74 }  ,draw opacity=1 ][line width=1.5]  (108.48,148) .. controls (112.72,146.02) and (115.88,143.53) .. (117.95,140.52) .. controls (116.96,144.04) and (117.07,148.05) .. (118.25,152.58) ;
\draw  [color={rgb, 255:red, 74; green, 74; blue, 74 }  ,draw opacity=1 ][line width=1.5]  (101.36,200.7) .. controls (100.38,205.28) and (100.45,209.29) .. (101.59,212.76) .. controls (99.39,209.85) and (96.13,207.5) .. (91.8,205.7) ;
\draw  [color={rgb, 255:red, 74; green, 74; blue, 74 }  ,draw opacity=1 ][line width=1.5]  (92.14,65.34) .. controls (92.39,68.61) and (93.25,71.3) .. (94.71,73.4) .. controls (92.65,71.88) and (89.99,70.96) .. (86.72,70.62) ;
\draw [color={rgb, 255:red, 255; green, 0; blue, 0 }  ,draw opacity=1 ]   (462.57,167.33) -- (468.8,173.56) ;
\draw [color={rgb, 255:red, 255; green, 0; blue, 0 }  ,draw opacity=1 ]   (468.8,167.67) -- (462.57,173.9) ;
\draw  [draw opacity=0][line width=1.5]  (474.39,100.91) .. controls (508.7,105.19) and (535.25,134.59) .. (535.25,170.23) .. controls (535.25,205.93) and (508.62,235.37) .. (474.23,239.57) -- (465.76,170.23) -- cycle ; \draw  [color={rgb, 255:red, 13; green, 107; blue, 219 }  ,draw opacity=1 ][line width=1.5]  (474.39,100.91) .. controls (508.7,105.19) and (535.25,134.59) .. (535.25,170.23) .. controls (535.25,205.93) and (508.62,235.37) .. (474.23,239.57) ;  
\draw  [color={rgb, 255:red, 74; green, 74; blue, 74 }  ,draw opacity=1 ][line width=1.5]  (393.44,148.34) .. controls (397.68,146.35) and (400.84,143.85) .. (402.91,140.86) .. controls (401.93,144.36) and (402.03,148.39) .. (403.21,152.92) ;
\draw  [color={rgb, 255:red, 74; green, 74; blue, 74 }  ,draw opacity=1 ][line width=1.5]  (521.03,107.33) .. controls (520.15,102.74) and (518.52,99.07) .. (516.12,96.32) .. controls (519.28,98.14) and (523.2,99.03) .. (527.88,99) ;
\draw  [color={rgb, 255:red, 74; green, 74; blue, 74 }  ,draw opacity=1 ][line width=1.5]  (377.73,65.74) .. controls (377.98,69.01) and (378.85,71.7) .. (380.31,73.8) .. controls (378.25,72.29) and (375.58,71.36) .. (372.32,71.02) ;

\draw (285.56,148.51) node [anchor=north west][inner sep=0.75pt]    {${\mathfrak{Re} \, z}$};
\draw (187.42,45.4) node [anchor=north west][inner sep=0.75pt]    {$\mathfrak{Im} \, z$};
\draw (70.29,139.45) node [anchor=north west][inner sep=0.75pt]    {$C$};
\draw (78.43,88.73) node [anchor=north west][inner sep=0.75pt]    {$z_{1}$};
\draw (55.51,24.31) node [anchor=north west][inner sep=0.75pt]   [align=left] {(a)};
\draw (570.49,148.54) node [anchor=north west][inner sep=0.75pt]    {${\mathfrak{Re} \, z}$};
\draw (472.35,45.43) node [anchor=north west][inner sep=0.75pt]    {$\mathfrak{Im} \, z$};
\draw (355.22,139.47) node [anchor=north west][inner sep=0.75pt]    {$C$};
\draw (363.36,88.76) node [anchor=north west][inner sep=0.75pt]    {$z_{1}$};
\draw (340.44,24.33) node [anchor=north west][inner sep=0.75pt]   [align=left] {(b)};

\end{tikzpicture}

   \caption{Contours of integration used to evaluate the contractions (a) $\expval{\texttt{A}_0\texttt{A}_{x}}$ and (b) $\expval{\texttt{A}_0\texttt{B}_{\pm x}}$ in the gapped phase. }
     \label{fig:contours}
\end{figure}

\subsubsection{\texorpdfstring{Contraction \textnormal{\(\expval{\texttt{A}_0\texttt{B}_{\pm x}}\)}}{Contraction}}

We can rewrite Eq.~\eqref{eq:ax6} as
\begin{equation}\label{eq:app8}
    \expval{\texttt{B}_0 \texttt{A}_{\pm x}} = \mathfrak{Re}\left\lbrace  \text{p.v.} \dfrac{1}{\gamma \pi} \int_{-\pi}^\pi \dd k \, ie^{-ikx}\left[1 \pm y(k)\right]\Lambda(k) \right\rbrace = \mathfrak{Re}\left\lbrace \mathcal{J}(x)\right\rbrace,
\end{equation}
where p.v. denotes the Cauchy principal value, the singular points are $0$ and $\pm \pi$, and 
\begin{equation}
    y(k) \coloneqq \dfrac{\gamma + i4(h-J \cos k)}{4J\sin k}.
\end{equation}
Let us consider the contour integral 
\begin{equation}\label{eq:cont1}
\int_{C}  \dd z\, z^{-x-1} F_{\pm }(z)  , 
\end{equation}
where
\begin{equation}\label{eq:app21}
    F_{\pm}(z) = \dfrac{i^x}{\gamma \pi}\left(1 \pm y(z)\right)\Lambda(z),
\end{equation}
and where $C=\bigcup_{i=1}^5 C_i$ is the contour shown in Fig.~\ref{fig:contours}(b) in blue.  $C_1$ and $C_3$ are the arcs with radii 1 located in the right and left-half planes, respectively; $C_2(\epsilon)$ and $C_4(\epsilon)$ are the upper and lower semicircles of radii $\epsilon$ located at $z=i$ and $z=-i$, respectively; $C_5(R)$ is the circle of radius $R$.

The explicit form of $F_{\pm }(z)$ is
\begin{equation}\label{eq:app22}
    F_{s}(z) = \begin{cases}
        \dfrac{4Ji^{x}\sqrt{z_1}}{\gamma \pi}\dfrac{z(z+z_1^*)}{z^2+1}\sqrt{\dfrac{(z-z_1)(z+1/z_1)}{z}} \, \dfrac{\abs{z}}{i z} \dfrac{(z-w)(z+w^*)}{\abs{z-w}\abs{z+w^*}}, \quad & s = +,\\
        \dfrac{-4Ji^{x}\sqrt{z_1}}{\gamma \pi} \dfrac{z_1^*(z-1/z_1^*)}{z^2+1}\,\sqrt{\dfrac{(z-z_1)(z+1/z_1)}{z}}\,   \dfrac{\abs{z}}{iz} \dfrac{(z-w)(z+w^*)}{\abs{z-w}\abs{z+w^*}}, \quad & s = -,
    \end{cases}
\end{equation}
where  
\begin{equation}\label{eq:app23}
     z_1 = \frac{-\gamma + i4h}{4J}, \quad w = \dfrac{\sqrt{J^2-h^2}+i h}{J}.
\end{equation}
Since Eq.~\eqref{eq:cont1} does not enclose any pole, it is equal to zero. By exploiting this fact setting $z = i e^{ik}$ in the integral over the contour $-C_1\cup -C_3$, it is evident that 
\begin{equation}
    \mathcal{J}(x) = -\int_{C_1\cup C_2}\dd z\, z^{-x-1}F_\pm(z)= \int_{C_2\cup C_4}\dd z\, z^{-x-1}F_\pm(z)  + \int_{C_5}\dd z\, z^{-x-1}F_\pm(z). 
\end{equation}
The integral over $C_2\cup C_4$ is equal to $-i\pi\, \text{Res}\left( z^{-x-1}F_\pm(z), z = i\right)-i\pi\,  \text{Res}\left( z^{-x-1}F_\pm(z), z = -i\right) \in i \mathbb{R}$. Therefore, 
\begin{equation}
    \expval{\texttt{B}_0 \texttt{A}_{\pm x}} = \mathfrak{Re}\left\lbrace \int_{C_5(R)}\dd z\, z^{-x-1}F_{\pm}(z) \right\rbrace,
\end{equation}
and, similarly as for the other contraction, we set the radius $R$ to infinity so the deformed contour surrounds the external branch cut starting at the branch point $z_1$. Subsequently, one performs a series expansion of $F_{\pm}(z_1 y)$ around $y = 1$, which leads to a divergent series, and one integrates term-by-term. After implementing these steps, one gets Eq.~\eqref{eq:b0asx}, 
where
\begin{align}
    \kappa_+(x)&\coloneqq 2\,\mathfrak{Re}\{z_1\}\,\mathfrak{Re}\left\lbrace \dfrac{z_1^{3/2}\sqrt{z_1+1/z_1}}{z_1^2+1}\exp\left( -i x\text{Arg}\, z_1 \right) \right\rbrace,\label{eq:kappa+} \\
    \kappa_-(x) &\coloneqq \left(\abs{z_1}^2-1\right)\mathfrak{Re}\left\lbrace \dfrac{\exp(-ix \text{Arg}z_1)}{\sqrt{z_1^2+1}}\right\rbrace. \label{eq:kappa-}
\end{align}

Recall that $\texttt{C}^{zz}(x) = -\expval{\texttt{B}_0\texttt{A}_x}\expval{\texttt{B}_0 \texttt{A}_{-x}}-\expval{\texttt{A}_0\texttt{A}_x}^2$. Hence, the asymptotic behavior of the correlation function as $x\rightarrow \infty $ in the gapped phase is given by Eq.~\eqref{eq:czz1}, where $\xi = (2\ln \abs{z_1})^{-1}$ is the correlation length. 

\subsection{Gapless phase}\label{ssec:gapless_phase}
In this section, we implement alternative methods to calculate the asymptotic behavior of $\texttt{C}^{zz}(x)$. Namely, we perform series expansion of the integrands of the contractions $\expval{\texttt{A}_0\texttt{A}_x}$ and $\expval{\texttt{B}_0 \texttt{A}_{\pm x}}$ not involving functions of $x$ around the special points $k=0,\pi$ and $k = q = \arccos(h/J)$. Then, we split the integrals at these points and introduce appropriate exponential regulators to find the leading contributions of the form $x^{-n}$ with $n$ integer or half-integer. We first treat the case $\gamma = \gamma_c(h\geq 0)$ and then $\gamma < \gamma_c(h).$ Once more, for all contractions, $x$  is even.

\subsubsection{\texorpdfstring{$\gamma = \gamma_c(h\geq 0)$}{\#0}}
We first treat $\expval{\texttt{A}_0\texttt{A}_x}$ and then $\expval{\texttt{B}_0 \texttt{A}_{\pm x}}$.

Let $\gamma = \gamma_c(h>0)$. Then, the integrand of Eq.~\eqref{eq:ax6.1}, $g(k)$, except $\sin k x$, simplifies to 
\begin{equation}
 g(k)\coloneqq \begin{cases}
        \dfrac{i}{\pi}\sqrt{\dfrac{J+h}{J-h}}\tan\dfrac{k}{2}, \quad & h-J\cos k \leq 0, \,\,(k\leq q.) \\[2ex]
        \dfrac{i}{\pi}\sqrt{\dfrac{J-h}{J+h}}\cot \dfrac{k}{2},\quad &h-J\cos k \geq 0. \,\, (k\geq q.)
    \end{cases} 
\end{equation}
If we perform a series expansion of this function around $k=0$ and $k=\pi$ and introduce exponential regulators, one finds that these points do not contribute because the series expansions are polynomials with odd powers in $k$ and $k-\pi$, respectively, 
e.g., at $k=0$, $g(k) = a_1 k+\mathcal{O}(k^3) $, and at $k=\pi$, $g(k) = b_1 (k-\pi) + \mathcal O (k-\pi)^3$. Let $a>0$. Then, the regulated integrals at these points and at leading order give
\begin{equation}
    \lim_{a\to 0}\int_{0}^\infty\dd k\, e^{-a k} k \sin kx = 0,  \quad \text{and} \quad   \lim_{a\to 0}\int_{-\infty}^\pi\dd k\, e^{a k} (k-\pi) \sin kx  = 0,
\end{equation}
and the same occurs at all orders in $k$ and $k-\pi$. Therefore, the only point that might contribute is $k = q$. Hence, have up to first order in $k-q$
\begin{equation}
    g(k) \approx \begin{cases}
        \dfrac{i}{\pi}\sqrt{\dfrac{J+h}{J-h}}\tan\dfrac{q}{2} - \dfrac{i}{2\pi}\sqrt{\dfrac{J+h}{J-h}}\left( 1 + \tan^2\dfrac{q}{2} \right)(q-k),\quad & k\leq q,\\
        \dfrac{i}{\pi}\sqrt{\dfrac{J-h}{J+h}}\cot\dfrac{q}{2} - \dfrac{i}{2\pi}\sqrt{\dfrac{J-h}{J+h}}\left(1 + \cot^2\dfrac{q}{2} \right)(k-q),\quad &k\geq q.
    \end{cases}
\end{equation}
Then, for $x \rightarrow \infty$, we have
\begin{subequations}
\begin{align}
    \expval{\texttt{A}_0 \texttt{A}_x}&\simeq \lim_{a\to 0}\int_{-\infty}^q \dd k\, e^{a k}\left[\dfrac{i}{\pi}\sqrt{\dfrac{J+h}{J-h}}\tan\dfrac{q}{2} - \dfrac{i}{2\pi}\sqrt{\dfrac{J+h}{J-h}}\left( 1 + \tan^2\dfrac{q}{2} \right)(q-k) \right]\sin kx \notag \\
    \quad &+ \lim_{a\to 0 }\int_{q}^\infty\dd k\, e^{-ak}\left[ \dfrac{i}{\pi}\sqrt{\dfrac{J-h}{J+h}}\cot\dfrac{q}{2} - \dfrac{i}{2\pi}\sqrt{\dfrac{J-h}{J+h}}\left(1 + \cot^2\dfrac{q}{2} \right)(k-q) \right]\sin k x+ \mathcal O (x^{-3})\notag \\
    &= \dfrac{i}{\pi}\dfrac{2J}{\sqrt{J^2-h^2}}x^{-2}\sin qx + \mathcal O (x^{-3}).
\end{align}
\end{subequations}

By repeating the same steps for $\expval{\texttt{B}_0 \texttt{A}_{\pm x}}$, one finds that $k = q$ is the only relevant point for performing the series expansion. 
From Eq.~\eqref{eq:x50}, let 
\begin{equation}
    f_c(k)\coloneqq \dfrac{1}{\pi}\dfrac{-\abs{u_k}^2+\abs{v_k}^2}{\abs{u_k}^+\abs{v_k}^2} 
\end{equation}
and
\begin{equation}
f_s(k)\coloneqq \dfrac{1}{\pi}\dfrac{(u_k + u_k^*)v_k}{\abs{u_k}^+\abs{v_k}^2}.
\end{equation}
The series expansion around $k=q$ of these functions are
\begin{equation}
    f_c(k) = \begin{cases}
        \dfrac{1}{\pi}\left(\dfrac{J+h}{J-h} \right)^{1/4}(q-k)^{1/2} - \dfrac{1}{4\pi}\dfrac{h}{(J-h)^{3/4}(J+h)^{1/4}}(q-k)^{3/2} + \mathcal{O}(q-k)^{3/2}, \quad&k\leq q,\\
         \dfrac{1}{\pi}\left(\dfrac{J-h}{J+h} \right)^{1/4}(k-q)^{1/2} + \dfrac{1}{4\pi}\dfrac{h}{(J+h)^{3/4}(J-h)^{1/4}}(k-q)^{3/2} + \mathcal{O}(q-k)^{3/2}, \quad&k\geq q,
    \end{cases}
\end{equation}
and
\begin{equation}
    f_s(k) = \begin{cases}
         \dfrac{1}{\pi}\left(\dfrac{J-h}{J+h} \right)^{1/4}(q-k)^{1/2} - \dfrac{1}{4\pi}\dfrac{h+4J}{(J+h)^{3/4}(J-h)^{1/4}}(q-k)^{3/2} + \mathcal{O}(q-k)^{3/2}, \quad&k \leq q,\\
           -\dfrac{1}{\pi}\left(\dfrac{J+h}{J-h} \right)^{1/4}(k-q)^{1/2} - \dfrac{1}{4\pi}\dfrac{h-4J}{(J-h)^{3/4}(J+h)^{1/4}}(q-k)^{3/2} + \mathcal{O}(k-q)^{3/2}, \quad&k\geq q.
    \end{cases}
\end{equation}
Thus, 
\begin{subequations}
\begin{align}
    \expval{\texttt{B}_0 \texttt{A}_x}&\simeq \lim_{a\to 0}\left(\int_{-\infty}^q\dd k\, e^{a k}\left[ \cos kx\, f_c(k) + \sin kx\, f_s(k) \right]+ \int_{q}^\infty\dd k\,  e^{-a k }\,\left[ \cos kx\, f_c(k) + \sin kx\, f_s(k) \right] \right)\\
    &= \lim_{a\to 0}\left(\int_{-\infty}^q \dd k\, e^{ak}(q-k)^{1/2} \left[  \dfrac{1}{\pi}\left(\dfrac{J+h}{J-h} \right)^{1/4}\cos k x +  \dfrac{1}{\pi}\left(\dfrac{J-h}{J+h} \right)^{1/4}\sin kx \right]\right) \notag\\
    &\quad + \lim_{a\to 0}\left(\int_{q}^\infty\dd k\, e^{-a k }(k-q)^{1/2}\left[ \dfrac{1}{\pi}\left(\dfrac{J-h}{J+h} \right)^{1/4} \cos k x   -\dfrac{1}{\pi}\left(\dfrac{J+h}{J-h} \right)^{1/4}\sin kx  \right] \right) + \mathcal{O}(x^{-5/2}) \\
    &= -\dfrac{1}{\sqrt{\pi}}x^{-3/2}\left[ \left(\dfrac{J+h}{J-h} \right)^{1/4}\cos(q x + \dfrac{\pi}{4}) + \left(\dfrac{J-h}{J+h} \right)^{1/4}\sin(q x + \dfrac{\pi}{4}) \right]  + \mathcal{O}\textbf{(}j_1(x)x^{-5/2}\textbf{)}, 
\end{align}
\end{subequations}
where $j_1(x)$ is an oscillatory function in $x$ not pertinent in our analysis. 
By implementing the same steps to $\expval{\texttt{B}_0 \texttt{A}_{-x}}$, one finds that the leading-order term is $\mathcal{O}(x^{-5/2})$:
    \begin{equation}
        \expval{\texttt{B}_0\texttt{A}_{-x}}\simeq \dfrac{3}{\sqrt{\pi}}x^{-5/2} \left[ \dfrac{J}{(J-h)^{1/4}(J+h)^{3/4}} \cos(q x + \dfrac{\pi}{4})+ \dfrac{J}{(J+h)^{1/4}(J-h)^{3/4}} \sin(q x + \dfrac{\pi}{4})\right]+ \mathcal O \textbf{(}j_2(x)x^{-7/2}\textbf{)}.
    \end{equation}
$j_2(x)$ is an oscillatory function not relevant to our analysis. 
Given the above asymptotic expansions, we get 
\begin{multline}\label{eq:yy1}
    \texttt{C}^{zz}(x) \simeq \dfrac{3}{\pi}x^{-4} \left[ \left(\dfrac{J+h}{J-h} \right)^{1/4}\cos(q x + \dfrac{\pi}{4}) + \left(\dfrac{J-h}{J+h} \right)^{1/4}\sin(q x + \dfrac{\pi}{4}) \right]\\\times  \left[ \dfrac{J}{(J-h)^{1/4}(J+h)^{3/4}} \cos(q x + \dfrac{\pi}{4})+ \dfrac{J}{(J+h)^{1/4}(J-h)^{3/4}} \sin(q x + \dfrac{\pi}{4})\right] \\+\dfrac{4}{\pi^2}\dfrac{J^2}{J^2-h^2}x^{-4}\sin^2 qx + \mathcal{O}\textbf{(}\theta(x)x^{-5}\textbf{)},
\end{multline}
where $\theta(x)$ is some function of $x$.
By setting
\begin{multline}\label{eq:mu}
    \mu(x)\coloneqq 3\left[ \left(\dfrac{J+h}{J-h} \right)^{1/4}\cos(q x + \dfrac{\pi}{4}) + \left(\dfrac{J-h}{J+h} \right)^{1/4}\sin(q x + \dfrac{\pi}{4}) \right]\\\times  \left[ \dfrac{J}{(J-h)^{1/4}(J+h)^{3/4}} \cos(q x + \dfrac{\pi}{4})+ \dfrac{J}{(J+h)^{1/4}(J-h)^{3/4}} \sin(q x + \dfrac{\pi}{4})\right] \\+\dfrac{4}{\pi}\dfrac{J^2}{J^2-h^2}x^{-4}\sin^2 qx,
\end{multline}
we arrive at Eq.~\eqref{eq:tcor1}.

\subsubsection{\texorpdfstring{$\gamma < \gamma_c(h)$}{\#0}}

Let us only consider half of the integration in Eq.~\eqref{eq:app6} and set
\begin{equation}\label{eq:f1}
    \tilde f_c(k) \coloneqq \dfrac{2i\Lambda(k)}{\gamma \pi}\quad \text{and}\quad  \tilde f_s(k)\coloneqq \dfrac{\gamma + 4 i (h-J \cos k)}{2\gamma \pi J \, \sin k}\Lambda(k). 
\end{equation}
The series expansions of these functions around $k=0, \pi$ and $q$ are, respectively, to leading order 
\begin{alignat}{2}
    \tilde f_c(k) &\approx \dfrac{-\gamma + i4(h-J)}{\gamma \pi}, \quad &&\tilde  f_s(k) \approx -i\dfrac{16(h-J)^2 + \gamma^2}{4\gamma \pi J \, k}\\
    \tilde f_c(k) &\approx \dfrac{\gamma - i4(h+J)}{\gamma \pi}, \quad &&\tilde  f_s(k) \approx i\dfrac{16(h+J)^2+\gamma^2}{4\gamma \pi J (k-\pi)}\\
    \tilde  f_c(k) &\approx \dfrac{i\sqrt{\gamma^2_c(h)-\gamma^2}}{\gamma \pi}\, \dfrac{q-k}{\abs{q-k}},\quad &&f_s(k) \approx \dfrac{\sqrt{\gamma_c^2(h)-\gamma^2}}{4J\pi \sqrt{J^2-h^2}}\dfrac{q-k}{\abs{q-k}}.
\end{alignat}
We have thus
\begin{align}\label{eq:app27}
    \int_{0}^\pi  \dd k\, \cos kx\, \tilde f_c(k)&\simeq \lim_{a\to 0}\biggl( \dfrac{-\gamma + i4(h-J)}{\gamma \pi}\int_0^\infty \dd k\, e^{-a k}\cos k x+\dfrac{\gamma - i4(h+J)}{\gamma \pi} \int_{-\infty}^\pi  \dd k\, e^{a k}\cos kx  \notag \\ \quad &+ \int_{-\infty}^\infty \dd k\, e^{-a\text{sgn}(k-q)} \cos kx \dfrac{q-k}{\abs{q-k}}\biggr) + \mathcal{O}\textbf{(}j_3(x)x^{-2}\textbf{)}\\
    &= \dfrac{2i\sqrt{\gamma_c^2(h)-\gamma^2}}{\gamma \pi}\dfrac{\sin qx}{x} + \mathcal{O}\textbf{(}j_3(x)x^{-2}\textbf{)}.
\end{align}

Following similar steps for $f_s(k)$, we get
\begin{equation}\label{eq:app28}
    \int_0^\pi \dd k\, \sin kx\, \tilde f_s(k)\simeq i\dfrac{8h}{\gamma} - \dfrac{\sqrt{\gamma_c^2(h)-\gamma^2}}{2\pi \sqrt{J^2-h^2}}\dfrac{\cos qx}{x} \left[ 1 + \mathcal{O}(x^{-1})\right].
\end{equation}
Adding Eqs.~\eqref{eq:app27} and \eqref{eq:app29} and taking the real part gives 
\begin{equation}\label{eq:app29}
     \expval{\texttt{B}_0\texttt{A}_{\pm x}} \simeq \mp \dfrac{1}{2\pi} \sqrt{\dfrac{16(J^2-h^2)-\gamma^2}{J^2-h^2}}\dfrac{\cos q x}{x}\left[1+\mathcal{O}(x^{-1})\right].
\end{equation}

By introducing similar regulators to the integrals in $\expval{\texttt{A}_0 \texttt{A}_x}$, one finds that for $x\gg 1$ and $\gamma < \gamma_c(h)$, the points $k=0,\pi$ produce the negative of the constant term and no $\mathcal{O}(x^{-2})$-terms. Hence, we obtain Eq~\eqref{eq:tcor}.

 \begin{figure}

\tikzset{every picture/.style={line width=0.75pt}} 

\begin{tikzpicture}[x=0.75pt,y=0.75pt,yscale=-1,xscale=1]

\draw    (252.82,288.89) -- (252.82,31.39) ;
\draw [shift={(252.82,29.39)}, rotate = 90] [color={rgb, 255:red, 0; green, 0; blue, 0 }  ][line width=0.75]    (10.93,-3.29) .. controls (6.95,-1.4) and (3.31,-0.3) .. (0,0) .. controls (3.31,0.3) and (6.95,1.4) .. (10.93,3.29)   ;
\draw    (126.26,184.8) -- (377.38,184.8) ;
\draw [shift={(379.38,184.8)}, rotate = 180] [color={rgb, 255:red, 0; green, 0; blue, 0 }  ][line width=0.75]    (10.93,-3.29) .. controls (6.95,-1.4) and (3.31,-0.3) .. (0,0) .. controls (3.31,0.3) and (6.95,1.4) .. (10.93,3.29)   ;
\draw  [color={rgb, 255:red, 255; green, 0; blue, 0 }  ,draw opacity=1 ] (253.03,143.05) .. controls (251.87,143.7) and (250.76,144.32) .. (250.76,145.05) .. controls (250.76,145.77) and (251.87,146.4) .. (253.03,147.05) .. controls (254.19,147.7) and (255.3,148.32) .. (255.3,149.05) .. controls (255.3,149.77) and (254.19,150.4) .. (253.03,151.05) .. controls (251.87,151.7) and (250.76,152.32) .. (250.76,153.05) .. controls (250.76,153.77) and (251.87,154.4) .. (253.03,155.05) .. controls (254.19,155.7) and (255.3,156.32) .. (255.3,157.05) .. controls (255.3,157.77) and (254.19,158.4) .. (253.03,159.05) .. controls (251.87,159.7) and (250.76,160.32) .. (250.76,161.05) .. controls (250.76,161.77) and (251.87,162.4) .. (253.03,163.05) .. controls (254.19,163.7) and (255.3,164.32) .. (255.3,165.05) .. controls (255.3,165.77) and (254.19,166.4) .. (253.03,167.05) .. controls (251.87,167.7) and (250.76,168.32) .. (250.76,169.05) .. controls (250.76,169.77) and (251.87,170.4) .. (253.03,171.05) .. controls (254.19,171.7) and (255.3,172.32) .. (255.3,173.05) .. controls (255.3,173.77) and (254.19,174.4) .. (253.03,175.05) .. controls (251.87,175.7) and (250.76,176.32) .. (250.76,177.05) .. controls (250.76,177.77) and (251.87,178.4) .. (253.03,179.05) .. controls (254.19,179.7) and (255.3,180.32) .. (255.3,181.05) .. controls (255.3,181.77) and (254.19,182.4) .. (253.03,183.05) .. controls (252.61,183.28) and (252.21,183.51) .. (251.85,183.74) ;
\draw  [color={rgb, 255:red, 255; green, 0; blue, 0 }  ,draw opacity=1 ][fill={rgb, 255:red, 255; green, 0; blue, 0 }  ,fill opacity=1 ] (252.28,141.83) .. controls (252.81,141.4) and (253.58,141.49) .. (254,142.02) .. controls (254.42,142.54) and (254.33,143.31) .. (253.81,143.74) .. controls (253.28,144.16) and (252.51,144.07) .. (252.09,143.54) .. controls (251.67,143.02) and (251.75,142.25) .. (252.28,141.83) -- cycle ;
\draw  [draw opacity=0][line width=1.5]  (262,253.47) .. controls (258.97,253.88) and (255.89,254.09) .. (252.76,254.09) .. controls (214.38,254.09) and (183.26,222.81) .. (183.26,184.23) .. controls (183.26,148.7) and (209.66,119.37) .. (243.81,114.95) -- (252.76,184.23) -- cycle ; \draw  [color={rgb, 255:red, 13; green, 107; blue, 219 }  ,draw opacity=1 ][line width=1.5]  (262,253.47) .. controls (258.97,253.88) and (255.89,254.09) .. (252.76,254.09) .. controls (214.38,254.09) and (183.26,222.81) .. (183.26,184.23) .. controls (183.26,148.7) and (209.66,119.37) .. (243.81,114.95) ;  
\draw  [color={rgb, 255:red, 255; green, 0; blue, 0 }  ,draw opacity=1 ] (252.52,51.39) .. controls (251.17,52.04) and (249.89,52.67) .. (249.89,53.39) .. controls (249.89,54.11) and (251.17,54.74) .. (252.52,55.39) .. controls (253.87,56.04) and (255.16,56.67) .. (255.16,57.39) .. controls (255.16,58.11) and (253.87,58.74) .. (252.52,59.39) .. controls (251.17,60.04) and (249.89,60.67) .. (249.89,61.39) .. controls (249.89,62.11) and (251.17,62.74) .. (252.52,63.39) .. controls (253.87,64.04) and (255.16,64.67) .. (255.16,65.39) .. controls (255.16,66.11) and (253.87,66.74) .. (252.52,67.39) .. controls (251.17,68.04) and (249.89,68.67) .. (249.89,69.39) .. controls (249.89,70.11) and (251.17,70.74) .. (252.52,71.39) .. controls (253.87,72.04) and (255.16,72.67) .. (255.16,73.39) .. controls (255.16,74.11) and (253.87,74.74) .. (252.52,75.39) .. controls (251.17,76.04) and (249.89,76.67) .. (249.89,77.39) .. controls (249.89,78.11) and (251.17,78.74) .. (252.52,79.39) .. controls (253.87,80.04) and (255.16,80.67) .. (255.16,81.39) .. controls (255.16,82.11) and (253.87,82.74) .. (252.52,83.39) .. controls (251.17,84.04) and (249.89,84.67) .. (249.89,85.39) .. controls (249.89,86.11) and (251.17,86.74) .. (252.52,87.39) .. controls (253.87,88.04) and (255.16,88.67) .. (255.16,89.39) .. controls (255.16,90.11) and (253.87,90.74) .. (252.52,91.39) .. controls (251.17,92.04) and (249.89,92.67) .. (249.89,93.39) .. controls (249.89,94.11) and (251.17,94.74) .. (252.52,95.39) ;
\draw  [color={rgb, 255:red, 255; green, 0; blue, 0 }  ,draw opacity=1 ][fill={rgb, 255:red, 255; green, 0; blue, 0 }  ,fill opacity=1 ] (252.01,94.79) .. controls (252.54,94.37) and (253.31,94.45) .. (253.73,94.98) .. controls (254.15,95.51) and (254.07,96.27) .. (253.54,96.7) .. controls (253.01,97.12) and (252.24,97.03) .. (251.82,96.5) .. controls (251.4,95.98) and (251.49,95.21) .. (252.01,94.79) -- cycle ;
\draw  [draw opacity=0][line width=1.5]  (260.4,95.6) .. controls (260.44,97.57) and (259.71,99.56) .. (258.23,101.07) .. controls (255.28,104.09) and (250.45,104.14) .. (247.44,101.19) .. controls (245.98,99.76) and (245.22,97.89) .. (245.15,96) -- (252.78,95.74) -- cycle ; \draw  [color={rgb, 255:red, 255; green, 0; blue, 0 }  ,draw opacity=1 ][line width=1.5]  (260.4,95.6) .. controls (260.44,97.57) and (259.71,99.56) .. (258.23,101.07) .. controls (255.28,104.09) and (250.45,104.14) .. (247.44,101.19) .. controls (245.98,99.76) and (245.22,97.89) .. (245.15,96) ;  
\draw [color={rgb, 255:red, 255; green, 0; blue, 0 }  ,draw opacity=1 ][line width=1.5]    (245.15,49.83) -- (245.15,96) ;
\draw [color={rgb, 255:red, 255; green, 0; blue, 0 }  ,draw opacity=1 ][line width=1.5]    (260.4,49.83) -- (260.4,95.6) ;
\draw [color={rgb, 255:red, 255; green, 0; blue, 0 }  ,draw opacity=1 ]   (249.57,181.33) -- (255.8,187.56) ;
\draw [color={rgb, 255:red, 255; green, 0; blue, 0 }  ,draw opacity=1 ]   (255.8,181.67) -- (249.57,187.9) ;
\draw  [draw opacity=0][line width=1.5]  (243.41,115) .. controls (246.47,114.59) and (249.59,114.38) .. (252.76,114.38) .. controls (291.14,114.38) and (322.25,145.65) .. (322.25,184.23) .. controls (322.25,219.93) and (295.62,249.37) .. (261.23,253.57) -- (252.76,184.23) -- cycle ; \draw  [color={rgb, 255:red, 13; green, 107; blue, 219 }  ,draw opacity=1 ][line width=1.5]  (243.41,115) .. controls (246.47,114.59) and (249.59,114.38) .. (252.76,114.38) .. controls (291.14,114.38) and (322.25,145.65) .. (322.25,184.23) .. controls (322.25,219.93) and (295.62,249.37) .. (261.23,253.57) ;  
\draw  [color={rgb, 255:red, 74; green, 74; blue, 74 }  ,draw opacity=1 ][line width=1.5]  (194.23,155.93) .. controls (190.4,158.63) and (187.73,161.64) .. (186.22,164.95) .. controls (186.57,161.33) and (185.77,157.38) .. (183.81,153.13) ;
\draw  [color={rgb, 255:red, 74; green, 74; blue, 74 }  ,draw opacity=1 ][line width=1.5]  (264.12,72.64) .. controls (262.01,75.16) and (260.75,77.68) .. (260.33,80.21) .. controls (259.91,77.68) and (258.65,75.16) .. (256.55,72.64) ;

\draw (357.49,162.54) node [anchor=north west][inner sep=0.75pt]    {${\mathfrak{Re}\, z}$};
\draw (261.35,31.76) node [anchor=north west][inner sep=0.75pt]    {$\mathfrak{Im}\, z$};
\draw (163.05,144.14) node [anchor=north west][inner sep=0.75pt]    {$C$};
\end{tikzpicture}
     \caption{Contour of integration used to evaluate the contraction $\expval{\texttt{B}_0\texttt{A}_{\pm x}}$ for $h>J$ and $\gamma = 0$.}
     \label{fig:hermitiancontour}
 \end{figure}

\subsection{Limiting Hermitian case}\label{ssec:hermitian_case}
In this section, we calculate the asymptotic behavior of $\texttt{C}^{zz}(x)$ in the limit of no dissipation, i.e., $\gamma \rightarrow 0$. To avoid spurious results (see the discussion in Sec.~\ref{subsec:3A}), this limit must be taken at the level of the non-Hermitian Hamiltonian \eqref{eq:x1}, which leads to a change of sign in the resulting real spectrum $E(k) = \lim_{\gamma \to 0} \Lambda(k)$. We first address the case $h>J$ for which we obtain the contractions \eqref{eq:asb0ax} and \eqref{eq:asb0xma} by implementing the Darboux method. Then, for $h=J$, we obtain the contraction \eqref{eq:asb0axc} by direct integration. Finally, by implementing regulators as in Sec.~\ref{ssec:gapless_phase}, we obtain the relevant contractions \eqref{eq:asb0baxgapless}

Let us consider the limit $\gamma \rightarrow 0$ of Eq.~\eqref{eq:x1}, so we keep  the sign convention of the spectrum, $\lim_{\gamma \to 0}\Lambda(k) = E(k) = 2\,\text{sgn}(J \cos k - h)\sqrt{h^2+J^2-2hJ \cos k}$, where $\text{sgn}(0) = 1$. The resultant state is an excited state of the true Hermitian ground state $H\vert_{\gamma \equiv 0}$, $\ket{\text{GS}}$, where the excitations are Hermitian quasiparticles with $\abs{k}>\pi/2.$ Later, for the sake of completeness, we calculate the same correlation functions in $\ket{\text{GS}}$.

\subsubsection{\texorpdfstring{$E(k) = \lim_{\gamma \to 0} \Lambda(k)$}{\#0}}
For our initial choice of the sign of $E(k)$, $\expval{\texttt{A}_0\texttt{A}_x} \equiv 0$, and 
\begin{equation}\label{eq:hermi}
    \expval{\texttt{B}_0 \texttt{A}_{\pm x}} = -\dfrac{1}{2\pi}\int_{-\pi}^\pi  \dd k\, \left( h\cos k x - J\cos [k(x\mp 1)] \right)\left[\text{sgn}(J \cos k - h)\sqrt{h^2+J^2-2hJ \cos k} \right]^{-1}
\end{equation}
The correlation function simplifies to $\texttt{C}^{zz}(x) = -\expval{\texttt{B}_0 \texttt{A}_x}\expval{\texttt{B}_0\texttt{A}_{-x}}.$ In what follows, we find the asymptotic behavior of Eq.~\eqref{eq:hermi} for $h>J$, $h>J$ and $h = J$. 

Let $h>J$. After some algebra,  Eq.~\eqref{eq:hermi} can be rewritten as
\begin{subequations}
\begin{align}
    \expval{\texttt{B}_0 \texttt{A}_{\pm x}} &= \dfrac{1}{2\pi}\int_{-\pi}^\pi \dd k\, e^{-ik x} \dfrac{ h-J e^{\pm ik } }{\text{sgn}(J \cos k - h)\sqrt{h^2+J^2-2hJ\cos k}}.
\end{align}
\end{subequations}
The sign function is equal to $-1$ for $h>J$. The next step is to analytically extend these integrals to the complex plane by defining $z = i e^{ik}$, 
\begin{equation}\label{eq:contra}
 \expval{\texttt{B}_0 \texttt{A}_{\pm x}} =  \dfrac{i^{x-3/2}}{2\pi}\int_{S^1}\dd z\, z^{-x-1}\left[\sqrt{h/J}-\sqrt{J/h}\,(z/i)^{\pm 1}\right]\left( \sqrt{\dfrac{(z-iJ/h)(z-ih/J)}{z}}\right)^{-1},
\end{equation}
where the integral is over the unit circle $S^1$. We choose the branch cuts such that one starts from $z=ih/J$ to infinity, and the other is in between the branch points $z =0$ and $z=iJ/h$ as shown in Fig.~\ref{fig:hermitiancontour}. We then deform the contour such that it surrounds the outer branch cut, as shown in Fig.~\ref{fig:hermitiancontour} in red. We call this contour $\tilde C$. In what follows, we consider each integral separately. 

Let us choose $+x$ in Eq.~\eqref{eq:contra}:
\begin{equation}
    \expval{\texttt{B}_0\texttt{A}_x} = \dfrac{i^{x-1/2}}{2\pi}\sqrt{\dfrac{J}{h}}\int_{\tilde C}\dd z\, z^{-x-1}(z-ih/J)\left( \sqrt{\dfrac{(z-iJ/h)(z-ih/J)}{z}}\right)^{-1}.
\end{equation}
As we did in the previous sections, make a change of variables  $z = (ih/J) y$ to get
\begin{equation}\label{eq:asb0ax}
    \expval{\texttt{B}_0 \texttt{A}_x} = -\dfrac{1}{\pi}\left(\dfrac{J}{h}\right)^x\int_{1}^\infty \dd y\, y^{-x-1/2}\sqrt{\dfrac{y-1}{y-J^2/h^2}}.
\end{equation}
Upon performing a series expansion around $y=1$ and performing the integral, it gives the asymptotic expansion 
\begin{equation}
    \expval{\texttt{B}_0 A_{x}}\simeq -\dfrac{1}{\sqrt{4\pi}}\dfrac{h}{\sqrt{h^2-J^2}}\left( \dfrac{J}{h}\right)^xx^{-3/2}\left[ 1 + \mathcal{O}(x^{-1})\right].
    \end{equation}
Under similar steps, the asymptotic expansion of $\expval{\texttt{B}_0 \texttt{A}_{-x}}$ is
\begin{subequations}\label{eq:asb0xma}
\begin{align}
    \expval{\texttt{B}_0\texttt{A}_{-x}} &= \dfrac{i^{x-3/2}}{2\pi}\sqrt{\dfrac{h}{J}}\int_{\tilde C}\dd z\, z^{-x-2}(z-iJ/h)\left(\sqrt{\dfrac{(z-iJ/h)(z-ih/J)}{z}} \right)^{-1}\\
    &= \dfrac{1}{\pi}\left(\dfrac{J}{h} \right)^x\int_1^\infty \dd y\,  y^{-x-3/2}\sqrt{\dfrac{y-J^2/h^2}{y-1}} \\
    &\simeq \dfrac{1}{\sqrt{\pi}}\dfrac{\sqrt{h^2-J^2}}{h}\left(\dfrac{J}{h}\right)^xx^{-1/2}\left[1  + \mathcal{O}(x^{-1})\right]
\end{align}
\end{subequations}
This yields Eq.~\eqref{eq:corrher} for $h>J$ and $\gamma \rightarrow 0$.

For  $h = J$, the integrals can be calculated analytically:
\begin{align}\label{eq:asb0axc}
    \expval{\texttt{B}_0 \texttt{A}_{\pm x}} &= -\dfrac{1}{\sqrt{2}\pi}\int_0^\pi \dd k\left( \cos kx \, \sqrt{1-\cos k}\mp \sin kx\, \sqrt{1+\cos k} \right) = \frac{2}{\pi}\dfrac{1}{1\pm 2 x}.
\end{align}
This yields Eq.~\eqref{eq:corrher} for $h = J$ and $\gamma \rightarrow 0$.

Let $h < J$. Here, we use regulators in $\expval{\texttt{B}_0\texttt{A}_{\pm x}}$ as it was done in the non-Hermitian case. The points of interest are $k = 0,\pi$ and $q  = \arccos(h/J)$. However, after a brief calculation, one finds that the only contributing point is $k = q$. Given this, the series expansion at $k=q$ of the functions multiplying $\cos k x$ and $\sin k$, except the sign function, in Eq.~\eqref{eq:hermi} are
\begin{equation}
 -\dfrac{1}{\pi}\dfrac{h-J\cos k}{\sqrt{J^2+h^2-2hJ\cos k}} =  -\dfrac{1}{\pi}(k-q)+\mathcal{O}\textbf{(}(k-q)^2\textbf{)}
\end{equation}
and
\begin{equation}
   \dfrac{1}{\pi}\dfrac{\sin k x}{\sqrt{J^2+h^2-2hJ\cos k}}= \dfrac{1}{\pi} - \dfrac{1}{2\pi}(k-q)^2 + \mathcal{O}\textbf{(}k-q)^3\textbf{)},
\end{equation}
respectively. The last term above is the one contributing to leading order in $1/x$. For $x\gg 1$ and $x$ even 
\begin{subequations}\label{eq:asb0baxgapless}
    \begin{align}
        \expval{\texttt{B}_0 \texttt{A}_{\pm x}} &= -\dfrac{1}{\pi}\int_0^\pi  \dd k\, \dfrac{\cos k x\,(h-J \cos k ) \mp J \sin k x\, \sin k}{\text{sgn}(J \cos k - h)\sqrt{J^2+h^2-2hJ\cos k}},\\
        &\simeq \mp \dfrac{2}{\pi}\lim_{a\to 0}\int_{-\infty}^\infty \dd k\, e^{a \,\text{sgn}(q-k)}\text{sgn}(q-k)\, \sin k x + \mathcal{O}\textbf{(}x^{-2}\cos qx\textbf{)}\\
        &\simeq \mp  \dfrac{2}{\pi}\dfrac{\cos q x}{x}\left[ 1+ \mathcal{O}(x^{-1})\right]. 
    \end{align}
\end{subequations}
Therefore, the asymptotic behavior of the correlation function is given by Eq.~\eqref{eq:corrher} with $h<J$ and $\gamma \rightarrow 0.$
Contributions containing exponential decay in $x$ are present but are subdominant and can be calculated via the Darboux method as it was done for $h>J.$

\subsection{Correlation function in the true Hermitian ground state}\label{ssec:correlation_function_standard}
For the sake of completeness, we calculate the asymptotics of $\texttt{C}^{zz}(x)$ as $x\rightarrow \infty$ in the Hermitian ground state of $H$ when $\gamma\equiv 0$; namely, we set $\gamma = 0$ \emph{before} imposing any sign-convention on the spectrum (cf. Eq.~\eqref{eq:x09}), and then we calculate the correlation function. We resort once more to the Darboux method to calculate it in the ferromagnetic ($h<J$) and paramagnetic phases ($h>J$) and obtain Eq.\eqref{eq:ph}, whereas, in the Ising critical phase ($h=J$), we perform a direct integration and obtain Eq.~\eqref{eq:critical}.

In the ground state of $H$ with $\gamma \equiv 0$ ab-initio, $E(k)\geq 0$ for all $k\in \mathcal K$. Then, the $z$-component of the spin-spin correlation function between the site $m = 0$ and $n = x$ is given by \cite{beginners,lieb_two_1961,PFEUTY197079}
\begin{equation}
    \texttt{C}^{zz}_0(x) = -4 \texttt{G}(x)^2 + 4\abs{ \texttt{F}(x)}^2,
\end{equation}
where
\begin{equation}
   \texttt{G}(x) = \dfrac{1}{2\pi}\int_0^\pi \cos kx \, \dfrac{h-J\cos k}{\sqrt{J^2+h^2-2hJ\cos k}}\dd k
    \label{C57}
\end{equation}
and
\begin{equation}
    \texttt{F}(x) = \dfrac{1}{2\pi}\int_0^\pi \sin kx \, \dfrac{J\sin k}{\sqrt{J^2+h^2-2hJ\cos k}}\dd k
    \label{C58}
\end{equation}
are the elementary Green's function between the sites $0$ and $x>0$ [not to be confused with Eq.~\eqref{eq:app21}]. 
\subsubsection{Critical point}
The Green's function can be analytically calculated at the critical point $h = J$. The functions \eqref{C57} and \eqref{C58} take then the simple form:
\begin{equation}
   \texttt{G}(x) = \dfrac{1}{\pi}\dfrac{1}{1-4x^2},
\end{equation}
and
\begin{equation}
    \texttt{F}(x) = -\dfrac{2}{\pi}\dfrac{1}{1-4x^2}.
\end{equation}
Hence, the exact correlation function is
\begin{equation}\label{eq:critical}
    \texttt{C}^{zz}_0(x) = \dfrac{4}{\pi^2}\dfrac{1}{4x^2-1} \simeq \frac{1}{\pi^2}x^{-2}\left[ 1+\mathcal{O}(x^{-2})\right].
\end{equation}
\subsubsection{Paramagnetic and ferromagnetic phases}
Once more, we implement the Darboux method to calculate the asymptotics of the correlation function in both the paramagnetic ($h>J$) and ferromagnetic ($h<J$) phases. Starting with $G(x)$, we extend the integration domain to $[-\pi,\pi]$ and write it in a slightly more convenient for making use of the differentiation under the integral sign:
\begin{subequations}
\begin{align}
    \texttt{G}(x) &= \pdv{h}\mathfrak{Re}\left\lbrace \dfrac{1}{4\pi}\int_{-\pi}^\pi\dd k\, e^{-ik x}\sqrt{J^2+h^2-2hJ\cos k} \right\rbrace\\
    &=\pdv{h}\mathfrak{Re}\left\lbrace \dfrac{i^{x-1/2}}{4\pi}\sqrt{hJ} \int_{S^1} \dd z z^{-x-1}\sqrt{\frac{(z-z_1)(z-z_2)}{z}}\right\rbrace.
\end{align}
\end{subequations}
In the last line, we analytically extended the integral to the complex plane by setting $z = ie^{ik}$, and the integration contour is the same as in Fig.~\ref{fig:hermitiancontour} with the external branch starting at $z_1= ih/J$ for the paramagnetic phase, and $z_1 = iJ/h$ for the ferromagnetic phase. Remark that for the former case, $z_2 = iJ/h$ and for the latter, $z_2 = i h/J$. For any of the two phases, the contour is then deformed radially to infinity so it surrounds the external branch cut. Making a change of variable $y = z/z_i$ and expanding the integrand, except the dominant term $y^{-x-3/2}$, around $y=1$ yields, for $x\rightarrow \infty$,
\begin{subequations}
\begin{align}
    \texttt{G}(x) &\simeq \pdv{h}\mathfrak{Re}\left\lbrace -\dfrac{i^{x-1/2}}{2\pi}\dfrac{\sqrt{hJ}}{z_1^{x-1/2}} \int_1^\infty \dd y\, y^{-x-3/2}\left[ \sqrt{1-\dfrac{z_2}{z_1}}(y-1)^{1/2} + \dfrac{1}{2}\dfrac{1}{\sqrt{1-z_2/z_1}}(y-1)^{3/2} + \mathcal{O}\textbf{(}(y-1)^{5/2}\textbf{)} \right]\right\rbrace\\
    &= \pdv{h}\mathfrak{Re}\left\lbrace -\dfrac{i^{x-1/2}}{2\pi}\dfrac{\sqrt{hJ}}{z_1^{x-1/2}}\left( \dfrac{1}{2}\sqrt{1-\dfrac{z_2}{z_1}}\sqrt{\pi}\,x^{-3/2} + \dfrac{3}{16}\dfrac{1+z_2/z_1}{\sqrt{1-z_2/z_1}}\sqrt{\pi}x^{-5/2}+ \mathcal{O}\left(x^{-7/2}\right)\right) \right\rbrace.
\end{align}
\end{subequations}
In the paramagnetic and ferromagnetic phases, we have
\begin{equation}\label{eq:gx0}
    \texttt{G}(x)\simeq \begin{cases}
        \dfrac{\partial}{\partial h}\left[ -\dfrac{1}{4\sqrt{\pi}}\sqrt{h^2-J^2}\,x^{-3/2}\left( \dfrac{J}{h}\right)^x - \dfrac{3}{16\sqrt{\pi}}\dfrac{h^2+J^2}{\sqrt{h^2-J^2}}x^{-5/2}\left(\dfrac{J}{h}\right)^x +\mathcal{O}\textbf{(}(J/h)^xx^{-7/2})\textbf{)}\right], \quad &h>J,\\[2ex]
        \dfrac{\partial}{\partial h}\left[ -\dfrac{1}{4\sqrt{\pi}}\sqrt{J^2-h^2}x^{-3/2}\left(\dfrac{h}{J}\right)^x - \dfrac{3}{16\sqrt{\pi}}\dfrac{h^2+J^2}{\sqrt{J^2-h^2}}x^{-5/2}\left(\dfrac{h}{J}\right)^x  +\mathcal{O}\textbf{(}(h/J)^xx^{-7/2})\textbf{)}\right], \quad& h<J.
    \end{cases}
\end{equation}
After taking the partial derivative, 
\begin{equation}
    \texttt{G}(x)\simeq \begin{cases}
       \dfrac{1}{4\sqrt{\pi}}\left(\dfrac{J}{h}\right)^x  \left[\dfrac{\sqrt{h^2-J^2}}{h}x^{-1/2} - \dfrac{5h^2+3J^2}{8h\sqrt{h^2+J^2}}x^{-3/2} + \mathcal{O}(x^{-5/2})\right],\quad&h>J,\\[2ex]
         -\dfrac{1}{4\sqrt{\pi}}\left(\dfrac{h}{J}\right)^x  \left[\dfrac{\sqrt{J^2-h^2}}{h}x^{-1/2} + \dfrac{3h^2-5J^2}{8h\sqrt{h^2+J^2}}x^{-3/2} + \mathcal{O}(x^{-5/2})\right],\quad&h<J.
    \end{cases}
\end{equation}

The Green's function $\texttt{F}(x)$ can also be rewritten in a more convenient form:
\begin{subequations}
\begin{align}
    \texttt{F}(x) &= \dfrac{1}{4\pi}\int_{-\pi}^\pi \sin kx \dfrac{1}{h}\pdv{k}\sqrt{J^2+h^2-2hJ\cos k}\,\dd k\\
    &= -\dfrac{x}{h}\dfrac{1}{4\pi}\int_{-\pi}^\pi \cos kx \, \sqrt{J^2+h^2-2hJ\cos k}\,\dd k,
\end{align}
\end{subequations}
and since this integral above is already calculated in Eq.~\eqref{eq:gx0}, we have
\begin{equation}
    \texttt{F}(x)\simeq \begin{cases}
       \dfrac{1}{4\sqrt{\pi}}\left(\dfrac{J}{h}\right)^x\left[\dfrac{h^2-J^2}{h}x^{-1/2} + \dfrac{3}{8}\dfrac{J^2+h^2}{h\sqrt{h^2-J^2}}x^{-3/2} + \mathcal{O}(x^{-5/2}) \right] \quad& h>J,\\[2ex]
         \dfrac{1}{4\sqrt{\pi}}\left(\dfrac{J}{h}\right)^x\left[\dfrac{J^2-h^2}{h}x^{-1/2} + \dfrac{3}{8}\dfrac{J^2+h^2}{h\sqrt{J^2-h^2}}x^{-3/2} + \mathcal{O}(x^{-5/2}) \right], \quad& h<J.
    \end{cases}
\end{equation}
The correlation function is then
\begin{equation}\label{eq:ph}
    \texttt{C}^{zz}_0(x) \simeq \begin{cases}
        \dfrac{1}{2\pi}\left(\dfrac{J}{h}\right)^{2x}x^{-2} \left[ 1 -\dfrac{1}{8}\dfrac{h^2-3J^2}{h^2-J^2}x^{-1} + \mathcal{O}(x^{-4}) \right], \quad& h>J,\\[2ex]
        \dfrac{1}{2\pi}\left(\dfrac{h}{J} \right)^{2x}x^{-2}\left[1 + \dfrac{1}{8}\dfrac{h^2-3J^2}{h^2-J^2}x^{-1} +\mathcal{O}(x^{-4}) \right], \quad &h<J,
    \end{cases}
\end{equation}
which coincides with previous results \cite{franchini_introduction_2017}.

\section{Stationary state in the thermodynamic limit: Krylov fidelity approach}\label{ssec:stationary_state_continuous}
In this Appendix, we show a detailed calculation of the time \eqref{eq:time} that defines the onset of the stationary-state regime in the thermodynamic limit via the Krylov fidelity \eqref{eq:fi1}, leading to Eqs.~\eqref{eq:t1}-\eqref{eq:t3}.

Let us recall the time-dependent Krylov spread complexity density (in the thermodynamic limit) and its infinite-time limit for a gapped $H$:
\begin{equation}
    \mathcal C(t) = \dfrac{1}{\pi}\int_0^\pi \dd k\, \dfrac{\abs{A_+(k;t)}^2}{1+\abs{A_+(k;t)}^2}  \quad \text{and} \quad \mathcal{C}(t\rightarrow\infty) = \mathcal{C}_\Omega = \frac{1}{\pi}\int_0^\pi \dd k\, \frac{\abs{\tau_k}^2}{1+\abs{\tau_k}^2}.
\end{equation}
Here,
\begin{equation}
    A_+(k;t) = -\dfrac{i[R_x(k)/\Lambda(k)]\sin \Lambda(k)t }{\cos\Lambda(k)t+i[R_z(k)/\Lambda(k)]\sin \Lambda(k)t } = -\dfrac{\tau_k}{1-l_k(t)},  \,\, \tau_k = \frac{v_k}{u_k} = \dfrac{R_x(k)}{\Lambda(k) + R_z(k)},
 \end{equation}
 and
 \begin{equation}
     l_k(t) = \dfrac{2}{f_k(1-e^{i2\Lambda(k) t})}, \quad f_k = 1 + \dfrac{R_z(k)}{\Lambda(k)} = \abs{f_k}e^{i\vartheta_k}.
 \end{equation}
 In what follows, set
 \begin{equation}
 \tilde l_k(t) = \left[1-l_k(t)\right]^{-1}\quad \text{and}\quad w_k(\Omega) = \dfrac{\abs{\tau_k}^2}{1+\abs{\tau_k}^2}.    
 \end{equation}
  For a given $\epsilon \in (0,1)$, we define the following fidelity based on the time-dependent Krylov spread and its infinite-time limit when $H$ is gapped as 
\begin{equation}\label{eq:in1}
\mathcal{F}(t) = \abs{\mathcal C(t)-\mathcal{C}_\Omega} =   \abs{ \dfrac{1}{\pi}\int_0^\pi \dfrac{\abs{\tau_k}^2}{1+\abs{\tau_k}^2}\dfrac{\abs{\tilde{g}_k(t)}^2-1}{1+ \abs{\tau_k}^2\abs{\tilde{g}_k(t)}^2}\,\dd k } = \abs{\dfrac{1}{\pi} \int_0^\pi w_k(\Omega) \, \mathcal{J}_k(t)\, \dd k} < \epsilon.
\end{equation}
The times for which this inequality is satisfied are denoted as $t^*(\epsilon)$, and the stationary state time in the thermodynamic limit is defined for the times $t>t^*(\epsilon).$ From $\mathcal{F}(t)$, we have more explicitly
\begin{multline}\label{eq:i1}
\mathcal{J}_k(t)\coloneqq \dfrac{{\abs*{\tilde{l}_k(t)}^2}-1}{1+\abs{\tau_k}^2\abs*{\tilde{l}_k(t)}^2} = \left( 4\left(\abs{f_k}\cos\vartheta_k - 1 \right) - 4\abs{f_k}e^{2\abs{\Gamma(k)}t}\cos[2E(k)t + \vartheta_k]\right) \\
\Biggl[\left(1+\abs{\tau_k}^2\right)\abs{f_k}^2e^{4\abs{\Gamma(k)}t} + 2\abs{f_k}e^{2\abs{\Gamma(k)}t} \left[2 \cos(2E(k)t + \vartheta_k)-\left(1+\abs{\tau_k}^2\right)\abs{f_k}\cos(2E(k)t) \right] + \left(1+ \abs{\tau_k}^2\right)\abs{f_k}^2\\ + 4(1-\abs{f_k}\cos\vartheta_k)\Biggr]^{-1}.
\end{multline}
For sufficiently long times, $\abs{\mathcal{J}_k(t)} \leq 1$  in a neighborhood of $k = 0$ and $k=\pi$; outside of these neighborhoods, $\mathcal{J}_k(t)$ is quickly suppressed until, in a loosely-defined way,  in the bulk of the integration region, $\mathcal{J}_k(t)$ has the form of a localized wave package with few oscillations and whose amplitude is smaller than unity. In what follows, we determine the approximated contributions of the endpoints and bulk region of the integration interval to  $\mathcal{F}(t)$.

Near $k=0,\pi$, $f_k$ behaves like
\begin{equation}
    f_k \approx \dfrac{8J^2k^2}{\left[4(J-h)-i\gamma\right]^2},\quad \text{and} \quad f_k\approx \dfrac{8J^2(k-\pi)^2}{\left[4(h+J) + i \gamma\right]^2},
\end{equation}
respectively. Around the same points, we have
\begin{equation}
    1+\abs{\tau_k}^2 \approx a_0 + \dfrac{a_{-2}}{k^2}, \quad \text{and} \quad   1+\abs{\tau_k}^2 \approx b_0 + \dfrac{b_{-2}}{(k-\pi)^2},
\end{equation}
respectively, where $a_{-2}, b_{-2} \in \mathbb{R}$.
Moreover,
\begin{equation}
   (1+\abs{\tau_k}^2)\abs{f_k}^2 \approx \dfrac{16J^2k^2}{16(h-J)^2+\gamma^2}. 
\end{equation}
and similarly for $k = \pi$. Hence,  $f_k\rightarrow 0$ and $\mathcal J_k(t)\rightarrow -1$ as $k\rightarrow 0,\pi$. For convenience, let us also note that
\begin{equation}
R_z(0) = i\dfrac{\gamma}{2} + 2(h-J)\, \quad R_z(\pi) = i \dfrac{\gamma}{2} + 2(h+J),
\end{equation}
and
\begin{equation}
    \Lambda(0) = -2(h-J) - i\dfrac{\gamma}{2}, \quad \Lambda(\pi) = -2(h+J) - i\dfrac{\gamma}{2}.  
\end{equation}

Given all the above, for large times, the series expansion of $\mathcal{J}_k(t)$ around $k=0$ is, to leading order in $k$,
\begin{equation}
    \mathcal{J}_k(t) \approx {\mathcal{J}}_{k,0}(t) = \dfrac{-4}{4+(1+\abs{\tau_k}^2)\abs{f_k}^2e^{4\abs{\Gamma(k)}t}} \approx -\dfrac{4}{4+\dfrac{16J^2k^2}{16(h-J)^2+\gamma^2}e^{2\gamma t}}.
\end{equation}
Around the same point, $w_k(\Omega)\approx 1$. Hence, the approximated contribution to $\mathcal{F}_k(t)$ from the points close to $k=0$ is
\begin{equation}\label{eq:an0.1}
    I_0(t) = -\dfrac{1}{\pi}\int_0^\infty  \dd k\,  \mathcal{J}_{k,0}(t)= -\dfrac{\sqrt{16(h-J)^2+\gamma^2}}{4J}e^{-\gamma t}.
\end{equation}
Similarly, the contribution from the points close to $k=\pi$ is
\begin{equation}\label{eq:an0.2}
    I_\pi(t) =  -\dfrac{\sqrt{16(h+J)^2+\gamma^2}}{4J}e^{-\gamma t}.
\end{equation}
We refer to $I_{0,\pi}(t)$ as the boundary contributions. 

Let us return to $\mathcal{J}_k(t)$. Again, for sufficiently long times, but now outside the neighborhoods of $k=0$ and $k=\pi$, we have
\begin{equation}
\mathcal{J}_k(t)\approx    \mathcal{J}_{b,k}(t)\coloneqq \dfrac{4\left(\abs{f_k}\cos \vartheta_k-1\right)e^{-4\abs{\Gamma(k)}t} - 4\abs{f_k} e^{-2\abs{\Gamma(k)}t}\cos(2E(k)t + \vartheta_k)}{(1+\abs{\tau_k}^2)\abs{f_k}^2}.
\end{equation}
After using Eq.~\eqref{eq:ax2} in $w_k(\Omega)\mathcal{J}_{b,k}(t)$, we get the two ``bulk'' contributions
\begin{equation}\label{eq:X}
      I_{b,1}(t) \coloneqq \int_{0}^\pi \dd k\, \left( \dfrac{1/2 - \Gamma(k)/\gamma}{1/2 + \Gamma(k)/\gamma}\right) \dfrac{2\Gamma(k)}{\gamma \pi}e^{4\Gamma(k)t} = \int_{0}^\pi \dd k\, X(k) e^{4\Gamma(k)t}B(k),
\end{equation}
and
\begin{equation}\label{eq:Y}
   I_{b,2}(t)\coloneqq -\mathfrak{Re}\left\lbrace 
    \int_{-\pi}^\pi \dd  k\, \dfrac{2}{\gamma \pi}\,\left(\dfrac{1}{2} - \dfrac{\Gamma(k)}{\gamma}\right) \left[ \dfrac{\gamma(h-J\cos k)-i\Gamma^2(k)}{2(h-J\cos k)- i\Gamma(k)} \right]  e^{i2E(k) t+ 2\Gamma(k) t}\right\rbrace = \mathfrak{Re}\left\lbrace 
    \int_{-\pi}^\pi \dd k \, Y(k) e^{i2E(k) t+ 2\Gamma(k) t}\right\rbrace.
\end{equation}
In $I_{b,1}(t)$, $B(k)$ is a well-behaved function that approaches zero faster than $\abs{X(k)}$ approaches infinity, so the overall contributions of the endpoints to the integral are negligible. 

We proceed to find the leading terms of the asymptotic expansions of the two preceding integrals by first noting that $\text{max}_{k \in \mathcal K^+}\, \Gamma(k) = \Gamma(\bar k)$, where
\begin{equation}
    \bar k = \arccos\left( \dfrac{16J h}{16h^2+\gamma^2} \right),
\end{equation}
and  $\Gamma''(\bar k)< 0$. Thus, 
\begin{align}\label{eq:an0.3}
    I_{b,1}(t) \approx \frac{1}{2}X(\bar k )e^{4\Gamma(\bar k )t}\int_{-\infty}^\infty \dd k\, e^{-4\abs*{\Gamma''(\bar k )}t(k-\bar k)^2}= \dfrac{1}{4}X(\bar k) e^{4\Gamma(\bar k)t}\sqrt{\dfrac{\pi}{\abs{\Gamma''(\bar k)}t}}.
\end{align}

The situation with $I_{b,2}(t)$ is a bit more delicate as the argument of the exponential inside the integral is complex. To implement the steepest-descent method, we analytically continue the integral with $k\rightarrow k +i\kappa$ and search for the $k+i\kappa\in \mathbb{C}$ such that the quadratic expansion of $2it\Lambda(k+i\kappa)^*$ is valid. This requirement translates into the following two conditions:  first, 
\begin{equation}\label{eq:an0}
    \abs{\left(\Lambda'(\bar k )\right)^*(\tilde k - \bar k) + \dfrac{1}{2}\left(\Lambda''(\bar k) \right)^*(\tilde k -\bar k)^2}< \abs{\dfrac{1}{3!}\left(\Lambda^{(3)}(\bar k)\right)^*(\tilde k -\bar k )^3},
\end{equation}
and second,
\begin{equation}\label{eq:an1}
    \abs{\dfrac{2t}{3!}\left(\Lambda^{(3)}(\bar k) \right)^*(k^*(t)-\bar k)^3}<1,
\end{equation}
where $\tilde k \in \mathbb{C}$ is the complex mode where $\nabla_{(k,\kappa)} \, \mathfrak{Re}\,i \Lambda(k+i\kappa )^* = 0$, and $k^*(t) \in \mathbb{C}$ is any mode satisfying Eq.~\eqref{eq:an1}, i.e., any point lying on the circle centered at $\tilde k$. This circle is the boundary of the disk containing the complex modes that contribute the most to the quadratic expansion of $2i\Lambda(k+i\kappa)^*t = 2\Gamma(k+i\kappa)t + 2 iE(k+i\kappa).$ It turns out that the region obtained from Eq.~\eqref{eq:an0} satisfies the condition \eqref{eq:an1}.  Given these points, we approximate $I_{b,2}(t)$ as
\begin{align}
    I_{b,2}(t) &\approx -2\, \mathfrak{Re}\, \left\lbrace Y(\bar k ) e^{2iE(\bar k )t + 2\Gamma(\bar k ) t}  \int_{-\infty}^\infty \dd k\, \exp\left(i2E'(\bar k )tk + [\Gamma''(\bar k ) + i E''(\bar k )]tk^2 \right)\right\rbrace\\
    &= -\mathfrak{Re}\left\lbrace 2Y(\bar k)\sqrt{\dfrac{\pi}{t \abs{\Lambda''(\bar k)}}}\exp(\gamma_Y t)\exp(i\omega_Y t)\exp[-\dfrac{1}{2}i\left( \Lambda''(\bar k) \right)^*]\right\rbrace,
\end{align}
where
\begin{equation}\label{eq:gammay}
    \gamma_Y = 2\Gamma(\bar k ) + \frac{E'(\bar k )^2\Gamma''(\bar k )}{\abs{\Lambda''(\bar k)}^2},\quad \text{and}\quad \omega_Y = 2\Gamma(\bar k ) -\frac{E'(\bar k )^2E''(\bar k )}{\abs{\Lambda''(\bar k)}^2}
\end{equation}
are the effective decay and frequency associated with the values of $\gamma$ and $h$ defined in Eq.~\eqref{eq:an0}. 

Overall, for sufficiently long times and small $\epsilon$, the fidelity has four main contributions
\begin{equation}\label{eq:ff1}
    \mathcal F (t) \approx \abs{I_0(t) + I_\pi(t) + I_{b,1}(t) + I_{b,2}(t)}< \epsilon,
\end{equation}
where each has a characteristic dissipation rate: $I_{0}(t)$ and $I_\pi(t)$ are characterized by $\gamma$ [see Eqs.~\eqref{eq:an0.1}-\eqref{eq:an0.2}]; $I_{b,1}(t)$ by $4\abs{\Gamma(\bar k )}$; and $I_{b,2}(t)$ by $\abs{\gamma_Y}$. We define the stationary state time in the continuum limit as follows: for $H$ gapped (i.e., $\Gamma(k)\neq 0$ for all $k\in \mathcal{K}$), for a given $\epsilon \in (0,1)$ and for sufficiently long times, there is a time $t(\epsilon)$ for which $\mathcal{F}(t(\epsilon)) < \epsilon$, such that for $t > t(\epsilon)$, $\mathcal{C}(t) \simeq \mathcal{C}_\Omega$ denotes the stationary state (or asymptotic) behavior of the spread complexity in the thermodynamic limit. By comparing the dissipation rates associated with each contribution in Eq.~\eqref{eq:f1}, one gets three different dynamical regions in $(h,\gamma)\subset [0,\infty)\times (\gamma_c(h),\infty)$ where one of the three dissipation rates are the slowest one. The decays characterizing the regions will coincide at the boundaries of such regions (except with $\gamma = \gamma_c(h)$ and $\gamma=0$).  Once we identify each of the regions, we define the $\epsilon$-independent time \eqref{eq:time},
which is a continuous function over $(h,\gamma)\subset [0,\infty)\times (\gamma_c(h),\infty)$, but its derivatives with respect to $h$ or $\gamma$ are discontinuous at the boundaries separating regions 1 to 3 shown in Fig.~\ref{fig:4}. Hence, we must understand $t^*$ in ``units'' of $\abs{\ln \epsilon}$. Region $1$ corresponds to $\gamma$; 2 to $4\abs{\Gamma(\bar k )}$; and 3 to $\gamma_Y$. We omit the formulas describing the boundaries, as they are solutions to multivariate polynomials of degrees greater than 10. Nevertheless, the times corresponding to the regions are
 \begin{equation}
        \mathcal{F}(t) \approx \abs{I_0(t)+I_\pi(t)}<\epsilon \ \Rightarrow \ t(\epsilon) > \dfrac{1}{\gamma}\ln\left[ \dfrac{4J\epsilon}{\sqrt{16(h+J)^2+\gamma^2}+\sqrt{16(h-J)^2+\gamma^2}} \right].
    \end{equation}
for region 1; 
 \begin{equation}
        \mathcal{F}(t) \approx \abs{I_{b,1}(t))} < \epsilon \ \Rightarrow \ t(\epsilon) > \dfrac{1}{8\abs{\Gamma(\bar k)}}W_0\left( \dfrac{\pi X^2(\bar k )\abs{\Gamma(\bar k)}}{2\epsilon^2\abs{\Gamma''(\bar k )}}\right).
    \end{equation}
for region 2; and
  \begin{equation}
        \mathcal{F}(t) \approx \abs{I_{b,2}(t)} < \epsilon \ \Rightarrow \ t(\epsilon) >\dfrac{1}{2\abs{\gamma_Y}}W_0\left( \dfrac{8\pi \abs{Y(\bar k )}^2\abs{\gamma_Y}}{\epsilon^2 \abs{\Lambda''(\bar k )}}\right).
    \end{equation}
for region 3. These times yield Eqs.~\eqref{eq:t1}-\eqref{eq:t3}, respectively.

\end{widetext}
\end{appendix}

\bibliography{articles.bib}

\end{document}